\documentclass[usenatbib]{mn2e}

\usepackage{graphicx}
\usepackage{amssymb}
\usepackage{amsmath}
\usepackage{color}
\usepackage{bm}
\usepackage{xspace}
\usepackage{pgffor}
\usepackage{upgreek}
\usepackage{nicefrac}
\usepackage[normalem]{ulem}
\usepackage[single=true]{acro}
\usepackage[colorlinks,urlcolor=blue,citecolor=blue,linkcolor=blue,draft]{hyperref}
\usepackage{wrapfig}

\renewcommand{\upi}{\uppi}
\newcommand{\be}{\begin{equation}}
\newcommand{\ee}{\end{equation}}
\newcommand{\nn}{\mbox{} \nonumber \\ \mbox{} }
\newcommand{\ba}{\begin{eqnarray}}
\newcommand{\ea}{\end{eqnarray}}

\newcommand{\kms}{\ensuremath{\rm km\,s^{-1}}}
\newcommand{\ergs}{\ensuremath{\rm erg\,s^{-1}}}

\newcommand{\B}{\bm{B}}

\renewcommand{\v}{\bm{v}}

\newcommand{\ve}{\ensuremath{\varepsilon}}

\newcommand{\br}{\ensuremath{\bm{r}}}

\newcommand{\bb}{\ensuremath{\bm{\beta}}}

\newcommand{\Db}{\delta}
\newcommand{\G}{\ensuremath{\Gamma}}

\newcommand{\curl}{{\rm curl\, }}

\newcommand{\ts}{\times}

\newcommand{\ort}[1]{\ensuremath{\bm{\hat #1}}}

\makeatletter
\def\dif{\@ifnextchar[{\@with}{\@without}}
\def\@with[#1]#2{
  \ensuremath{\frac{\foreach \x in {#2}{\mathrm{d}\x\,}}{\foreach \x in {#1}{\mathrm{d}\x\,}}}
}
\def\@without#1{
  \ensuremath{%
    \ifx\hfuzz#1\hfuzz
    \mathrm{d}
    \else
    \foreach \x in {#1}{\mathrm{d}\x\,}
    \fi
    }
}
\makeatother

\newcommand\etal{{\itshape et al.}}
\newcommand{\eg}{{\itshape e.g.}\xspace}
\newcommand\cf{{\itshape cf.}\xspace}
\newcommand\ie{{\itshape i.e.}\xspace}

\newcommand{\fer}{{\itshape Fermi} LAT\xspace}

\newcommand{\Bf}{{magnetic field}\xspace}
\newcommand{\Bfs}{{magnetic fields}\xspace}
\newcommand{\Ef}{{electric  field}\xspace}

\newcommand{\xray}{X-ray\xspace}
\newcommand{\xrays}{X-rays\xspace}

\newcommand{\LC}{light cylinder\xspace}
\newcommand{\Lf}{Lorentz factor\xspace}

\DeclareAcronym{mhd}{
  short = MHD ,
  long  = magnetohydrodynamics ,
  class = hydro ,
  first-style = default 
}
\DeclareAcronym{rmhd}{
  short = RMHD ,
  long  = \acifused{mhd}{relativistic MHD}{relativistic magnetohydrodynamics} ,
  class = hydro ,
  first-style = default
}
\DeclareAcronym{3d}{
  short = 3D ,
  long  = three dimensional ,
  class = hydro ,
  first-style = default
}
\DeclareAcronym{2d}{
  short = 2D ,
  long  = two dimensional ,
  class = hydro ,
  first-style = default
}

\DeclareAcronym{cd}{
  short = CD ,
  long  = contact discontinuity ,
  class = hydro ,
  first-style = default
}
\DeclareAcronym{ts}{
  short = TS ,
  long  = termination shock ,
  class = hydro ,
  first-style = default
}
\DeclareAcronym{eos}{
  short = EOS ,
  long  =  equation of state ,
  long-plural-form = equations of state ,
  class = hydro ,
  first-style = default
}
\DeclareAcronym{pwn}{
  short = PWN,
  long  = pulsar wind nebula,
  short-plural = e,
  long-plural = e,
  class = astro,
  first-style = default
}
\DeclareAcronym{cr}{
  short = CR ,
  long  = cosmic ray ,
  class = astro ,
  first-style = default
}
\DeclareAcronym{ns}{
  short = NS ,
  long  = neutron star ,
  class = astro ,
  first-style = default
}
\DeclareAcronym{ism}{
  short = ISM ,
  long  = interstellar medium ,
  class = astro ,
  first-style = default
}
\DeclareAcronym{cmb}{
  short = CMB ,
  long  = cosmic microwave background ,
  class = astro ,
  first-style = default
}
\DeclareAcronym{ic}{
  short = IC ,
  long  = inverse Compton ,
  class = astro ,
  first-style = default
}
\DeclareAcronym{mec}{
  short = MEC ,
  long  = monochromatic emission coefficient ,
  class = astro ,
  first-style = default
}
\DeclareAcronym{gem}{
  short = Geminga ,
  long  =  PSR~B0633$+$1,
  class = pwn ,
  first-style = default
}
\DeclareAcronym{j1509}{
  short = PSR~J1509$-$5850 ,
  long  =  PSR~J1509$-$5850 ,
  class = pwn ,
  first-style = empty
}
\DeclareAcronym{j1741}{
  short = PSR~J1741$-$2054 ,
  long  =  PSR~J1741$-$2054 ,
  class = pwn ,
  first-style = empty
}
\DeclareAcronym{j1135}{
  short = PSR~J1135$-$6055 ,
  long  =  PSR~J1135$-$6055 , 
  class = pwn ,
  first-style = empty
}
\DeclareAcronym{mouse}{
  short = Mouse ,
  long  =  PSR~J1747$-$2958 ,
  class = pwn ,
  first-style = default
}
\DeclareAcronym{mushroom}{
  short = `Mushroom' ,
  long  =  PSR~B0355$-$54 ,
  class = pwn ,
  first-style = default
}

\newcommand{\caselong}[1]{``#1''}
\DeclareAcronym{rb}{
  short = rifle-bullet ,
  long  =  Rifle Bullet, 
  class = case ,
  long-format = \caselong , 
  first-style = empty
}
\DeclareAcronym{cw}{
  short = cart-wheel ,
  long  =  Cart Wheel, 
  class = case ,
  long-format = \caselong , 
  first-style = empty
}
\DeclareAcronym{fb}{
  short = frisbee ,
  long  =  Frisbee, 
  class = case ,
  long-format = \caselong , 
  first-style = empty
}

\DeclareAcronym{naoj}{
  short = NAOJ ,
  long  =  National Astronomical Observatory of Japan  ,
  single = National Astronomical Observatory of Japan (NAOJ) ,
  class = institution ,
  first-style = default
}

\newcommand{\TS}{\ac{ts}\xspace}
\newcommand{\CD}{\ac{cd}\xspace}

\newcommand{\ISM}{\ac{ism}\xspace}
\newcommand{\NS}{\ac{ns}\xspace}
\newcommand{\MEC}{\ac{mec}\xspace}
\newcommand{\Rs}{\acp{pwn}\xspace}
\newcommand{\R}{\ac{pwn}\xspace}
\newcommand{\mhd}{\ac{mhd}\xspace}
\newcommand{\rmhd}{\ac{rmhd}\xspace}

\newcommand{\ns}{\textsc{ns}}
\newcommand{\ism}{\textsc{ism}}
\newcommand{\syn}{\textsc{syn}}
\newcommand{\obs}{\textsc{obs}}
\newcommand{\tsh}{\textsc{ts}}
\newcommand{\cd}{\textsc{cd}}
\newcommand{\nt}{\textsc{nt}}

\newcommand{\apj}{ApJ}
\newcommand{\apjs}{ApJS}
\newcommand{\apjl}{ApJ}
\newcommand{\mnras}{MNRAS}

\newcommand{\aap}{A\&A}
\newcommand{\nat}{Nature}

\newcommand{\araa}{ARA\&A}
\newcommand{\apss}{Ap\&SS}

\newcommand{\ssr}{SSRv}

\newcommand{\aplett}{Astrophys. Lett.}
\newcommand{\rikkyo}{{Department of Physics, Rikkyo University, Nishi-Ikebukuro 3-34-1, Toshima-ku, Tokyo 171-8501, Japan}}
\newcommand{\Sec}{Section~}

\title[3D dynamics and morphology of bow-shock PWN\lowercase{e}]{3D dynamics and morphology of bow-shock Pulsar Wind  Nebulae}
\author[Barkov \etal]{Maxim V. Barkov$^{1,2,3}$\thanks{Correspondence author: mbarkov@purdue.edu (MVB)},
Maxim Lyutikov$^{1}$ and Dmitry Khangulyan$^{4}$
 \\
$^{1}$ Department of Physics and Astronomy, Purdue University, West Lafayette, IN 47907-2036, USA\\
$^{2}$ Astrophysical Big Bang Laboratory, RIKEN, 351-0198 Saitama, Japan \\
$^{3}$ Space Research Institute of the Russian Academy of Sciences (IKI), 84/32 Profsoyuznaya Str, Moscow, Russia, 117997 \\
$^{4}$\rikkyo
}

\begin{document}
\date{Received/Accepted}
\maketitle

\begin{abstract}
  Bow-shock \Rs show a variety of morphological shapes. We attribute this diversity to the geometrical factors: relative
  orientations of the pulsar rotation axis, proper velocity, and the line of sight (magnetic inclination angle may also
  have certain influence on the morphology). We identify three basic types of bow-shock nebulae: (i) a \acl{rb} (pulsar
  spin axis and proper velocity are aligned); (ii) a \acl{fb} (pulsar spin axis and proper velocity are orthogonal with
  the spin axis lying in the plane of the sky), and (iii) a \acl{cw} (like \ac{fb} but the spin axis is perpendicular to
  the plane of the sky).  Using \acs{3d} \acs{rmhd} simulations, as well as analytical calculations, we reproduce
  the key morphological features of the bow-shock \Rs, as well as variations seen across different systems. Magnetic
  stresses within the shocked pulsar wind affect the overall structure strongly, producing ``whiskers'', ``tails'',
  ``filled-in'' and ``mushroom'' shapes, as well as non-symmetric morphologies. On the other hand, the interstellar medium
  inhomogeneities and the anisotropy of the energy flux in the pulsar wind have only a mild impact of the \R
  morphology.  In a few cases, when we clearly identify specific morphological structures, our results do not favor alignment
  of the pulsar spin axis and proper velocity. Our calculations of the underlying emission processes explain the low
  synchrotron \xray efficiency (in terms of the spin-down luminosity) and imply an energetically subdominant
  contribution of the \acl{ic} process.
  % \glsresetall
  \acresetall
\end{abstract}

\begin{keywords}
ISM -- magnetic fields: ISM -- jets and outflows: magnetic reconnection: MHD: pulsars -- individual: Geminga
\end{keywords}

\maketitle

%%%%%%%%%%%%%%%%%%%%%%%%%%%%%%%%%%%%%%%%%%%%%%%%%%
\section{Introduction}
%%%%%%%%%%%%%%%%%%%%%%%%%%%%%%%%%%%%%%%%%%%%%%%%%%
\label{intro}
Pulsars produce  relativistic magnetized winds that create {\aclp{pwn} \citep[\acsp{pwn},][]{reesgunn,2006ARA&A..44...17G,2008AIPC..983..171K,2015SSRv..191..391K,2017SSRv..207..175R}}. A distinct type of \Rs is produced by  
fast moving pulsars that quickly escape from the supernova remnant  \citep[for a recent review see][]{kpkr17}.
Typical pulsar velocities of hundreds kilometers per second are much higher than the typical sound speeds in the \ISM, $c_{s, \ism} = 10-100 \,\kms$ - pulsars 
are moving with highly supersonic velocities.  The interaction of the pulsar wind with the \ISM produces a bow-shock nebula with an extended tail.

\begin{figure*}
\includegraphics[width=55mm,angle=-0]{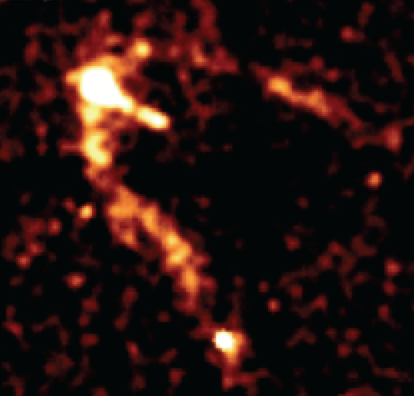}
\includegraphics[width=55mm,angle=-0]{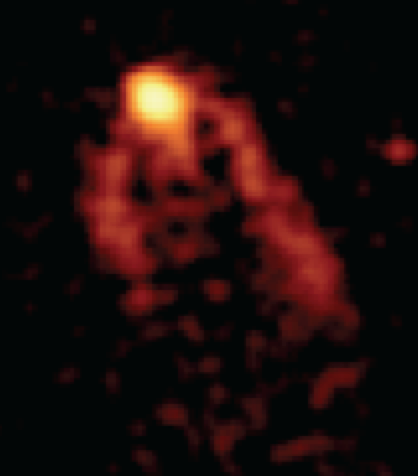}
\includegraphics[width=55mm,angle=-0]{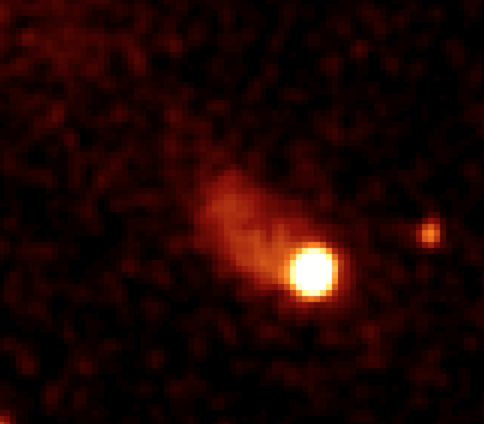}
\includegraphics[width=55mm,angle=-0]{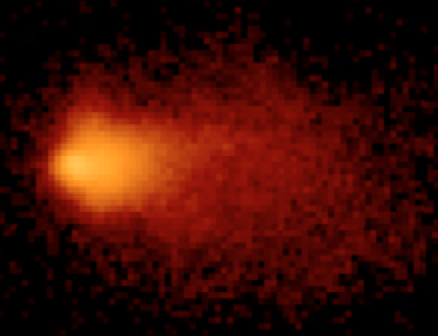}
\includegraphics[width=55mm,angle=-0]{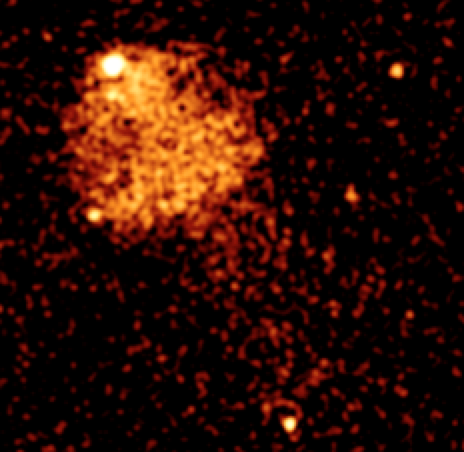}
\includegraphics[width=55mm,angle=-0]{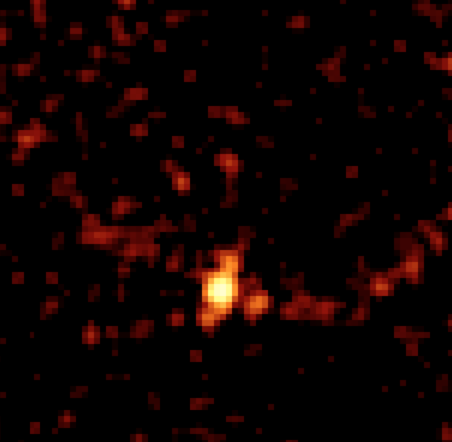}
\caption{ Examples of morphological variations of bow-shock \Rs. Top row: \ac{gem} (left) and \ac{j1509} (center) \citep{kkr16}  \Rs show ``whiskers with tail'' 
morphology (these are prototypes of \ac{fb} geometry ---see Fig.~\ref{fig:Y_Ebd}, models fs3a45, fs1a10 and fs1a45---);  \ac{j1741} \citep[right panel][]{asr15} 
show filled-in morphology (this is a prototype of \ac{cw} geometry (see Fig.~\ref{fig:Z_Ebd}, models fs3a45, fs1a10 and fs1a45).
Bottom row:  wide-head outflows in  \ac{mouse}  \citep[left,][]{kpkr17} and \ac{mushroom} \citep[center,][]{krkp16}, which can be prototype of \ac{rb} geometry (see Fig.~\ref{fig:Z_Ebd}, 
model bs1a45; and Fig.~\ref{fig:b_Ebd}, Y-axis projection of bs01a45). The cap of \ac{mushroom} is formed by the inverted front ``jet'' and the equatorial outflow, the stalk is formed by the back ``jet''. Asymmetrical jets  \ac{j1135}  \citep[right,][]{mar12}, which can be prototype of \ac{fb}~--~\ac{rb} geometry (see Fig.~\ref{fig:Z_Ebd}, 
model fbs1a45).  }
\label{fig:gem_1509}
\end{figure*}

Analytical and \ac{2d} hydromagnetic models (in what we call \ac{rb} configuration, see below) \citep[\eg][]{Wilkin96,Bucciantini02a,2005MNRAS.358..705B,bad05,2018arXiv180306240T} or \ac{3d} hydrodynamic simulation
\citep{2007MNRAS.374..793V} of bow-shock \Rs
 predict the formation of a smooth two-shock structure: a forward shock in the \ISM separated by a \CD from a \TS in the pulsar wind.  
Contrary to these expectations, the observed \Rs show large variations in  morphologies -  sometimes  filled-in tails and sometimes edge-brightened bow shocks  (``whiskers''), 
sometimes jet-like feature extend from a pulsar along the tail \citep[][and Figs. \ref{fig:gem_1509}]{kpkr17}.
\footnote{
In addition,  some bow-shock \Rs show ``kinetic jets'' - elongated feature extending well beyond the shock-confined \R. \cite{blkb18}  
interpreted these as kinetic flow of particles that escaped the \R via reconnection between the internal and external \Bfs\citep[see also][]{2008A&A...490L...3B}. In this paper we concentrate on the \acs*{mhd} and will not discuss the kinetic jets further.
}
One of the first attempts to build a radiation model for such systems was done by \citet{2017SSRv..207..235B}, who developed a sophisticated non-thermal particle acceleration and radiation model relying on a basic analytical 
hydrodynamic description.  Also the morphology of the bow shock was reproduced by simplified analytical model by 
\cite{2017ApJ...851...61R}.

Observationally, we can distinguish four characteristic morphological classes of \Rs created by fast moving pulsars: ({\it i}) \ac{gem} and  \ac{j1509} 
are prototypes of ``three jets''  (or ``whiskers with a tail'')
\Rs, Fig.~\ref{fig:gem_1509}, left and center panels in top row; ({\it ii}) \ac{j1741} is a  ``head - thin jet'' \R, Fig.~\ref{fig:gem_1509}, right panel in top row; ({\it iii}) 
\ac{mouse} and \ac{mushroom}
show  ``wide head  - thin tail'' outflow, Fig.~\ref{fig:gem_1509},  left and center panels in bottom row; and  ({\it iv}) \ac{j1135} can be a prototype of more general class of 
\Rs with asymmetrical 
jets, Fig.~\ref{fig:gem_1509},  right panel in bottom row.

%fffffffffffffffffffffffffffffffffffffffffffffffffffffffffffffffff
%fffffffffffffffffffffffffffffffffffffffffffffffffffffffffffffffff 

%fffffffffffffffffffffffffffffffffffffffffffffffffffffffffffffffff
\begin{figure}
\includegraphics[width=.99\linewidth]{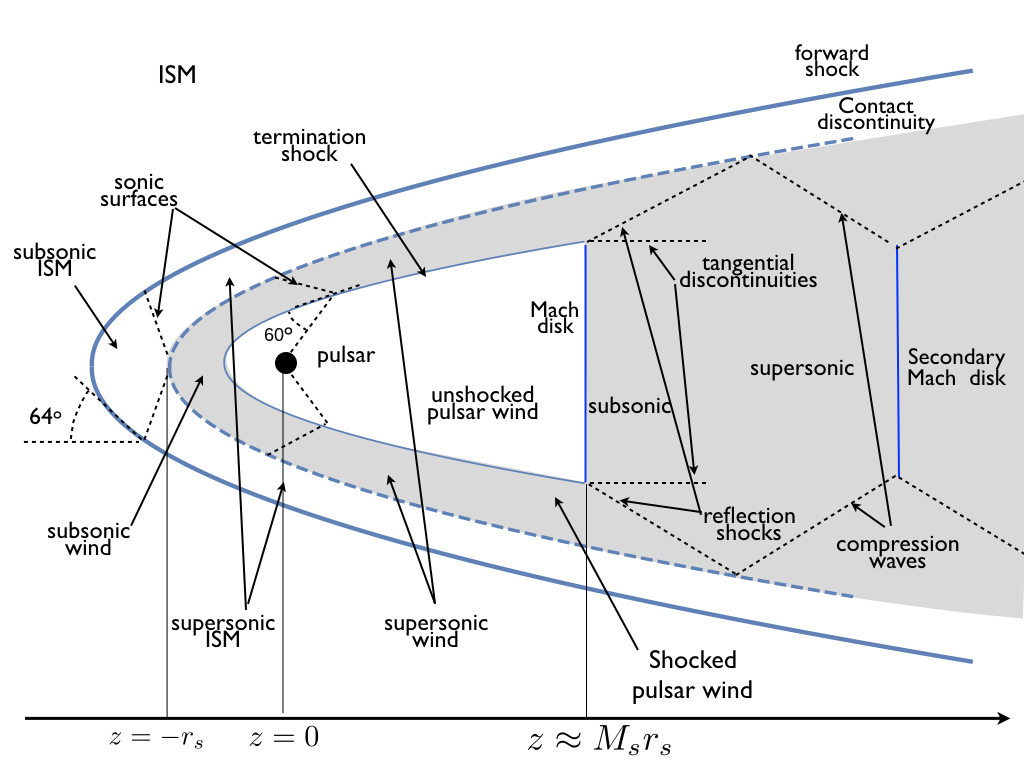}
\caption{Expected internal structure of ram pressure-confined \Rs. Pulsar wind is shocked at the elongated termination shock, which forms a Mach disk at the back end. Near the apex point the post-shock velocity is subsonic, while at larger distances the flow always remains supersonic. The flow accelerates within the conical shells outside of the termination shock to mildly relativistic velocities. Behind the Mach disk the flow is highly subsonic and over-pressurized with respect to the flow along the sides. Over-pressurized central part expands sideways, bounded by reflection shocks. It overshoots the equilibrium pressure; reflection shocks turn into converging compression waves, that over-pressurize the central part of the flow, and lead to the formation of  another Mach disk and so on. %Formation of the Mach disks can be repetitive.
}
\label{fig:headflow}
\end{figure}
%fffffffffffffffffffffffffffffffffffffffffffffffffffffffffffffffff

%{\bf BMV: Maxim, could you rewrite this paragraph?}
These peculiar tail shapes have been interpreted as the result of density inhomogeneities in the \ISM \citep{Romani97,2007MNRAS.374..793V}.  However, we find this explanation unsatisfactory: ({\it i}) variations in the external density can affect the structure of the forward shock and the \CD, yet it is hard to see how they can change the internal structure of \Rs~--~\eg, some tails show filled-in morphology, while others show a short central tail surrounded by edge-brightened halos; ({\it ii}) as we demonstrate in \Sec\ref{Kompaneets}, in order to produce appreciable variations of the overall shape of bow-shock \Rs the external density variations should have large amplitudes and occur on scales much smaller than the stand-off distance; ({\it iii}) different \Rs have similar morphological features, \eg, around \ac{gem} and \ac{j1509} pulsars, Fig.~\ref{fig:gem_1509}.  % many  morphological variations  repeat from \R to \R

From these observations we conclude that the {\it  peculiar morphological features result from the internal dynamics of the pulsar wind}, rather than through inhomogeneities in the \ISM. 
In this paper, we  demonstrate that magnetic stresses in the shocked \R flow strongly modify the structure of the flow and the morphology of \Rs. The structure of the \Bf within the shocked pulsar wind depends on the relative orientation of pulsar velocity, rotation axis, and magnetic inclination. Change of these parameters, as well as the direction of the line of sight, lead to a diversity of morphologies, that, generally, reproduces the observations. 

 In \Sec\ref{analytical} we describe the internal structure of bow-shock \Rs, and in \Sec\ref{Kom} we calculate  expected shapes of \CD analytically for a variety of approximations for anisotropic pulsar wind and density variations in the \ISM.    
We discuss various possible orientation of the pulsar rotation axis relative to the proper motion and the structures of current in 
the pulsar magnetopause tail in \Sec\ref{sc:geom}. 
Results of the \ac{3d} \ac{rmhd} numerical simulation are presented in \Sec\ref{sc:numset}. In \Sec\ref{s:em} we compute synchrotron emission and produce synthetic brightness maps. We compare our numerical results  with 
{\it Chandra} \xray observations in \Sec\ref{particular}.  In \Sec\ref{Discussion} we discuss limitations of the approach and possible extensions of the model and 
in \Sec\ref{sc:conclusion} we summarize our results.

\section{Bow-shock \Rs: analytical considerations}
\label{analytical}
\subsection{Overall properties}

%{\bf BMV: I think this section can have place in the introduction. That do you think?}

Pulsars eject relativistic winds with power
\be 
L_w \sim B_\ns^2 R_\ns^2 c  ( \Omega R_\ns/c)^4\,,
\ee
where \(B_\ns\), \(R_\ns\), and \(\Omega\) are pulsar surface \Bf, radius, and angular velocity, respectively; \(c\) is the light velocity \citep{GoldreichJulian,1969ApJ...158..727M,Spitkovsky06}. 
If a pulsar, producing such a wind, moves through the \ISM with a proper velocity, \(V_\ns\), exceeding  the sound speed in \ISM,  \(c_s\), significantly, \ie $V_\ns\gg c_s $, a two-bow-shock structure is formed, Fig.~\ref{fig:headflow}. The space between the bow shocks is filled with shocked gas and a \CD separates relativistic gas that originates in the pulsar from the shocked \ISM. Let us discuss the salient properties of this configuration. Since in the pulsar reference frame the flow is steady, we determine sizes of characteristic features as their separation from the pulsar. % measure the distances from the pulsar.  % use this frame, thus we assume that at large distances the \ISM moves with constant velocity \(V_\ns\).

\begin{itemize}
\item {\it The head part}.

  In the head part the \ISM ram pressure confines
the pulsar wind  producing two shocks -- forward shock in the \ISM and the \TS in the pulsar wind -- 
separated by the \CD.   The shocked pulsar wind forms a pulsarsheath (in analogy with heliosheath) - 
a region of shocked  pulsar wind material bounded by the \CD.  On the outside of \CD there is an \ISM sheath - a region
of shocked \ISM plasma bounded by the forward shock. 
Qualitatively this picture resembles interaction of the 
Solar wind with the Local \ISM \citep[see][for review]{zank99}. 
The pressure balance between the pulsar wind and the \ISM gives the stand-off distance $r_s$
\be
r_s =\sqrt{\frac{L_w }{ 4 \upi c  \rho_\ism V_\ns^2}} = 4 \times 10^{16}  L_{w,36}^{1/2} \, n_{\ism,0}^{-1/2} V_{\ns,7.5}^{-1} \, {\rm cm}\,,
\label{rs}
\ee
where \(\rho_\ism=m_pn_\ism\) is \ISM mass and number  density (here \(m_p\) is mass of proton). We use the following normalization agreement: \(A=10^xA_x\rm\,cgs\,units\).

For highly supersonic pulsar proper velocities  $V_\ns /c_s \equiv M_s \gg 1$
(here $M_s$ is the Mach number with respect to the external medium) 
the forward shock is perpendicular to the flow velocity only at the apex point (as a result the shocked flow there is always subsonic in the pulsar frame).   Away from the apex point, 
where the forward shock front makes a sufficiently small angle with the flow velocity, 
$\phi < \arcsin \sqrt{(\gamma_\ism +1)/(2 \gamma_\ism)}$,  \citep[as can be obtained for the case \(M_s\gg1\) from the shock polar equation, see][here $\gamma_\ism$ is 
the adiabatic index of the \ISM]{LLVI} the shocked \ISM flow
remains subsonic (although the shock is still strong).  %Further away the \ISM  flow  remains supersonic.

Similarly, the pulsar wind goes through the \TS and becomes subsonic close to the apex point. The high pressure near the
apex re-accelerates the flow to supersonic (mildly relativistic) velocities \citep[similar effect is seen in simulations
of binary pulsar systems, see \eg][]{2008MNRAS.387...63B,2012A&A...544A..59B}.  Further way from the apex point, the
relativistic pulsar wind passes through an oblique shock and may remain supersonic \citep[for conditions at relativistic
magnetized oblique shocks see][]{2002AstL...28..373B,2016MNRAS.456..286L}, Fig.~\ref{fig:headflow}.  For example, for a
weakly magnetized ultrarelativistic pulsar wind the post shock flow bulk Lorentz factor is
\(\Gamma = 3/(\sqrt{8} \sin\phi)\) (here \(\phi\) is angle between the unshocked wind velocity and the \TS).  Thus, if the angle between the radially moving pulsar wind and the \TS front
becomes smaller than $60^\circ$, the post-\TS flow remains supersonic.

\item  {\it Tailward Mach disk}.  In the tailward region, 
the \TS forms  a closed surface streatching far behind the pulsar.
 For strongly supersonic pulsar motion, $M_s \gg 1$,
 the ram pressure of the wind $L_w$ equals the \ISM pressure $p_\ism$ at
\be
r_\textsc{m} \approx  r_s M_s \approx 4  \times 10^{17}  \, L_{w,36}^{1/2} \, n_{\ism,0}^{-1/2} c_{s,6.5}^{-1} \, {\rm cm}\,.
\label{rM}
\ee
% where 
% $
% M_s = {  V_\ns/  c_s}
% $
% is the sonic Mach number of the pulsar motion in the \ISM. 
Distance $r_\textsc{m}$ provides an estimate for the location of the Mach disk in the tail region.  Thus, the \TS in the pulsar wind locates further from the pulsar (by a factor $\sim M_s \gg 1$) in the tail region as compared to the head region.

%{Note, that  the formation of the tail-ward Mach disk in expanding winds is unavoidable, due to the decreasing pressure in the wind - this is qualitatively different from jet flows from the airplane nozzles or from the case when a collimated flow is  injected into the medium.}

\item  {\it  Post-Mach disk expansion}.
After passing through  the tailward  Mach disk the corresponding part of the flow is strongly heated. At the same time, the part of the flow at the edges of the \R has low pressure - it has been spent on flow bulk acceleration. 
As a result, the post  Mach disk flow is  under-expanded. The pressure balance is reached through reflection shocks (Prandtl-Meyer expansion waves), Fig.~\ref{fig:headflow}. (The reflection shock may, in principle, affect the shape of the \CD as well.)  Thus, the
overall evolution of the flow in the tail resembles behavior
 of under-expanded plume in  rocket exhaust nozzle 
 \citep[\eg,][]{thom71}. Regions of flow expansion, 
 mediated by Prandtl-Meyer expansion waves, are  followed by the flow
 compressions,  Fig.~\ref{fig:headflow}.
In the compression regions the pressure  ``overshoots'' the ambient pressure, so the  flow becomes
  over-pressurized again. 
    The process of expansion-compression wave formation begins anew,
  until the dissipation damps the oscillations. 
Thus, we expect  a repeated formation of Mach disks in the tail.
Overall, the flow  remains mildly relativistic, changing from weakly subsonic
  to weakly supersonic. 

\end{itemize}

\section{Head structure:   anisotropic winds and external density  gradients}
\label{Kom}
\subsection{ The Kompaneets approximation}
\label{Kompaneets}

In this section we consider the shape of the bow-shock \Rs in the limit of highly supersonic proper velocity of the pulsar, $M_s \gg1 $.
In this case,  one can use  the thin-shell approximation  --- an expansion
of hydrodynamic equations in the  small parameter:  the inverse of the compression
ratio $\eta_{\textsc{comp}}=(\gamma_\ism+1)/(\gamma_\ism-1)$ \citep{bkk71,Wilkin96}. In the limit $\eta_{\textsc{comp}}\gg1$, the thickness of the shocked 
layer is negligible, \ie  the  forward shock coincides with the \CD. 

In the limit, $M_s \gg 1 $,  the head structure of the forward shock  is well understood in the case of non-relativistic  spherically symmetric winds \citep{bkk71,1975Ap&SS..35..299D,Wilkin96}. The pressure balance on the \CD involves both the ram pressure of the \ISM and that of the wind, as well as centrifugal corrections due to  the motion of shocked material, both from the \ISM and the wind, along the curved \CD.  Typically, the centrifugal corrections  are  minor \citep{1975Ap&SS..35..299D}. Although due to the relativistic \ac{eos} the pulsar wind may have a large inertia imposing significant dynamical differences as compared to stellar winds \citep{2012A&A...544A..59B}, for sake of simplicity we neglect the centrifugal contribution here.

\begin{figure}
\includegraphics[width=.99\linewidth]{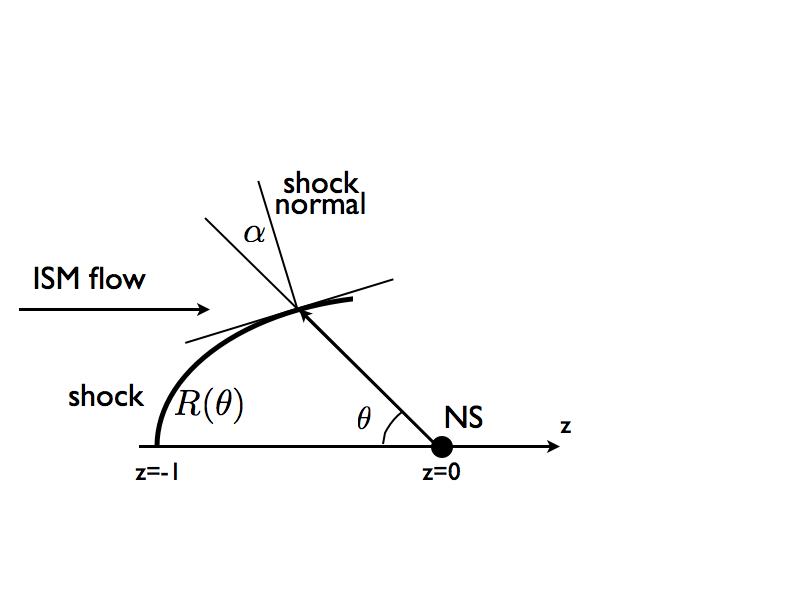}
\caption{Geometry of the thin-shell (Kompaneets) approximation. Note that for the shown configuration the sign of the angle \(\alpha\) is negative: \(\alpha<0\).}
\label{Kompaneets1}
\end{figure}

Let $R(\theta)$ be the shape of the \CD. Then the angle between radial direction and normal to the \CD is (Fig.~\ref{Kompaneets1})
\be
\tan \alpha = \partial_\theta \ln R
\ee
%
%fffffffffffffffffffffffffffffffffffffffffffffffffffffffffffffffff
%fffffffffffffffffffffffffffffffffffffffffffffffffffffffffffffffff 
  
Equating normal pressures \citep[the Kompaneets approximation,][]{Komp} 
\be
\frac{ L_w (\theta) }{ 4  \upi c R^2} \cos^2 \alpha = \rho_\ism V_\ns^2 \cos^2 ( \alpha - \theta)
\label{kkom}
\ee
we find
\be
 \cos \theta  R+  \sin \theta \partial_\theta R = \sqrt{f(\theta)/g(\theta)}
 \label{compa}
\ee
where we assume anisotropic wind power $L_w= L_0 f(\theta)$, $\int  f(\theta) { d \Omega /( 4\upi)} =1$, possible variations of the external density along the shock, $
\rho_\ism =g(\theta)\rho_{\ism,0} $ and normalize all the distances to the stand-off distance given by  Eq.~(\ref{rs}) (where \(L_w\) is substituted by \(L_0\)).

Equation \ref{compa} determines the head structure under the thin-shell approximation. 
We use this equation to obtain the head structures for various anisotropies of the pulsar wind and external density gradients.

\subsection{Bow-shock shapes}

\subsubsection{Isotropic pulsar wind}
For an isotropic wind and constant-density medium one has  $f=g=1$, and Eq.~(\ref{compa}) gives
\be
R= \frac{ \theta }{ \sin \theta}
\label{Riso}
\ee
\citep{1975Ap&SS..35..299D}.
In this case, the shape at apex,  $\theta \rightarrow 0$, is $R \approx  1 + \theta^2 /6$ \citep[this result is only slightly different form the case of two colliding non-relativistic winds  considered by][,  with the centrifugal corrections taken into account; in that case for small angles $R \approx  1 + \theta^2 /5$]{Wilkin96}.

\subsubsection{Anisotropic pulsar wind}

%When the spin of the pulsar is aligned with the velocity (what we call ``bullet'' geometry, see below) angle $\theta_p$  equals $\theta$.

\begin{figure}
% \begin{center}
\includegraphics[width=.49\linewidth]{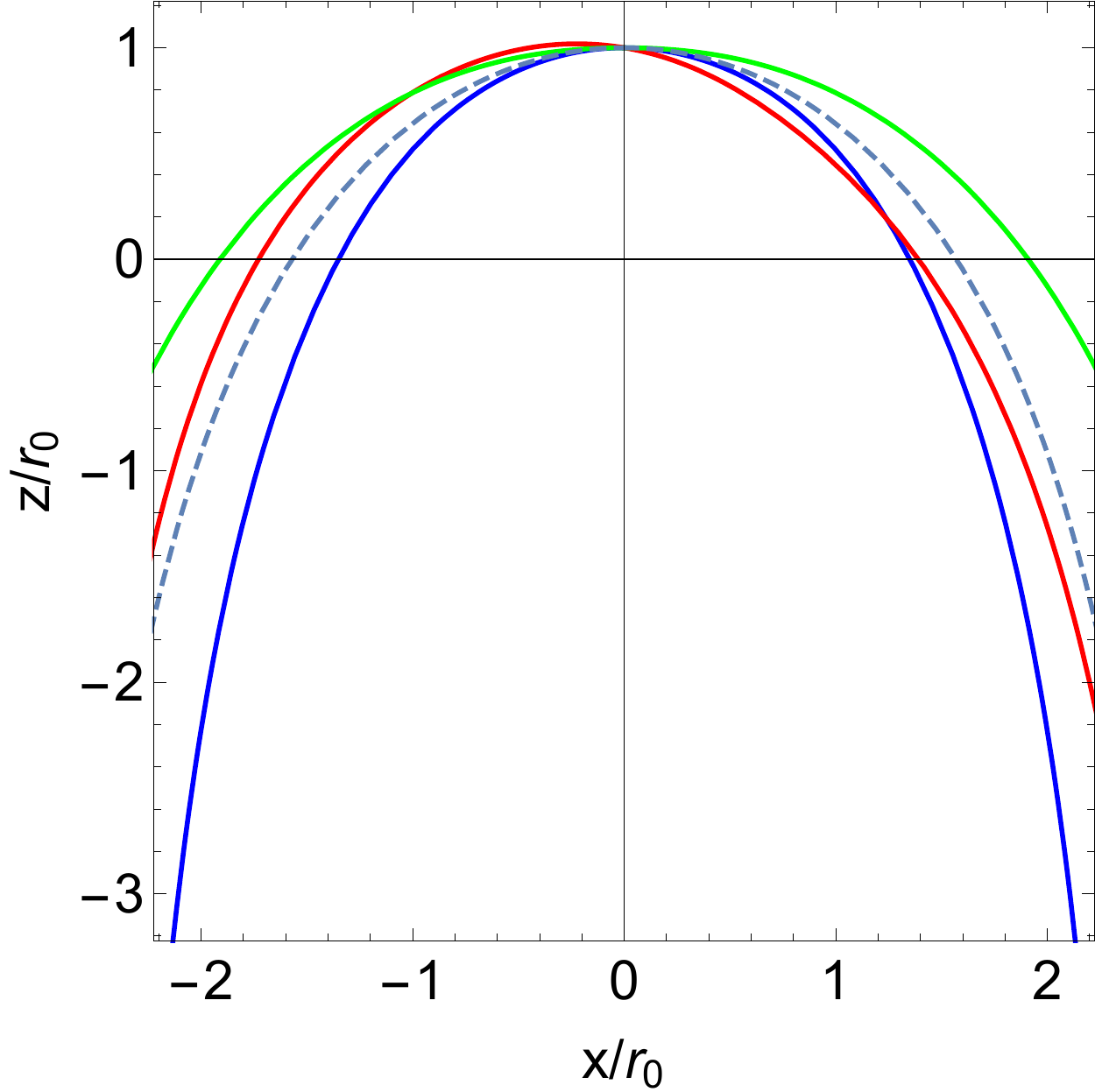}
\includegraphics[width=.49\linewidth]{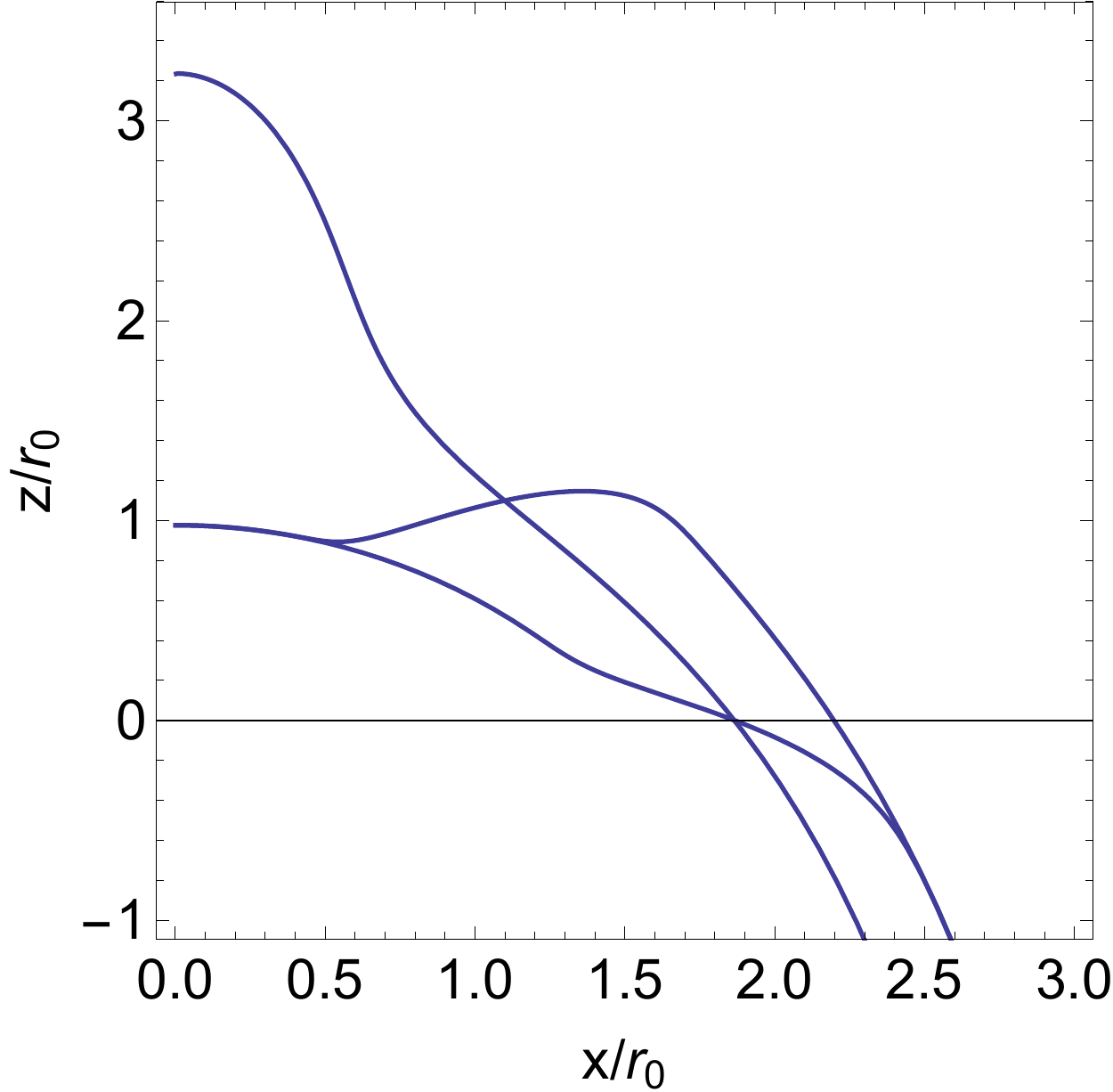}
%\end{center}
%\caption{Analytic shapes  of the wind-\ISM boundary for anisotropic wind  power $f= 1+  A \sin ^2 (\theta - \theta_j) $  for $A=1$  (Left Panel) and $A=2$ (right panel) and three orientations of the  axis, $\theta_j = 0, \upi/4, \upi/2$, plus a case of isotropic wind $A=0$. Distances are normalized to the stand-off distance of the isotropic wind.}
\caption{Left panel: Analytic shapes  of the wind-\ISM boundary for anisotropic pulsar wind with  $f\propto 1+  A \sin ^2 (\theta - \theta_j) $  for $A=1$ and three orientations of the pulsar rotation axis, $\theta_j = 0, \upi/4, \upi/2$ {(green, red, blue correspondingly)}, plus a case of isotropic wind $A=0$ (dashed lines).
Right panel: Shapes  of the wind-\ISM boundary for wind with a jet. The jet peak power is 10 times the isotropic value, opening angle is $0.1$ rad,    
and three orientations of the pulsar rotation axis, $\theta_j = 0, \upi/4, \upi/2$, are shown. Even a very narrow and powerful jet produces only a mild deformation of the \CD shape.}
\label{head-jet-analytic-jet}
\end{figure}

Pulsars produce anisotropic winds with energy flux presumably depending as $\propto\sin ^2 \theta_p$, where $\theta_p$ is the polar angle with respect to the pulsar rotation axis \citep{1969ApJ...158..727M,2002MNRAS.336L..53B}. In order to estimate the resulting shock deformations
we assume that the pulsar wind is anisotropic with the energy flux determined by 
\ba &&
f= C (1+ A \sin^2(\theta-\theta_j))
\nn &&
C=  \frac{1 }{ 1 + 2A/ 3}  
\ea
where $\theta_j$ is the projection of the angle between the \NS velocity and the spin on the plane of the sky. Results of calculations for different parameters $A$ and  $\theta_j$ are presented in  Fig.~\ref{head-jet-analytic-jet}, left panel.
Overall, the  pulsar wind anisotropy produces only mild variations of the bow-shock shape.

To illustrate the point that the wind anisotropy cannot produce sharp bow-shock features, let us assume that the pulsar wind consists of an isotropic outflow plus a jet directed in the plane of the sky at the angle $\theta_j$ and having a Gaussian
profile with width $\Delta \theta$ and relative power at the maximum $A_j$ (in terms of isotropic angular power), Fig.~\ref{head-jet-analytic-jet}. The point is that even a very narrow and powerful jet produces only a mild variation of the \CD shape.

\subsubsection{External density gradient}

Let us next assume that a pulsar, which ejects an isotropic wind, propagates across medium with a density gradient. We consider two cases,  (i) the gradient is perpendicular to the pulsar velocity (so that in Eq. (\ref{kkom}) $\rho \equiv \rho ( R(\theta) \sin \theta) $ - this results in a non-axial-symmetric \CD shape)
and (ii) the gradient is along the velocity (so that in Eq. (\ref{kkom}) $\rho \equiv \rho ( R(\theta) \cos \theta) $ -  this produces kinks in the shape of the \CD).

Since we are interested in the overall impact of the density inhomogeneity, for convenience we first consider a perpendicular gradient given by
\be
\rho = \rho_0 \left(1+ \frac{\sqrt{\eta_\rho}-1}{\sqrt{\eta_\rho}+1} \tanh \left(\frac{R \sin (\theta )}{x_{\rho }}\right) \right)^2\,. 
\label{rr}
\ee
The density contrast, from its minimum at large negative \(R\sin(\theta)\) to a maximum at large positive \(R\sin(\theta)\), is given by \(\eta_\rho\). Parameter \(x_\rho\) is the characteristic length over which the density changes.  

For such density profile the shape of the \CD in the $x-y$ plane is
\be
y=x \cot \left(x+ \frac{\sqrt{\eta_\rho}-1}{\sqrt{\eta_\rho}+1} x_{\rho } \log \left(\cosh \left(\frac{x}{x_{\rho }}\right)\right)\right)\,. 
\ee

\begin{figure*}
 \begin{center}
\includegraphics[width=.49\linewidth]{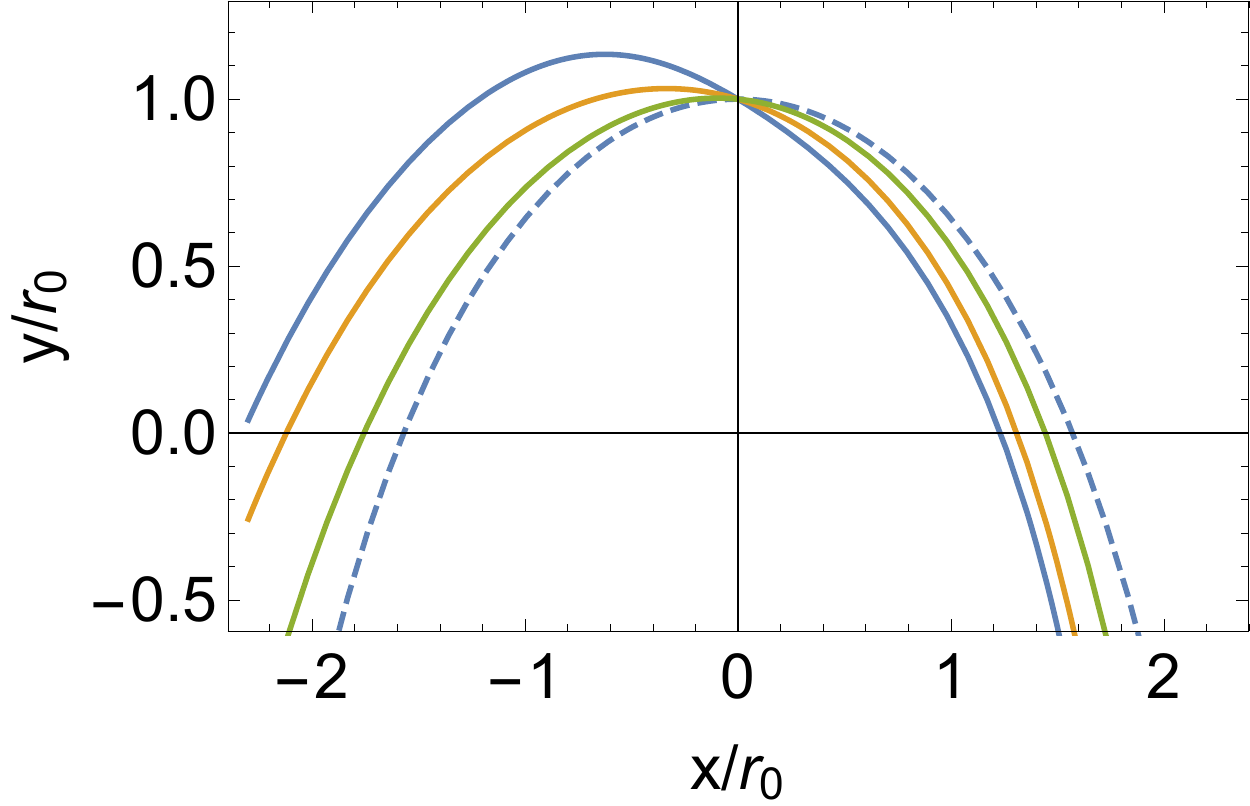}
\includegraphics[width=.49\linewidth]{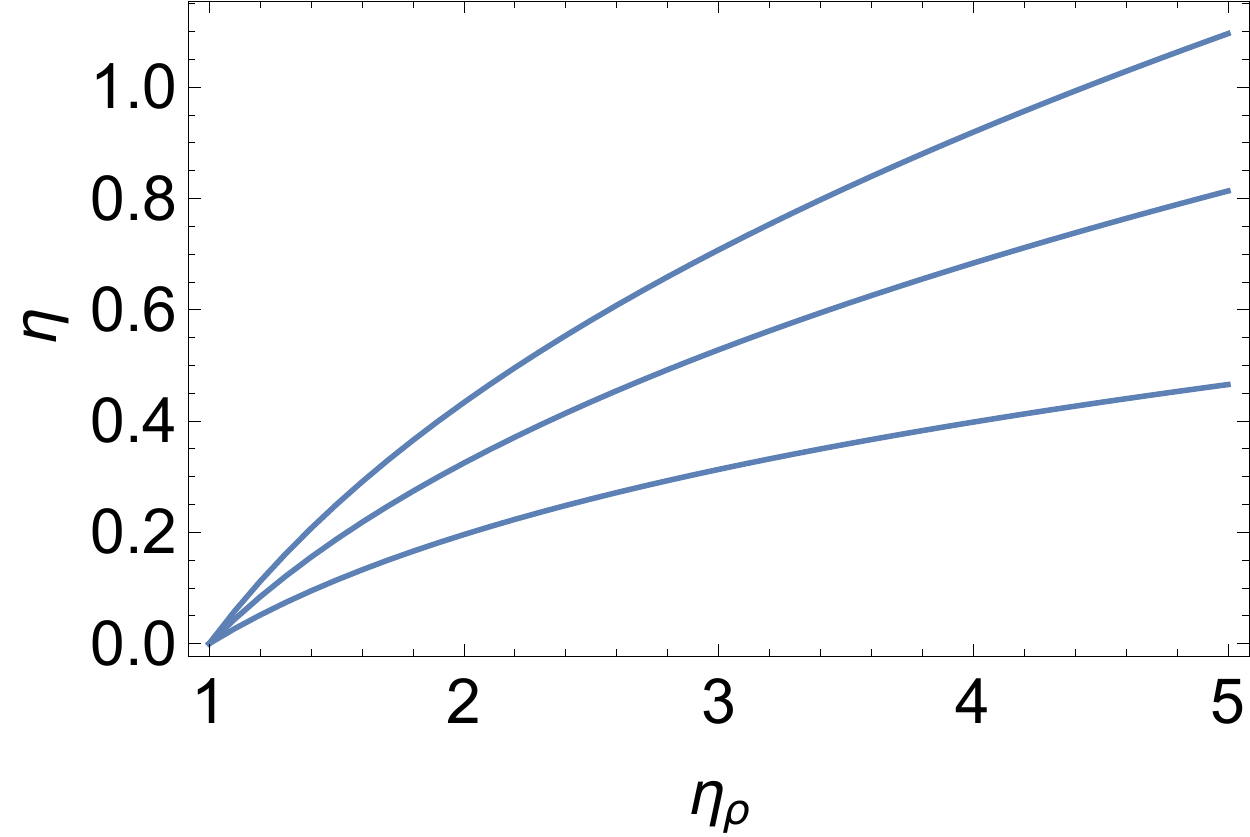}
\end{center}
\caption{{\it Left Panel}. Shapes  of \CD for pulsar propagating in \ISM with the density given by Eq.~(\ref{rr}). The density contrast was $\eta_\rho=5$ and typical scales 
$x_{\rho } = 0.5, \, 1\, 2$ (smaller $x_{\rho }$ correspond to more skewed shapes). {\it Right  Panel}. Skewness parameter $\eta$, Eq.~(\ref{eta}),  
as a function of the density ratio $\eta_\rho$. }
\label{fig:thead-jet-1}
\end{figure*}

To  characterize the anisotropy  of the shock we use the following parameter:
\be
\eta  =  \left|\frac{ R( \upi/2) - R( -\upi/2) }{ R(0)} \right|
\label{eta}
\ee
(see Fig.~\ref{fig:thead-jet-1}).

\begin{figure}
 \begin{center}
\includegraphics[width=.99\linewidth]{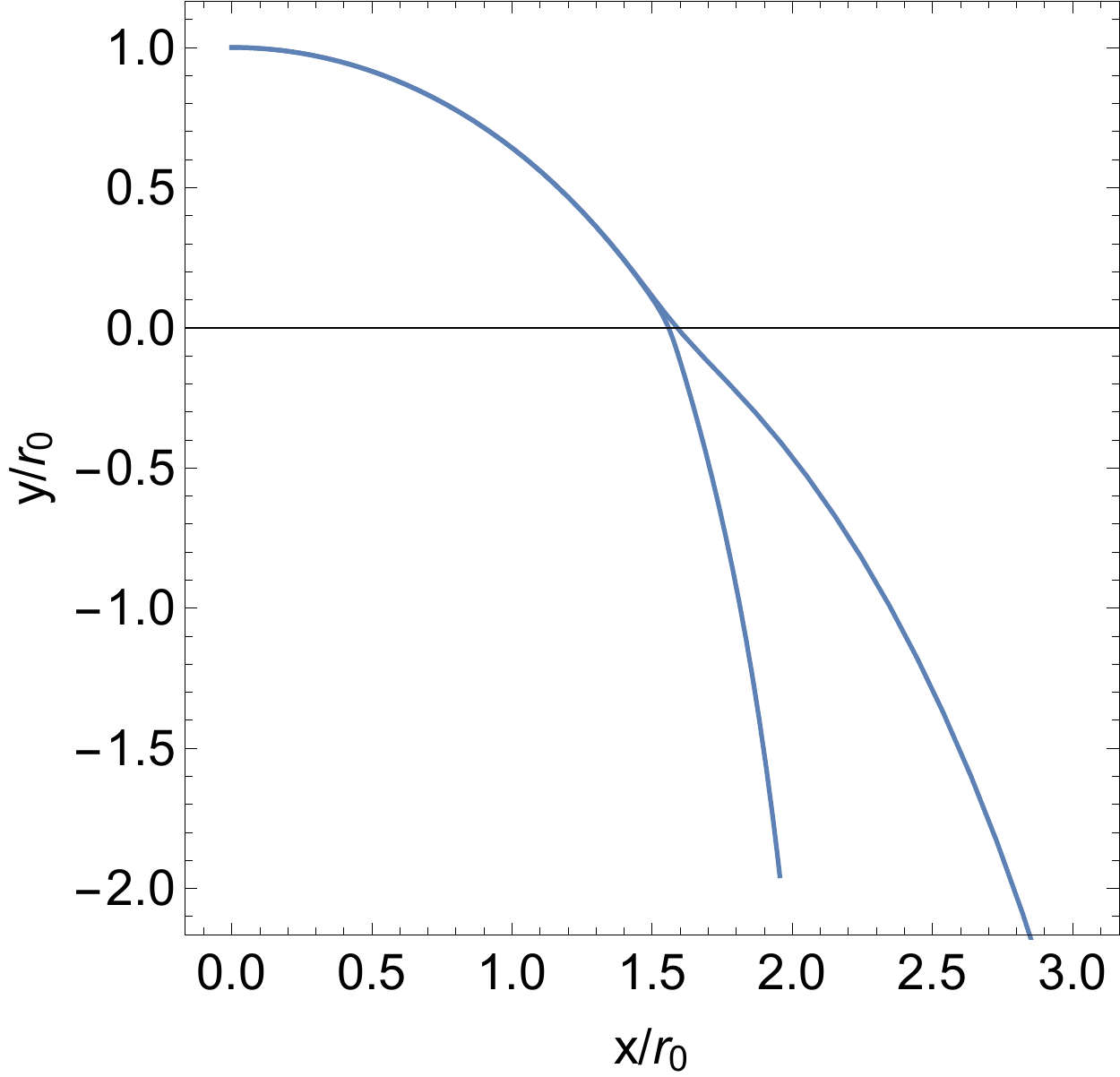}
\end{center}
\caption{Shape of \CD for a pulsar propagating along the \ISM density gradient. At the point $y=0$ the density experiences a jump by a factor of $2$ (two curves correspond to increasing and decreasing density) on the scale of $0.1$ of the stand-off distance. For larger scales of density variations the   change in the shape of the boundary is almost unnoticeable.}
\label{fig:thead-jet-2}
\end{figure}

We also perform similar calculations for the density gradient along the pulsar motion, Fig.~\ref{fig:thead-jet-2}. In this case, a sharp variation of the density, on a scale much smaller than the stand-off distance, can produce ``kinks'' in the shapes of the wind-\ISM boundary.

We conclude that in order to produce appreciable distortions of the form of the wind-\ISM boundary due to wind/external density anisotropy,  it is required that either a very large density variation occurs on scales of the order of the stand-off distance, 
or a variation by a factor of $\sim 2$ occurring on much smaller spacial scales. Otherwise,  variations of the external density produce  only mild distortions of the shape of the bow shock, which would be indistinguishable given the typical observational uncertainties. 
If the \ISM density varies by a factor of two on the scale of the stand-off distance 
(typically $10^{16} $ cm), the resulting variations of the shape are only $15\%-25\%$.  This seems to be the upper limit since we assumed that the pulsar moves in the plane of the sky.  
Thus, we conclude that neither the intrinsic wind anisotropies nor the external density variations can explain the observed variations in the bow-shock \Rs morphologies.

\section{Magnetic fields in the bow-shock \Rs}
\label{sc:geom}

\subsection{Properties of pulsar wind before the \TS: anisotropy and  the equatorial sheath}

The structure of \Bfs in the  bow-shock \Rs is, generally, a complicated transformation the wind \Bf (which is determined by processes in the pulsar magnetosphere and wind zone) by the interaction of the wind with the \ISM.  
The formation of wind occurs on the scale of a  light cylinder  radius, which is much smaller than the stand-off distance, and the formation of pulsar wind in the bow-shock \Rs is not affected by the processes taking place in the \R. Thus, one can expect that the properties of pulsar winds in \Rs formed by fast- and slow-moving pulsars are similar.

% bow-shock \Rs and \Rs formed aare similar to nebula should be the same as pulsar proceeds in a similar way to the case of isolated pulsar. {\bf Mitya: there is a logical caveat here. In bow-shock \Rs pulsars get a big kick which is determined by processes close to the pulsar, thus there could be some differences between winds of slow and fast moving pulsars.}

The structure of pulsar winds is a topic in itself. Let us briefly  describe our  current understanding. On scales larger than the  light cylinder  radius the wind power 
(which is mostly Poynting power) scales as $L (\theta_p) \propto  \sin^2 \theta_p$, where $\theta_p $ is the angle to the pulsar rotation axis  \citep{Michel73,bogovalov_99,2002MNRAS.336L..53B}, and the \Bf is torroidal in respect to the pulsar rotation axis.   In the equatorial region, occupying a section of the polar angles of 
the order of the pulsar-magnetic axis inclination angle, the \Bf is reversing polarity each half a period, while at larger latitude it is unidirectional \citep{bogovalov_99,KomissarovLyubarsky}. 
  As the wind propagates from the \LC, the \Bf is dissipated in the region of field reversals \citep{Coroniti90,lub03}.  Thus, the magnetization 
$\sigma$  in the wind is small near the equator \citep[this is confirmed by the modeling of the inner knot of Crab Nebula, see][]{2016MNRAS.456..286L}  and can reach large values at intermediate latitudes 
\citep[where flares are presumably generated, see \eg][]{2017JPlPh..83f6302L}.
% (behavior of $\sigma$ near the axis is not consternated - toroidal  \Bf must be zero, but density might be zero as well).  
%Importantly, the high-$\sigma$ parts of the wind are not radiatively efficient in \xrays, as demonstrated by the toroidal structure of the Crab Nebula.
%The angular size of the low-magnetized region of the wind equals the magnetic inclination angle -  the angle between the rotational axis and the magnetic moment. This is one of the parameters in our modeling.

\subsection{ \acl{rb}, \acl{fb}, and \acl{cw} geometry}

In addition to the fairly complected structure of the pulsar winds, for bow-shock \Rs there is an additional complication due to the pulsar proper motion. This introduces another special direction -- along the pulsar velocity -- and generally it makes the whole structure to be non-axisymmetric. Thus,
 the structure of the \Bf depends on two geometrical factors:  (i) the angle between the pulsar rotation axis and the magnetic moment; 
(ii) the angle between the pulsar rotation axis and the direction of the proper motion. Obviously, the observational appearance of the formed complex \ac{3d} structure depends strongly on the  line-of-sight direction. 
 
\begin{figure*}
 \begin{center}
\includegraphics[width=.32\linewidth]{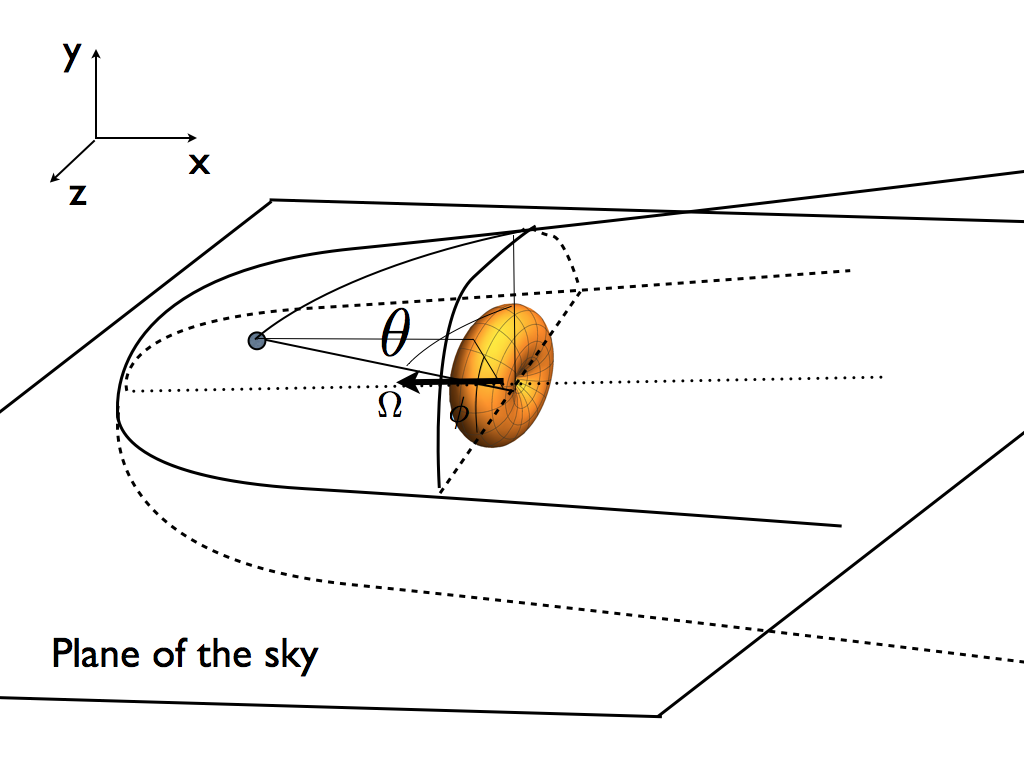}
\includegraphics[width=.32\linewidth]{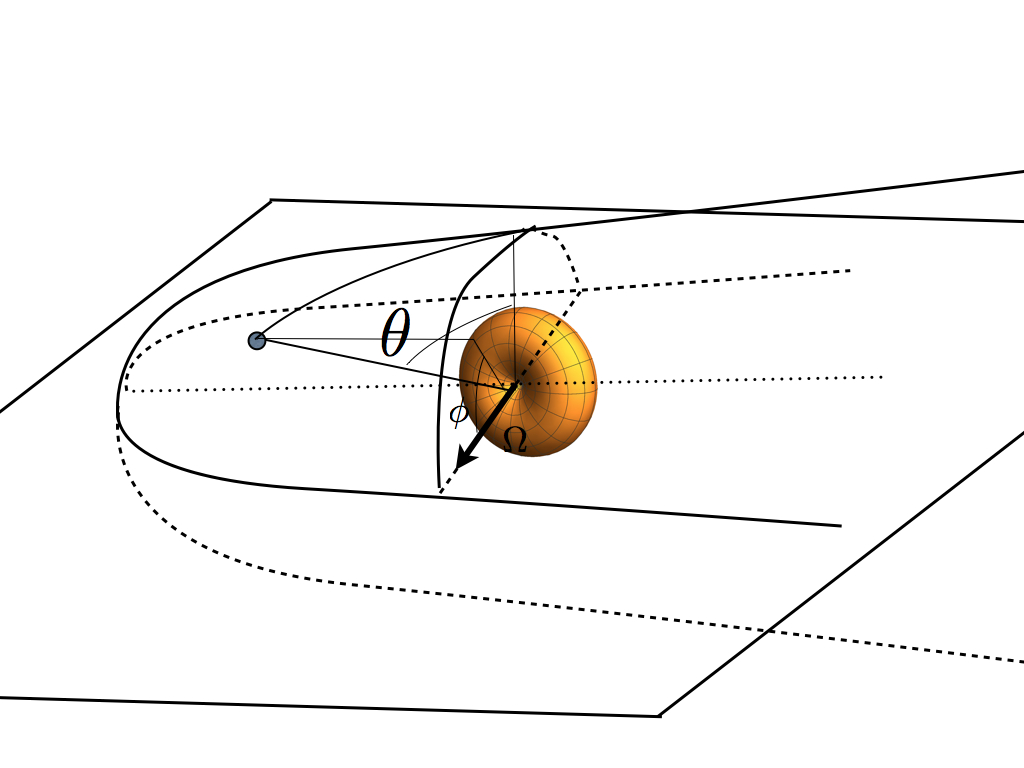}
\includegraphics[width=.32\linewidth]{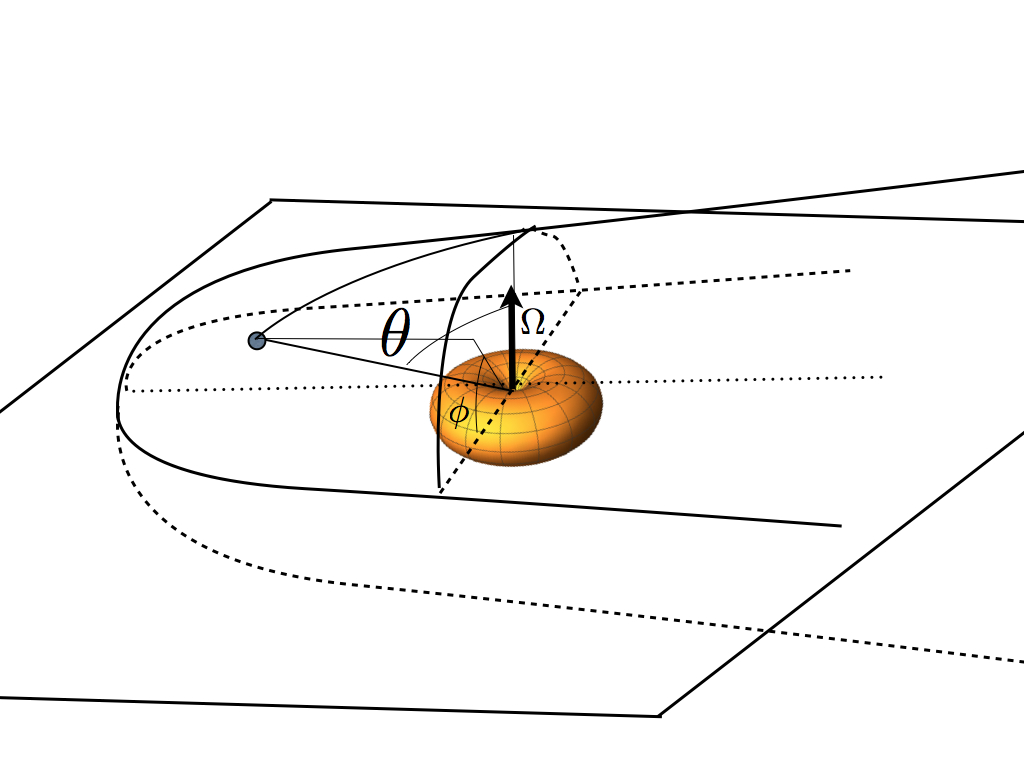}
\end{center}
\caption{Basic geometries: \ac{rb}  (with the spin of the \NS aligned with the velocity), \ac{fb}   (with the spin of the \NS perpendicular  to the  velocity but lying in the plane of the sky),  and 
\ac{cw}   (with the spin of the \NS perpendicular both to the  velocity and the plane of the sky). The central doughnut-like structure indicates the distribution of wind power, 
$\propto \sin^2 \theta_p$, where $\theta_p$ is the polar angle; $\theta$ is a polar angle with respect to the velocity.}
\label{fig:bfc}
\end{figure*}

 To simplify 
the discussion, we introduce three distinct cases (see Fig.~\ref{fig:bfc}): (i) \acl{rb} geometry -- when the rotation axis is aligned with 
the direction of  motion. In  this  case the whole system has a cylindrical symmetry with concentric areas of toroidal fields of changing polarity. 
(ii) \acl{fb} geometry -- when the rotation axis is perpendicular  to the direction of motion and is in the plane of the sky; 
(iii) \acl{cw} geometry -- when the rotation axis is perpendicular both  to the direction of motion and the plane of the sky. 
\Acl{fb} and \acl{cw} are intrinsically identical, but differ by the  line-of-sight direction. 
The two physically distinct geometries, \ac{rb} and \ac{fb}~--~\ac{cw}, will have very different \Bfs in the  tail, Fig.~\ref{fig:tail-Bfield}. 

In Appendix \ref{Bfhead} we consider analytically the structure of the \Bf\ in the simpler  \acl{rb} geometry. We point out that in the case of small magnetization, $\sigma \leq 1$, there is a narrow highly magnetized layers on the inside of the contact discontinuity due to effects of magnetic  draping \citep{Lyutikovdraping}.

\begin{figure}
\includegraphics[width=0.97\linewidth]{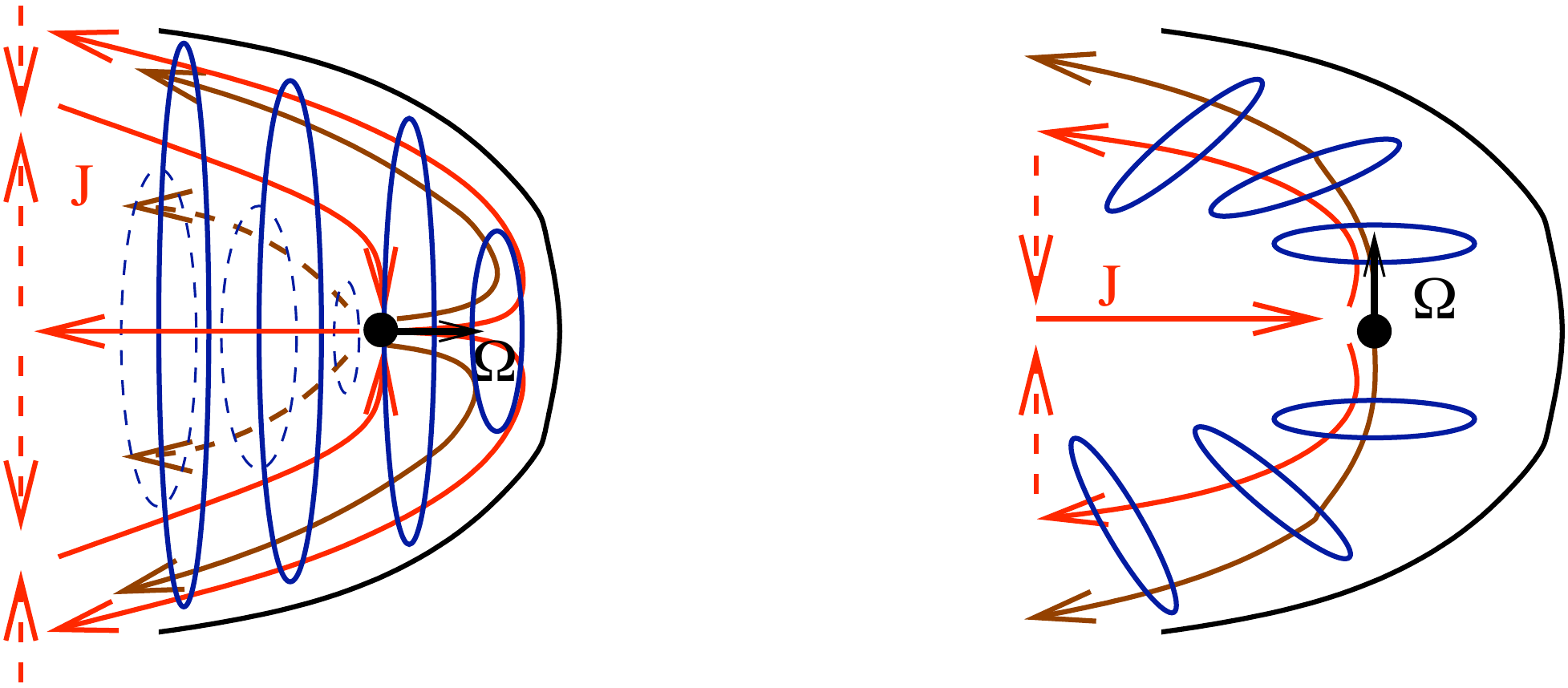}
\includegraphics[width=0.97\linewidth]{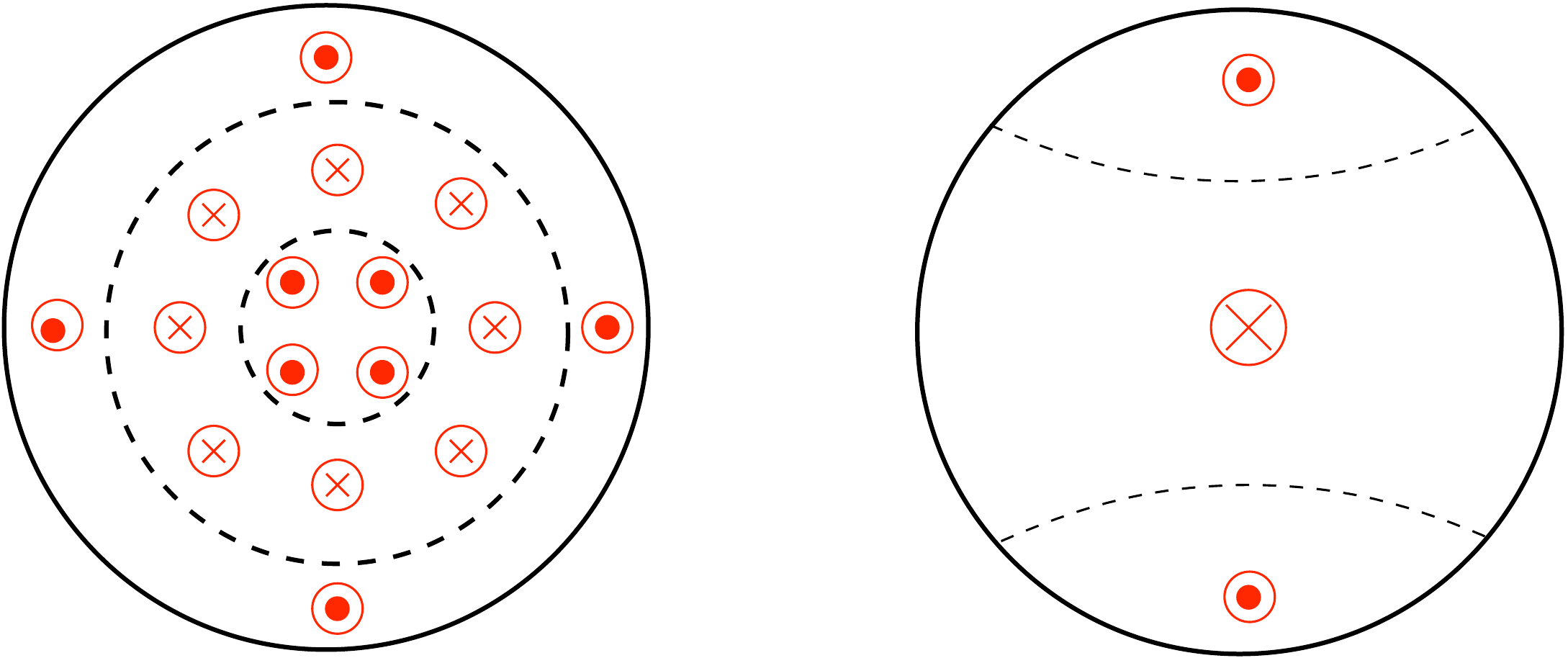}
\caption{Magnetic field in the  \ac{rb} (left)  and \ac{fb}/\ac{cw} (right) geometries. In case of a \ac{rb} geometry (\NS spin  oriented along the  velocity), 
the pulsar produces axisymmetric current flows and \Bfs. Far in the tail the electric currents form a triple sequence of  concentric regions with oppositely directed current flows 
(low left). For \ac{fb}, (\NS spin is orthogonal to the velocity; Sun is in  a \ac{fb}/\ac{cw} configuration),  the pulsar produces three adjacent  regions of opposite current flows (low right). }
\label{fig:tail-Bfield}
\end{figure}

\section{3D \rmhd simulations of bow-shock \Rs} 
\label{sc:numset}

\subsection{Numerical  Setup}
We performed a number of \ac{3d} \ac{rmhd} simulations of the interaction of magnetized pulsar wind with magnetized
external medium.  The simulations were performed using a \ac{3d} geometry in Cartesian coordinates using the {\it PLUTO}
code\footnote{Link http://plutocode.ph.unito.it/index.html} \citep{mbm07}.  Spatial parabolic interpolation, a
3\({}^{\rm rd}\) order Runge-Kutta approximation in time, and an HLL Riemann solver were used \citep{HLL83}.  {\it
  PLUTO} is a modular Godunov-type code entirely written in C and intended mainly for astrophysical applications and
high Mach number flows in multiple spatial dimensions.  The simulations were performed on CFCA XC30 cluster of
\ac{naoj}.  The flow has been approximated as an ideal, relativistic adiabatic gas, one particle species, and polytropic
index of 4/3. The size of the domain is $x \in [-4, 10]$, $y \mbox{ and } z \in [-5, 5]$ (the initial \ISM velocity is
directed along \(x\)-axis, we note that the coordinate system orientation is different from the one adopted above).  To
have a good resolution in the central region and long tail zone we use non uniform resolution in the computational domain
with the total number of cells $N_{\rm X} = 468$, and $N_{\rm Y} = N_{\rm Z} = 336$, see Table~\ref{tab:grid} for details.

\begin{table*}
\caption{Parameters of the Grid}
\begin{tabular}{lccccccc}
\hline
\hline
  Coordinates         &  Left    &  $N_{\rm l}$   &   Left-center  &   $N_{\rm c}$ &   Right-center & $N_{\rm r}$ & Right  \\
\hline
&&&&&\\[-5pt]
 $ X  $     & \quad $-4$ & \quad $72$ & \quad $-1$ &\quad 144 & \quad $1$ &\quad 252 & \qquad $10$ \\
 $ Y  $     & \quad $-5$ & \quad $96$ & \quad $-1$ &\quad 144 & \quad $1$ &\quad 96  & \qquad $5$ \\
 $ Z  $     & \quad $-5$ & \quad $96$ & \quad $-1$ &\quad 144 & \quad $1$ &\quad 96  & \qquad $5$ \\
\hline
&&&&&\\[-5pt]
\end{tabular}
%\tablecomments{Parameters of the Grid}
\label{tab:grid}
\end{table*}

%\subsection{Pulsar wind setup} \label{s:pws}

In this work we use  the prescription of pulsar wind  similar to what was used by \cite{2014MNRAS.438..278P}.
The pulsar with radius 0.2 is placed at the origin.\footnote{Properties of \ISM were adjust in a way that the stand-off-distance, Eq.~\eqref{rs} is \(1\).} The pulsar produces supersonic magnetized pulsar with
 toroidal magnetic field that change its polarity in northern an southern hemispheres.
   For the total\footnote{It includes the kinetic and electromagnetic energy fluxes.} energy flux density of the wind we adopt the 
monopole model \citep{Michel73,bogovalov_99}
\be
f_{\rm tot}(r,\theta_p)=L_0 \left(\frac{1}{r}\right)^2\left(\sin^2\theta_p+g\right).
\label{eq:wpow}
\ee
where  we added the parameter $g=0.03$
to avoid vanishing energy flux at the poles.

In the wind the energy is distributed between the magnetic $f_m$  component
\be
f_m(r, \theta_p) = \frac{\sigma(\theta_p)f_{tot}(r, \theta_p)}{1+\sigma(\theta_p)} 
\label{eq:fm}
\ee
and kinetic $f_k$ one
\be
f_k(r, \theta_p) = \frac{f_{tot}(r, \theta_p)}{1+\sigma(\theta_p)}, 
\label{eq:fk}
\ee
where $\sigma(\theta_p)$ is wind magnetization, which depends on latitude, the angle $\theta_p$ is counted from 
the pulsar spin axis.

As one adopts a torroidal geometry of the \Bf, the numerical  stability of the code requires vanishing of the magnetic field close to the polar axis. This is achieved by introducing the following dependence of the pulsar wind magnetization:
\be
\sigma_0(\theta_p) = \sigma_0 \min\left(1, \theta_p^2/\theta_0^2, (\upi-\theta_p)^2/\theta_0^2\right)\,
\label{eq:sigvan}
\ee
where $\theta_0$ is a small parameter,  which was set to \(0.2\).

\begin{figure}
\includegraphics[width=0.97\linewidth]{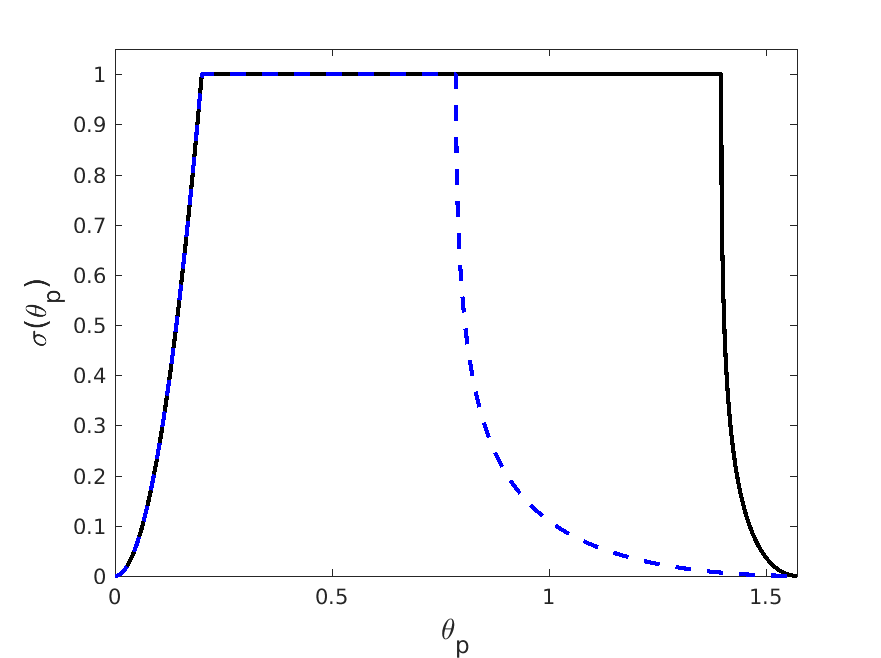}
\includegraphics[width=0.97\linewidth]{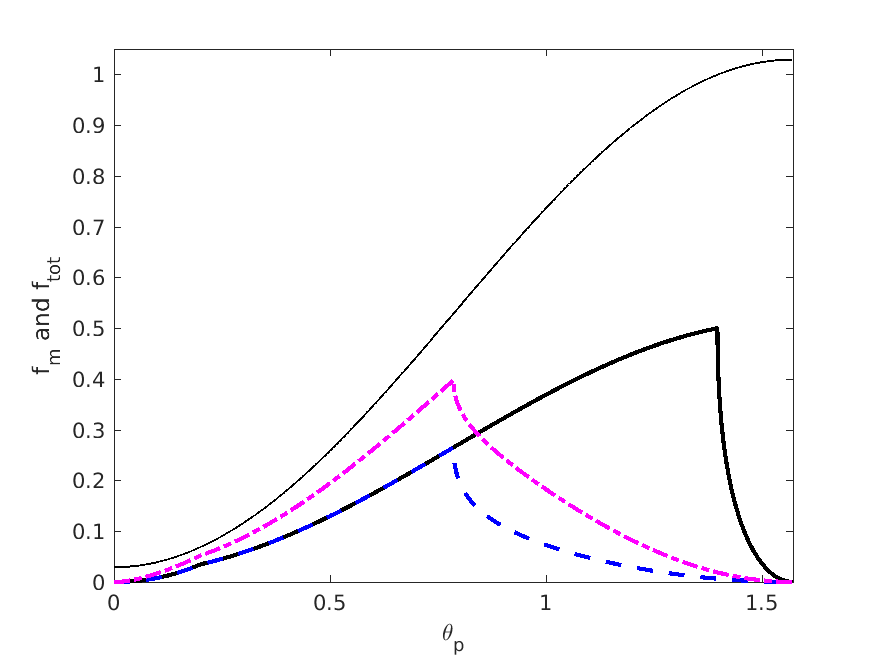}
\caption{Top panel: the polar angle dependence of the wind magnetization, $\sigma(\theta_p)$, for $\alpha = \pi/18$ (solid line) and $\alpha = \pi/4$ (dashed line). Bottom panel:
the polar angle dependence of the total energy flux $f_{\rm tot}$ (thin solid line), and Poynting flux,  $f_m$, for three considered cases: \(\sigma_0=1\) and \(\alpha=\upi/4\) (dashed line, models fs1a45, bs1a45, and fbs1a45); \(\sigma_0=1\) and \(\alpha=\upi/18\)  (solid line, model fs1a10); and \(\sigma_0=3\) and \(\alpha=\upi/4\) (dash-dotted line, model fs3a45).}
\label{fig:sigma}
\end{figure}

Near the equator  the alternating components of \Bf are assumed to annihilate, leaving a low-magnetized equatorial sector with magnetization varying according to
\be
\sigma(\theta_p) = \frac{\sigma_0(\theta_p)\chi_\alpha(\theta_p)}{1+\sigma_0(\theta_p)(1-\chi_\alpha(\theta_p))}\,,
\label{eq:sigeq}
\ee
where
\be
\chi_\alpha(\theta_p) = \left\{
\begin{array}{ll}
(2\phi_\alpha(\theta_p)/\upi-1)^2, & |\upi/2-\theta_p|< \alpha\\
1 & \mbox{otherwise}
\end{array}
\right. \,,
\label{eq:chi}
\ee
and $\cos \phi_\alpha(\theta_p)=-\cot(\theta_p)\cot(\alpha)$. The angle $\alpha$ is an angle between 
magnetic axis and pulsar rotation axis, see Fig.~\ref{fig:sigma} \citep[see ][for more detail]{2013MNRAS.428.2459K}.

The pulsar wind  was injected with the initial Lorentz factor, $\Gamma = 2.9$, which corresponds to the initial Mach number of \(25\). 
%HOW MACH NUMBER IS DEFINED?
% <1%, I put maximally cold wind which not crush the simulations. 

% \begin{table}
% \caption{Parameters of the models}
% \begin{tabular}{lccccc}
% \hline
% \hline
%   Model              & $\theta$     & $\psi$   &   $\sigma_0$  &   $\alpha$  \\
% \hline
% &&&&&\\[-5pt]
%   fs01a45       & \quad0 & \quad 0 & \quad 0.1 &\quad $\upi/4$  \\
%   fs1a45       & \quad0 & \quad 0 & \quad 1 &\quad $\upi/4$ \\
%   fs3a45       & \quad0 & \quad 0 & \quad 3 &\quad $\upi/4$  \\
%   fs1a10       & \quad0 & \quad 0 & \quad 1 &\quad $\upi/18$  \\
%   bs01a45       & \quad$\upi/2$ & \quad 0 & \quad 0.1 &\quad $\upi/4$  \\
%   bs1a45       & \quad$\upi/2$ & \quad 0 & \quad 1 &\quad $\upi/4$  \\
%   fbs1a45       & \quad$\upi/4$ & \quad $\upi/4$ & \quad 1 &\quad $\upi/4$  \\
% \hline
% &&&&&\\[-5pt]
% \end{tabular}
% \label{tab:models}
% \end{table}

\begin{table*}
\caption{Parameters of the models}
\begin{tabular}{lccccccc}
\hline
\hline
  Model              & $\theta$     & $\psi$   &   $\sigma_0$  &   $\alpha$   & Figures & Sections\\
\hline
&&&&&\\[-5pt]
  bs01a45       & \quad$\upi/2$ & \quad 0 & \quad 0.1 &\quad $\upi/4$  & Figs.~\ref{bullet},~\ref{bullet1},~\ref{fig:b_Ebd} & \S~\ref{s:dmw},~\ref{synch}\\
  bs1a45       & \quad$\upi/2$ & \quad 0 & \quad 1 &\quad $\upi/4$  & Figs.~\ref{overall},~\ref{bullet},~\ref{bullet1},~\ref{fig:b_J},~\ref{fig:b_Ebd} & \S~\ref{s:dmw},~\ref{synch}\\
\hline
  fs01a45       & \quad0 & \quad 0 & \quad 0.1 &\quad $\upi/4$  & Figs.~\ref{fig:Compare},~\ref{fig:Y_Ebd} & \S~\ref{fff},~\ref{synch}\\
  fs1a45       & \quad0 & \quad 0 & \quad 1 &\quad $\upi/4$ & Figs.~\ref{fig:frisbee},~\ref{fig:Compare},~\ref{fig:X_Ebd},~\ref{fig:Y_Ebd} & \S~\ref{fff},~\ref{synch}\\
  fs3a45       & \quad0 & \quad 0 & \quad 3 &\quad $\upi/4$  & Figs.~\ref{fig:Compare},~\ref{fig:X_Ebd},~\ref{fig:Y_Ebd} & \S~\ref{fff},\ref{synch}\\
  fs1a10       & \quad0 & \quad 0 & \quad 1 &\quad $\upi/18$  & Figs.~\ref{fig:Compare},~\ref{fig:X_Ebd},~\ref{fig:Y_Ebd},~\ref{fig:Z_Ebd} & \S~\ref{fff},~\ref{synch}\\
\hline
  fbs1a45       & \quad$\upi/4$ & \quad $\upi/4$ & \quad 1 &\quad $\upi/4$  & Figs.~\ref{fig:mixed},~\ref{fig:fb_Ebd} & \S~\ref{fff},~\ref{fff}\\
\hline
&&&&&\\[-5pt]
\end{tabular}
\label{tab:models}
\end{table*}

\subsection{Initial setup}
\label{s:inc}

We start our simulation with a non-equilibrium configuration and evolve it until a  
quasi-stationary solution is settled. From the left edge $(X=-4)$ we inject \ISM. 
To reduce computational expenses we set the initial \ISM speed to $v_\ism=0.1 c$, which corresponds to the Mach number of $M_s=85$. 
The density of the \ISM was 
adopted so that in the case of non-magnetized  spherical pulsar wind the stand-off distance given by Eq.~\eqref{rs} equals \(1\). The adopted initial \ISM speed is not realistic, but
it does not affect significantly the region inside the \CD \citep[][]{blkb18}. 
The \ISM flow carries a weak \Bf with  magnetization $\sigma_\ism=0.01$; the \ac{ism} \Bf  is directed along $z$-axis. 

We run three sets of simulations depending on the orientation of the pulsar spin with respect to velocity: 
(i) \ac{rb}; (ii) \ac{fb}~--~\ac{cw}; (iii) mixed  \ac{fb}~--~\ac{rb} configuration.
We study a few specific peak magnetization values, \(\sigma_0\), and pulsar magnetic inclination angles, $\alpha$, see Table~\ref{tab:models} for detail.

The magnetization of pulsar wind is present by three values $\sigma_0 = 0.1; 1; 3$ and $\alpha  = \upi/4$, also we check one model (\ac{fb}) for $\alpha = \upi/18$. 

We choose three cases of pulsar orientation \ac{fb}, \ac{rb}, and the intermediate one, \ac{fb}~--~\ac{rb}.
The orientation is determined by two angles  $\theta$ (clockwise turn around Y axis) and angle $\psi$ (clockwise turn around axis Z),
In the case of the \ac{fb} and \ac{cw} geometry, the pulsar rotation axis is parallel to axis Z ($\theta=0, \psi=0$), 
in the case of the \ac{rb} geometry, the pulsar rotation axis is parallel 
to axis X ($\theta=\upi/2, \psi=0$), the intermediate case was formed by clockwise turn of the \ac{fb} configuration around 
Y axis on angle $\theta=\upi/4$ and after that clockwise turn around axis Z on angle $\psi=\upi/4$.  
The parameters of the models are presented in Table~\ref{tab:models}.

%\section{3D relativistic \gls{mhd} simulations of bow shock \Rs: Results}
%\label{sc:res}

\subsection{Overall comparison with theory}

\begin{figure*}
\includegraphics[width=0.99\linewidth]{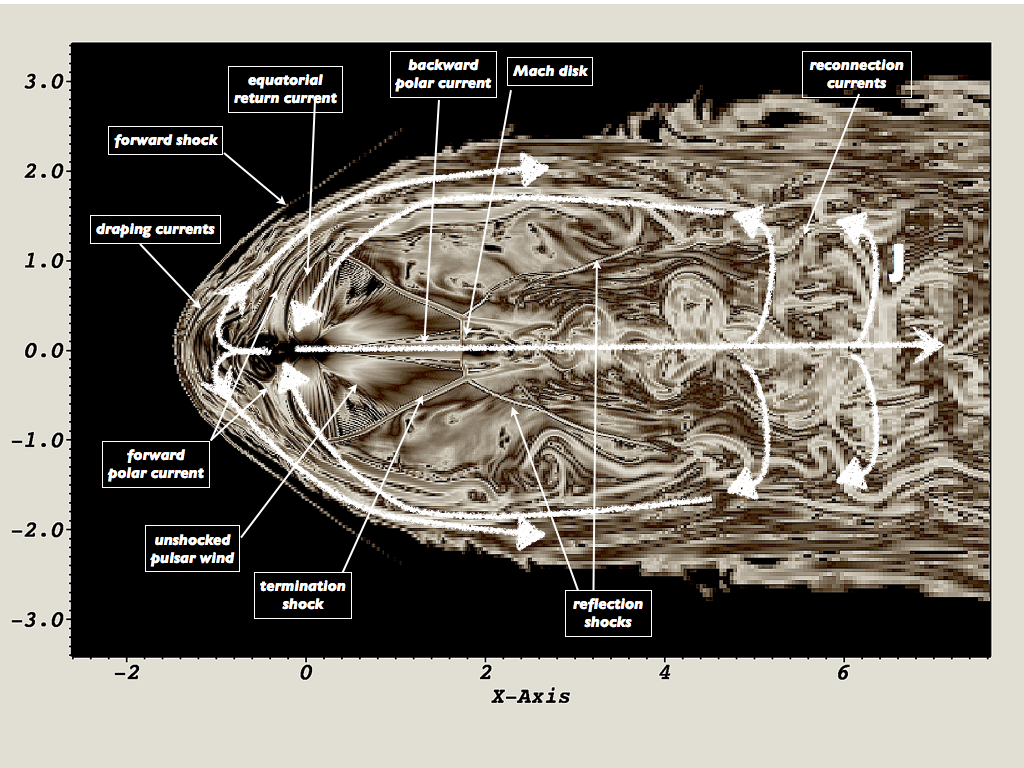}
\caption{The current density map for the \ac{rb} configuration (bs1a45, $\sigma = 1$, $\alpha = \upi/4$) with the key structural elements marked, see \S~\ref{analytical} for detail.}
\label{overall}
\end{figure*}

Many details of the theoretical expectations described above can be clearly seen in the simulations maps.  In
Fig.~\ref{overall} we show the current density map  with the key features highlighted
for the \ac{rb} configuration. Upstream of the \R there is a clear signature of a bow shock, where the external \Bf is
amplified. Since \Bf is zero in the pulsar wind close to the pulsar rotation axis, the \Bf is small on the symmetry
axis. Closer to the apex point magnetic hoop stresses lead to increased magnetization. (In the axial-symmetric \ac{2d} case,
the magnetic collimation may result in unphysical behavior
\citep[\eg notice a cut-out triangle near the apex point in figures in][]{bad05}. In the \ac{3d} case,  development of instabilities 
allows the flow to relax.)  On the inner side of the \CD the effects of magnetic draping (see Appendix~\ref{Bfhead}) lead to the
formation of highly magnetized layers \citep[``draping currents''; previous \ac{2d} low-$\sigma$ simulations also shown effects of
 magnetic draping, see, {\eg}, last panel in figure~1 in][]{bad05}. Unshocked pulsar wind is extended
``sideways'' since the wind energy flux $\propto \sin^2 \theta_p$, where for \ac{rb} configuration $\theta_p $ is the angle
with respect to the direction of \ISM initial velocity.  The pulsar produces a quadrupolar-type structure of currents: two
outgoing currents propagating straight ahead (``forward polar current'') and towards the tail (``backward polar
current''), and two currents return to the  pulsar equator  (in \ac{3d} - an axis-symmetric current layer).
  
Non-spherically symmetric \TS, as well as tailward Mach disk, are clearly seen. (Structures visible in the unshocked pulsar wind 
are mostly due to numerical artifacts.)  Since the post-Mach disk flow is over-pressurized with respect to the  sideway flows, a reflection shock is formed. 
Due to lack of resolution far down the  tail, resistive effects lead to dissipative reconnection currents and onset of turbulence. 
We hypothesize that  in future higher resolution simulations the second Mach disk may appear visible.

\subsection{\acl{rb} configuration}
\label{s:dmw}
\begin{figure*}
\includegraphics[width=80mm,angle=-0]{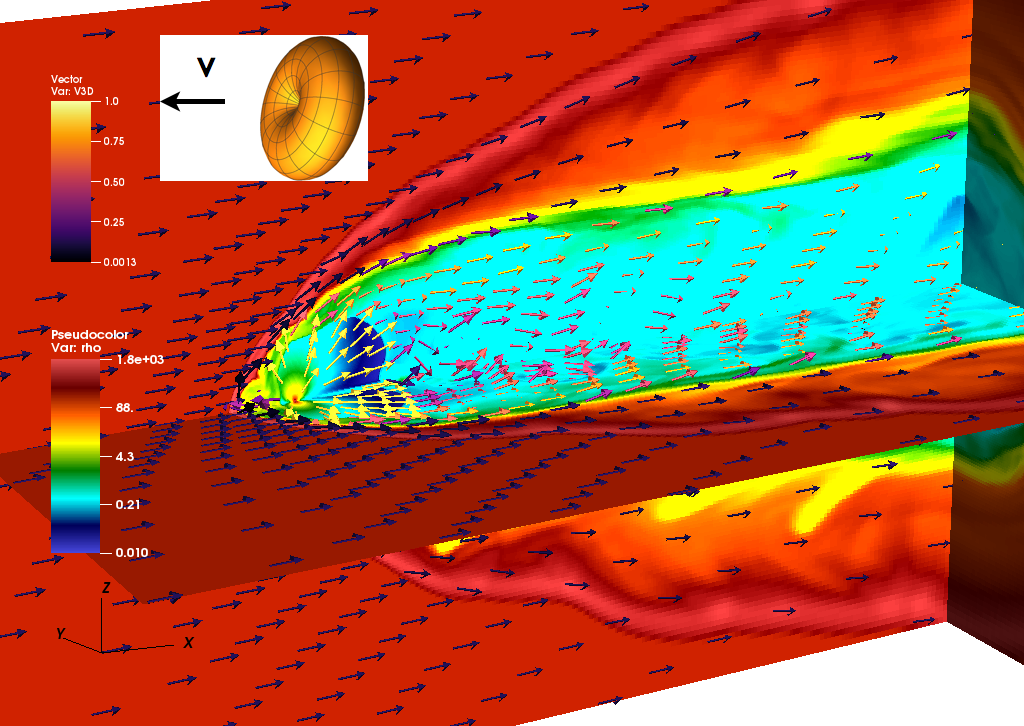}
\includegraphics[width=80mm,angle=-0]{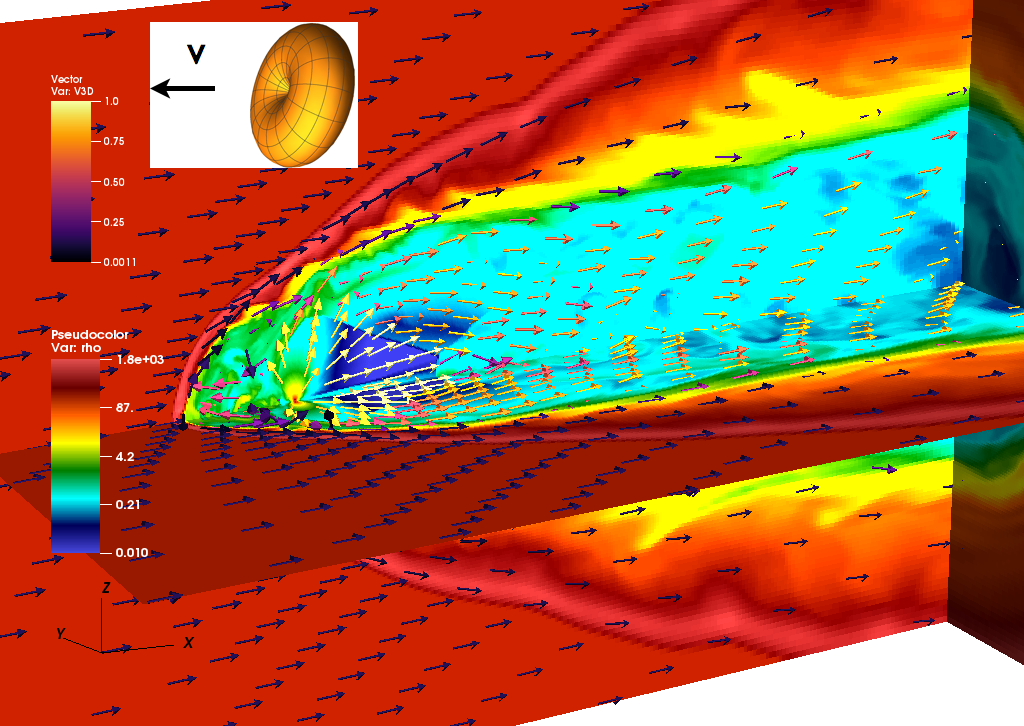}
\includegraphics[width=80mm,angle=-0]{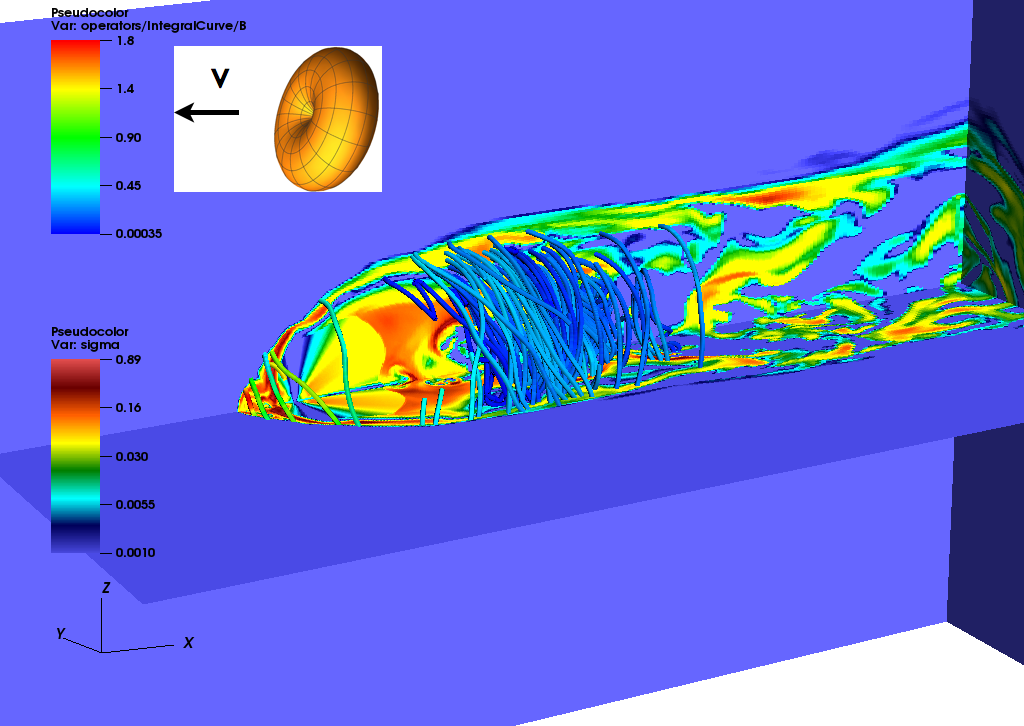}
\includegraphics[width=80mm,angle=-0]{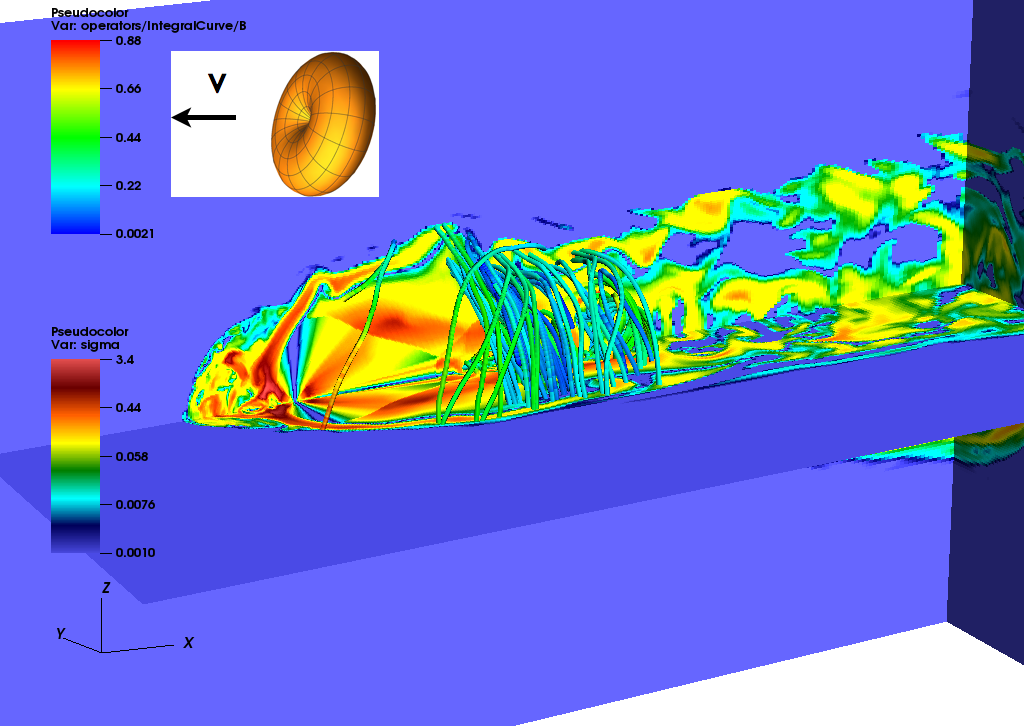}
\includegraphics[width=80mm,angle=-0]{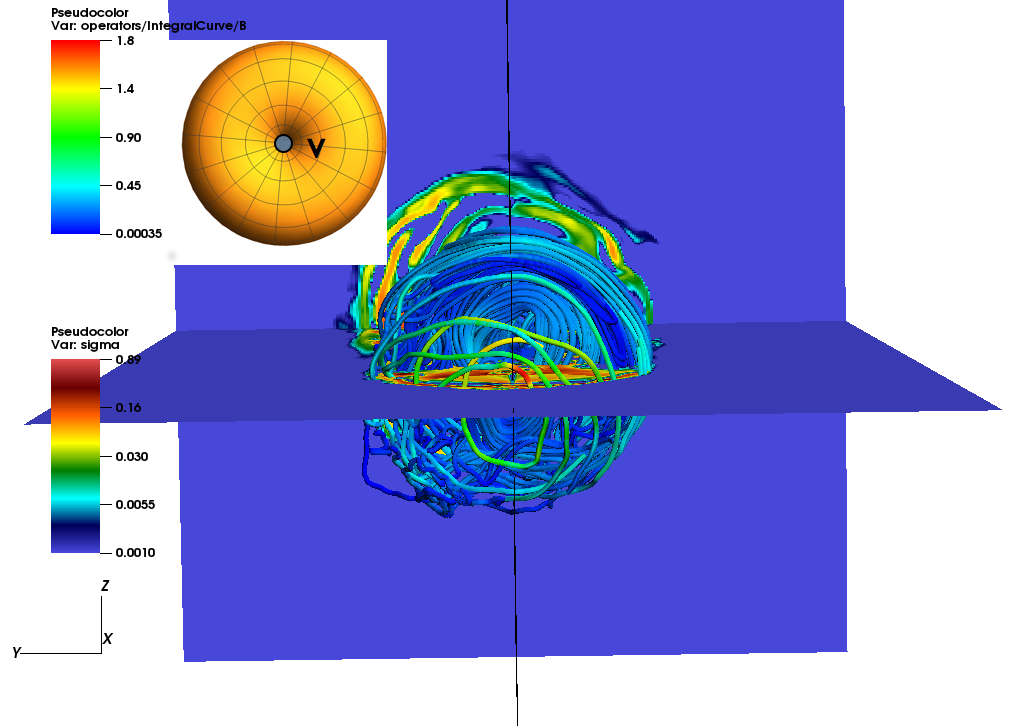}
\includegraphics[width=80mm,angle=-0]{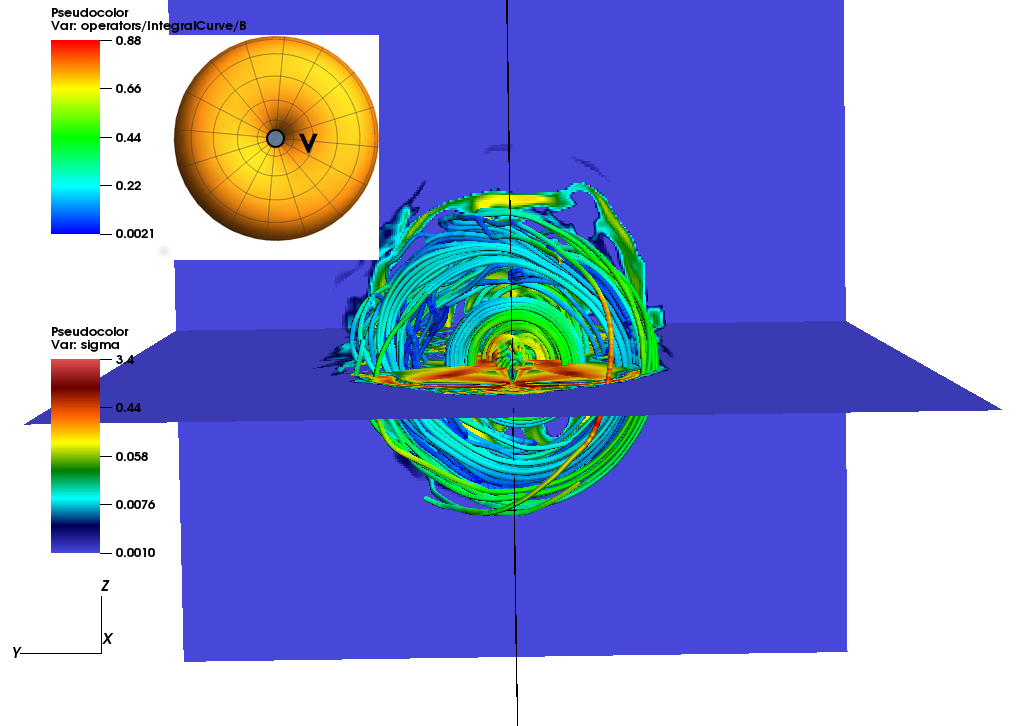}
\caption{\Acl{rb} configuration: \ac{3d} rendering.   Top row: Plasma density (color logarithm, lower colour bar) and  the  velocity field 
(arrows, upper colour bar) for the models  
bs01a45 ($\sigma_0= 0.1$, left column),  bs1a45 ($\sigma_0= 1$, right column) (see Table~\ref{tab:models}).
Middle row: Plasma magnetization  (color logarithm,  lower colour bar) and magnetic field lines (upper colour bar).
Bottom row: same as middle row but projected on X axis. We should note that the twist in inner parts and in outer parts of magnetic field is different,
and we can see turn of magnetic field lines in a middle region at the reconnection sites.}
\label{bullet}
\end{figure*}

\begin{figure*}
\includegraphics[width=80mm,angle=-0]{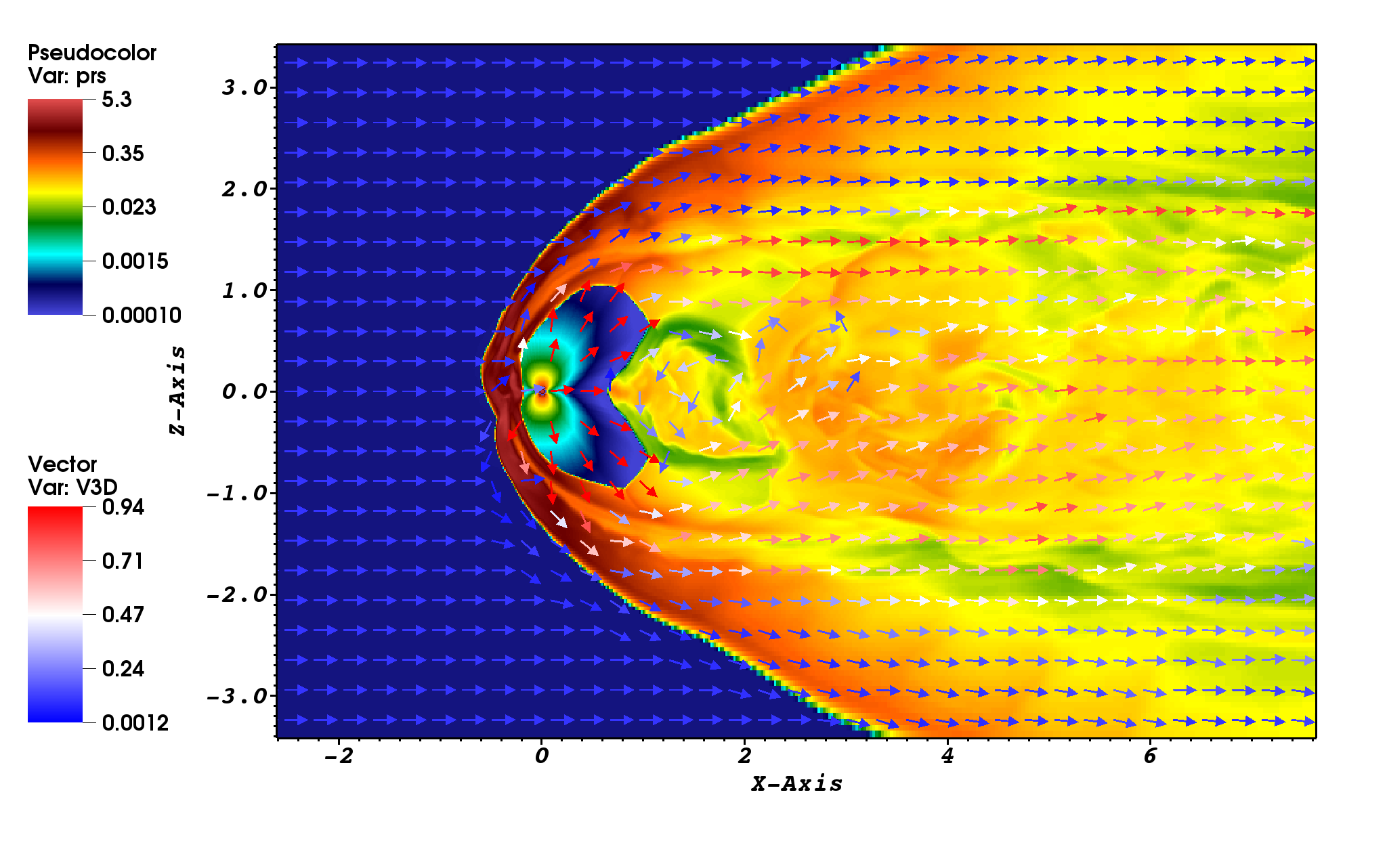}
\includegraphics[width=80mm,angle=-0]{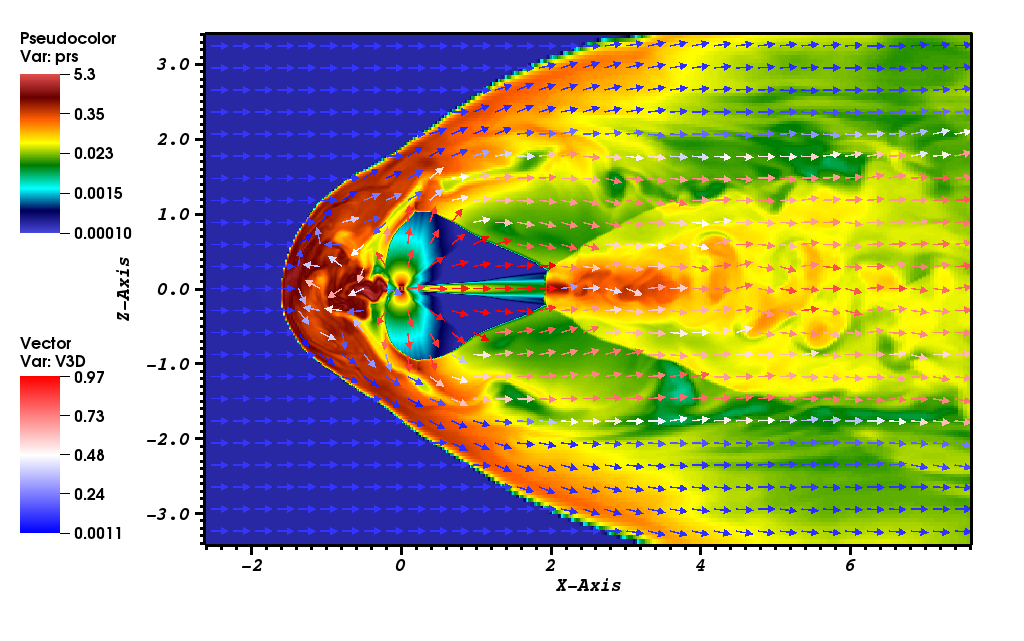}
\caption{\Acl{rb} configuration:  XZ-slice  of pressure (color logarithm) and  velocity field for the models  bs01a45 (left, $\sigma=1$)  and bs1a45 (right, $\sigma=0.1$).}
\label{bullet1}
\end{figure*}

\begin{figure*}
\includegraphics[width=80mm,angle=-0]{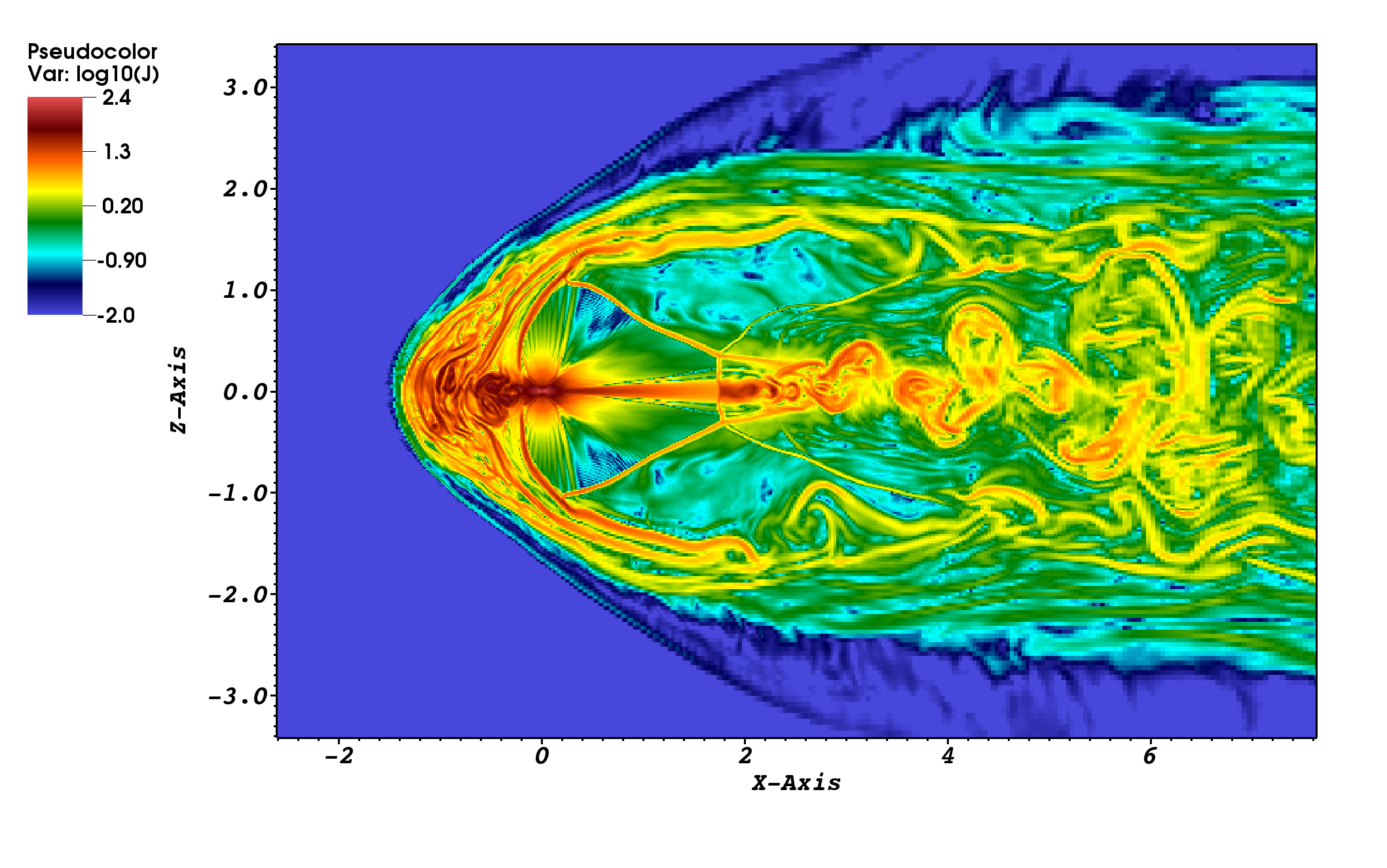}
\includegraphics[width=80mm,angle=-0]{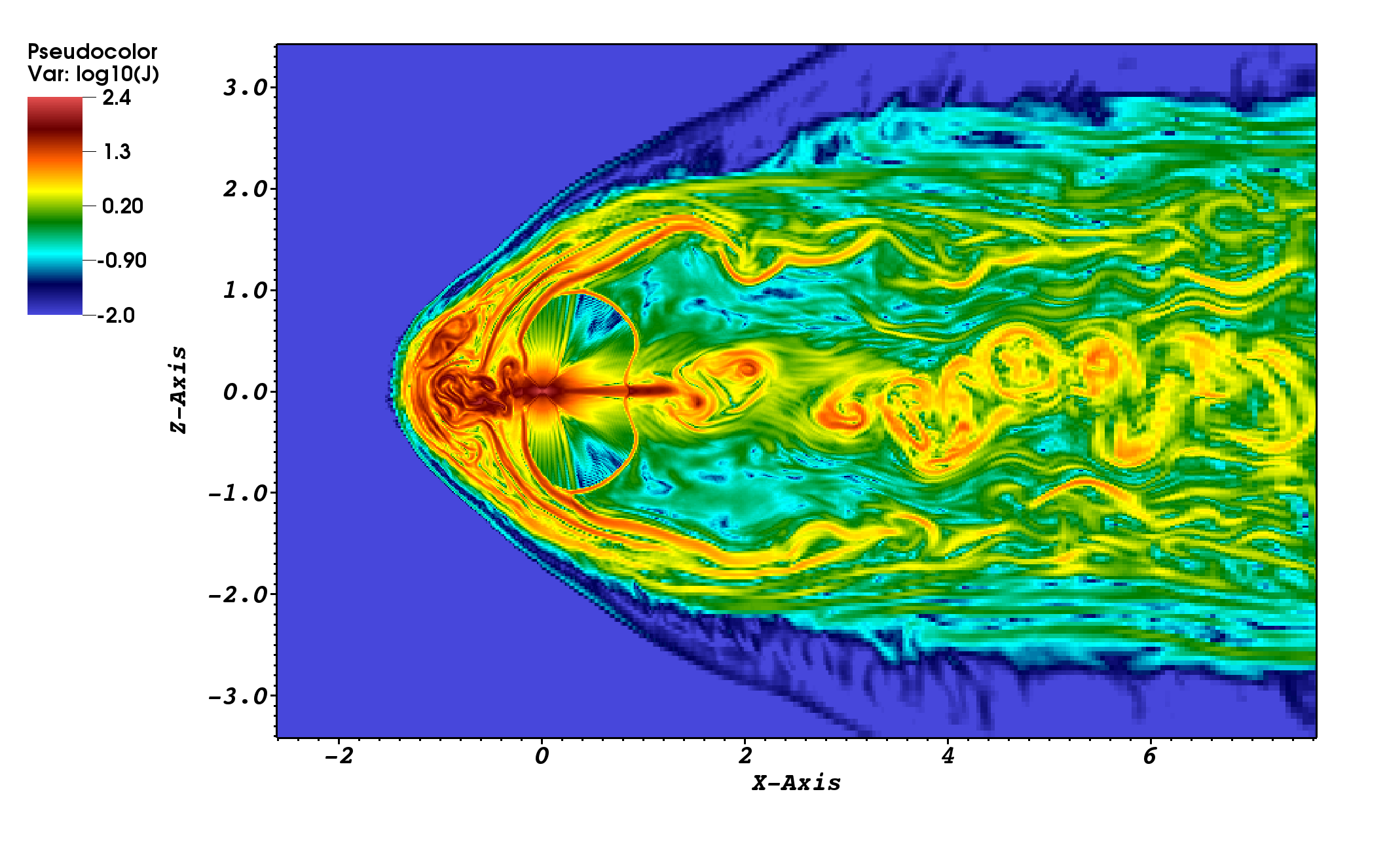}
\caption{\Acl{rb} configuration, model bs1a45:  XY-slice of  current density (color logarithm) at  different times. 
Two panels show the amplitude of the long term fluctuation in the position of the Mach disk.}
\label{fig:b_J}
\end{figure*}

We performed  two types of  simulations in the \ac{rb} configuration with different wind 
magnetization, $\sigma_0 = 0.1$ and  $\sigma_0 =1$. The inclination angle was fixed at   $\alpha= \upi/4$, see Figs.~\ref{bullet},~\ref{bullet1},~\ref{fig:b_J}.

%fffffffffffffffffffffffffffffffffffffffffffffffffffffffffffffffff
%fffffffffffffffffffffffffffffffffffffffffffffffffffffffffffffffff 

%fffffffffffffffffffffffffffffffffffffffffffffffffffffffffffffffff
%fffffffffffffffffffffffffffffffffffffffffffffffffffffffffffffffff 

%fffffffffffffffffffffffffffffffffffffffffffffffffffffffffffffffff
%fffffffffffffffffffffffffffffffffffffffffffffffffffffffffffffffff 

In models with the \ac{rb} geometry  a bow shock with approximately  axial-symmetric geometry is formed (in contrast, \ac{cw} and \ac{fb} models are essentially \ac{3d}, see \Sec\ref{fff}). Simulations for \ac{rb} geometry show the formation of the headward  and the  tailward jets. 
In the case of 
high magnetization ($\sigma_0=1$),  the headward jet pushes away the forward shock significantly farther as compared to the low magnetization case ($\sigma_0=0.1$).   
This is the effect of magnetic hoop stresses in the shocked pulsar wind - the magnetic field then tries to keep the plasma closer to the axis 
\citep[see \eg][]{2002MNRAS.329L..34L,2003AstL...29..495K}, increasing the local pressure and, as a result, a stand-off distance. In case of \ac{2d} 
simulations this effect becomes dominant, leading to unphysical results, see discussion above.

In the both cases ($\sigma_0 = 0.1$ and \(1\)),  the backward  Mach disk and forward bow shock are formed at a similar distances from the pulsar ($r_{\textsc{m}}\sim\mbox{1-2}\times r_s$). 
However, while the forward bow shock location is steady, the position of the  Mach disk changes with time, showing a large-amplitude oscillation,  Fig.~\ref{fig:b_J}. We infer  
two  distinguish types of the tail oscillation,  one on a long ($\sim80 r_s/c$) and another on a short ($\sim 20 r_s/c$) time scale. The origin of these oscillations is not clear, probably they are triggered by the kink instabilities in the back ``jet''. 

In the tail,  the \Bf is predominantly toroidal with the polarity reversing between the axial, intermediate and border regions, as expected 
(left panel in Fig.~\ref{fig:tail-Bfield}). In the intermediate region we reveal zones suitable for \Bf reconnection (see Fig~\ref{overall}). 

% THIS IS NOT CLEAR:
There are two apparent processes that lead to formation of the reconnection zones. First one is related to the
interaction of the forward ``jet'' with the \ISM at the head of the bow shock. This interaction results in formation of
a complex structure of the \Bf in the head region. Advection of the plasma from this region eventually results in
numerous sites suitable for \Bf reconnection. The second effect is related to the structure of the currents in the \R. The
forward current from the head region and reverse current streaming to the pulsar equatorial region appear to be
compressed in a relatively narrow outer layer of the \R. The currents' mixing region extends to a significant distance,
$\sim 4r_s$, tailward, see Fig.~\ref{fig:b_J}. We anticipate that the characteristic mushroom (or umbrella) morphology seen in several \Rs, \eg \R created by \acl{mushroom}, might be caused by the \Bf reconnection in the outer layer of the \R.

% The current in forward region mixing with current in the tail has place at a distance $\sim 4r_s$ . 
% This forms region associated with fast 
% dissipation zones of magnetic energy, so we expect the formation of mushroom or umbrella-like structures for \Rs in the \ac{rb} configuration.

% I DID NOT UNDERSTAND THE TEXT BELOW
%Fig.s~\ref{fig:prs_V} shows gas pressure distribution by color and velocity field by arrows.  On the pressure maps, we see free pulsar wind 
%zone which surrounded the pulsar. The shape of the free wind zones significantly depends on the pulsar orientation respect to its direction of 
%motion. Billet models show more or less axisymmetric configurations. The tail of wind zone in the case of 
%strong magnetization (bs1a45) is strongly variable and shows states with the subsonic flow after termination wind shock (see Fig.~\ref{fig:logJ_bul} 
%upper panel) and supersonic flow (lower panel).
%At the Fig.~\ref{fig:logJ_bul} we see the distribution of the logarithm of a current density magnitude in arbitrary units. These plots trace regions 
%where magnetic field changes. It can be the shocks, the magnetic field reconnection sites, the jets, etc.

The \ac{rb} configuration is the only one that can be studied under \ac{2d} approximation \citep[see \eg][]{bad05}. However, even for this geometry \ac{2d} and \ac{3d} simulations provide considerably different results. The front ``jet'' revealed with \ac{3d} simulations \citep[this paper; see also][]{2014MNRAS.438..278P} appears to be unsteady, which apparently allows to avoids the computational problems seen in \ac{2d} simulations. The key difference is probably related to the suppression of the kink instability in \ac{2d} simulations. To avoid the strong unphysical magnetic collimation near the axis expected in \ac{2d} case,  \cite{bad05} suppressed magnetization of the pulsar wind in a cone
near the pulsar rotation axis. This approach, however,  affects significantly the shape of  pulsar wind \TS and 
the \R flow in general. 

\subsection{\acl{fb} and \acl{cw} configurations}
\label{fff}
\begin{figure}
\includegraphics[width=88mm,angle=-0]{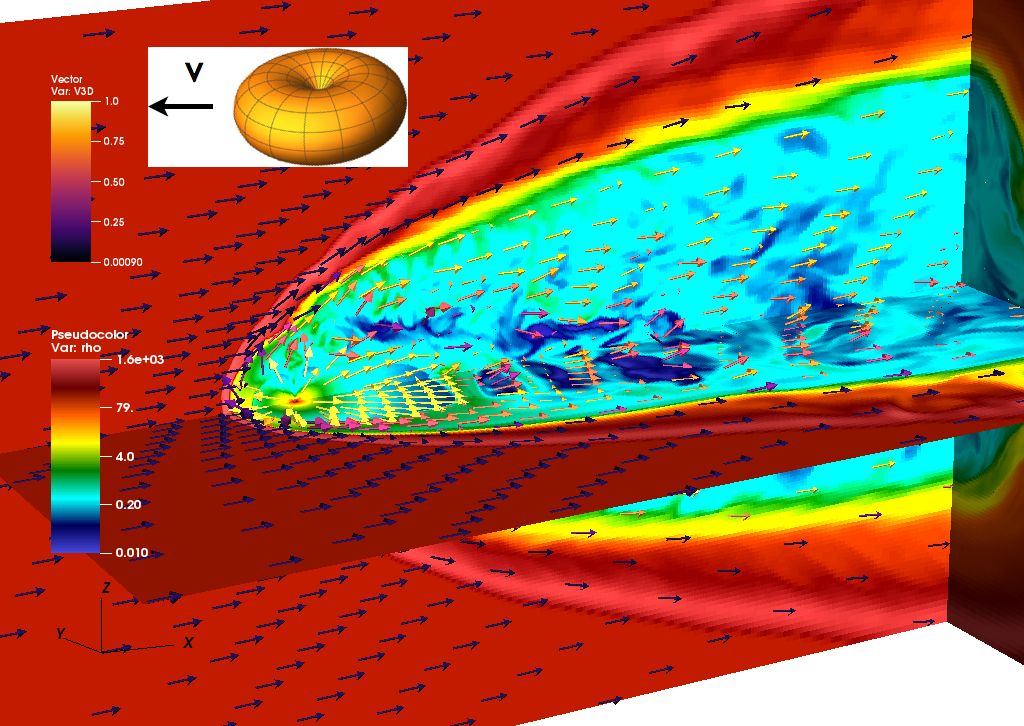}
\includegraphics[width=88mm,angle=-0]{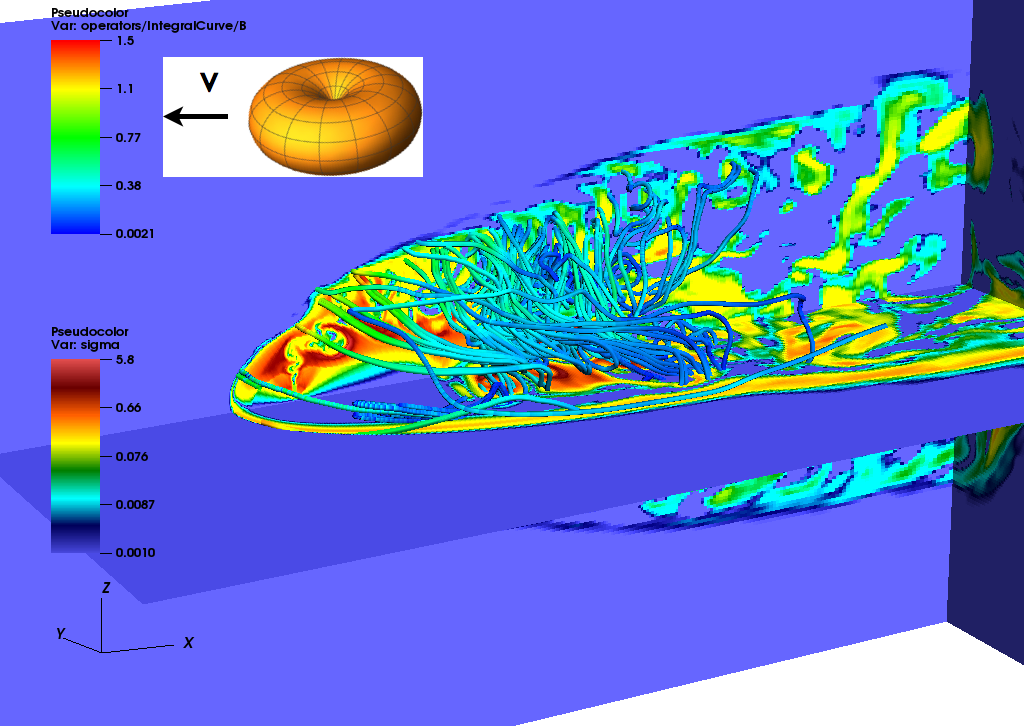}
\includegraphics[width=88mm,angle=-0]{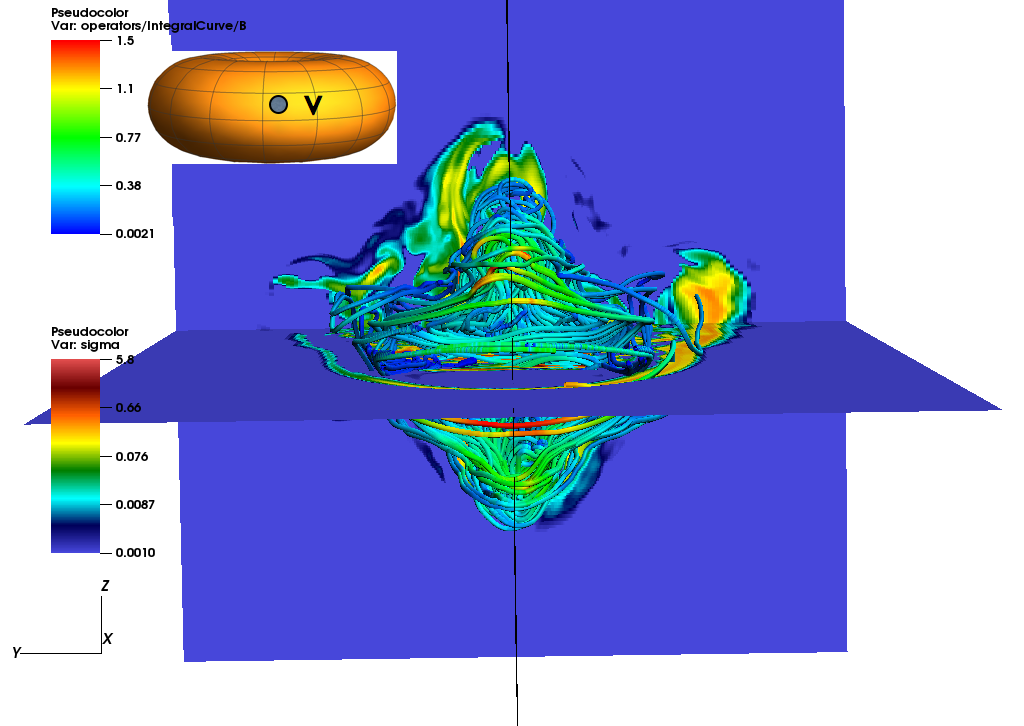}
\caption{\Acl{fb}/\acl{cw} configuration, model fs1a45. \Ac{3d} rendering of plasma density and velocity field for the  (top panel) 
and plasma magnetization,  projection on Y axis (middle panel), and  on X axis (bottom panel). The notation is similar to Fig.~\ref{bullet}.}
\label{fig:frisbee}
\end{figure}

\begin{figure*}
\includegraphics[width=88mm,angle=-0]{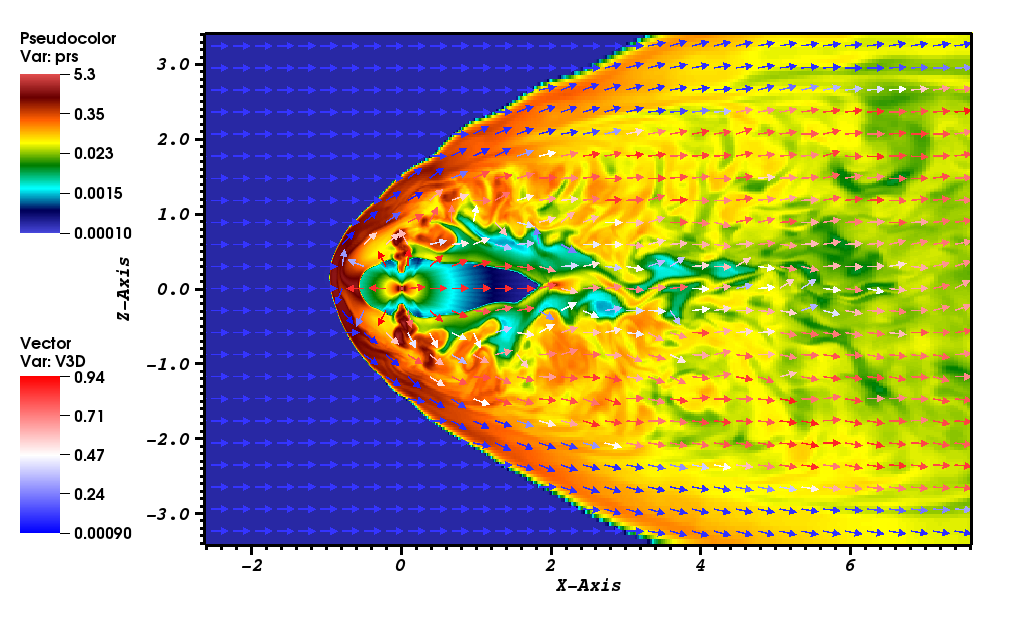}
\includegraphics[width=88mm,angle=-0]{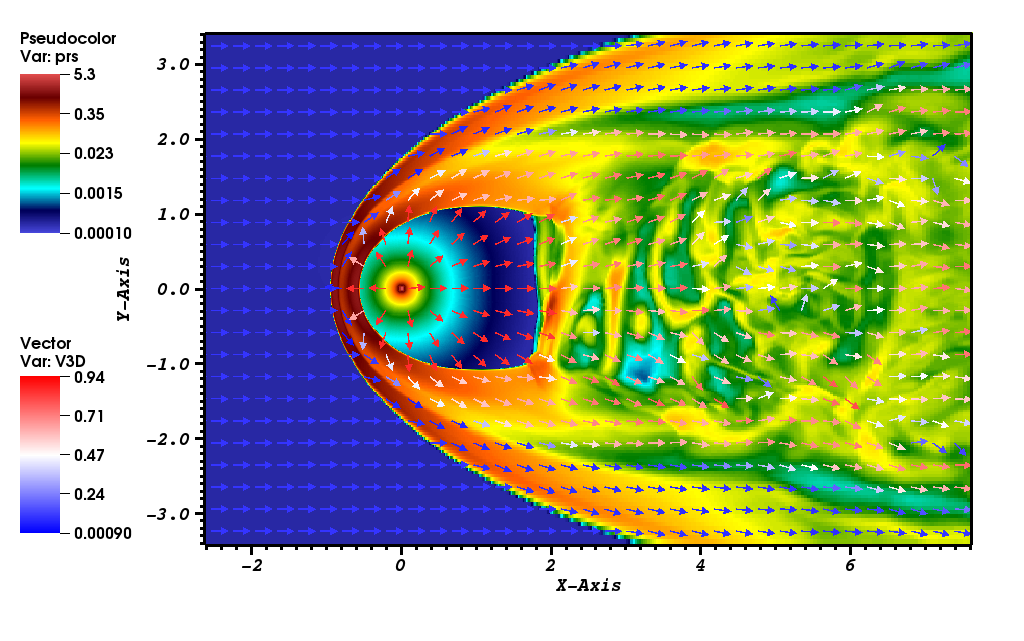}
\includegraphics[width=88mm,angle=-0]{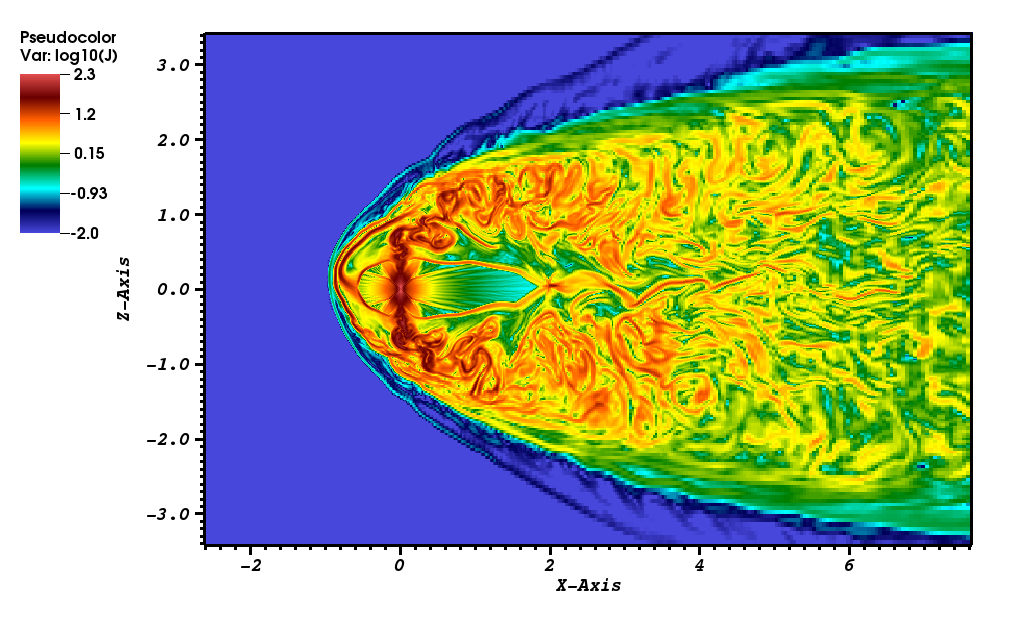}
\includegraphics[width=88mm,angle=-0]{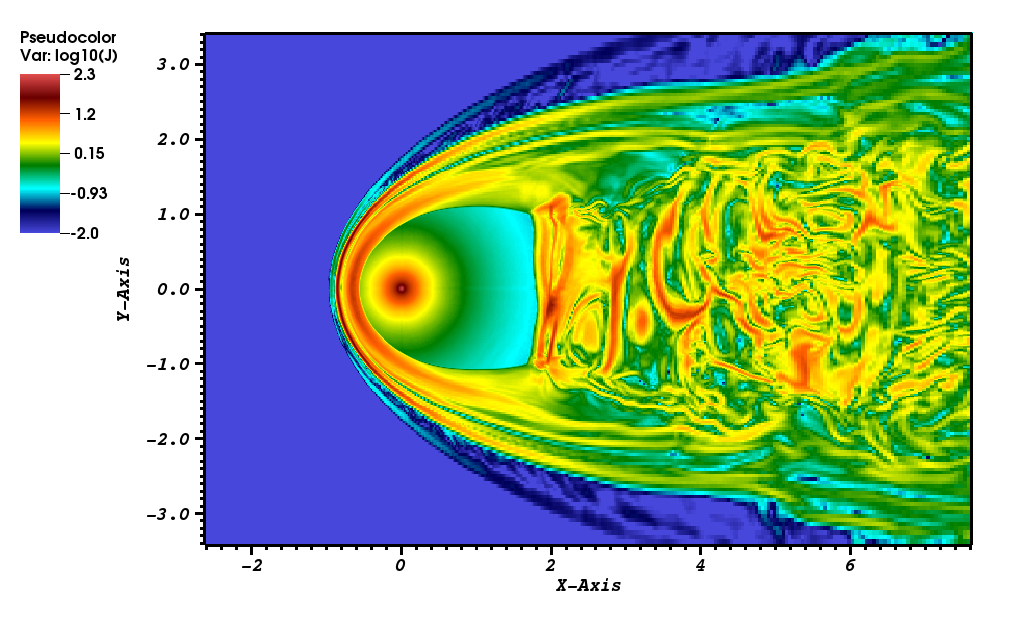}
\includegraphics[width=88mm,angle=-0]{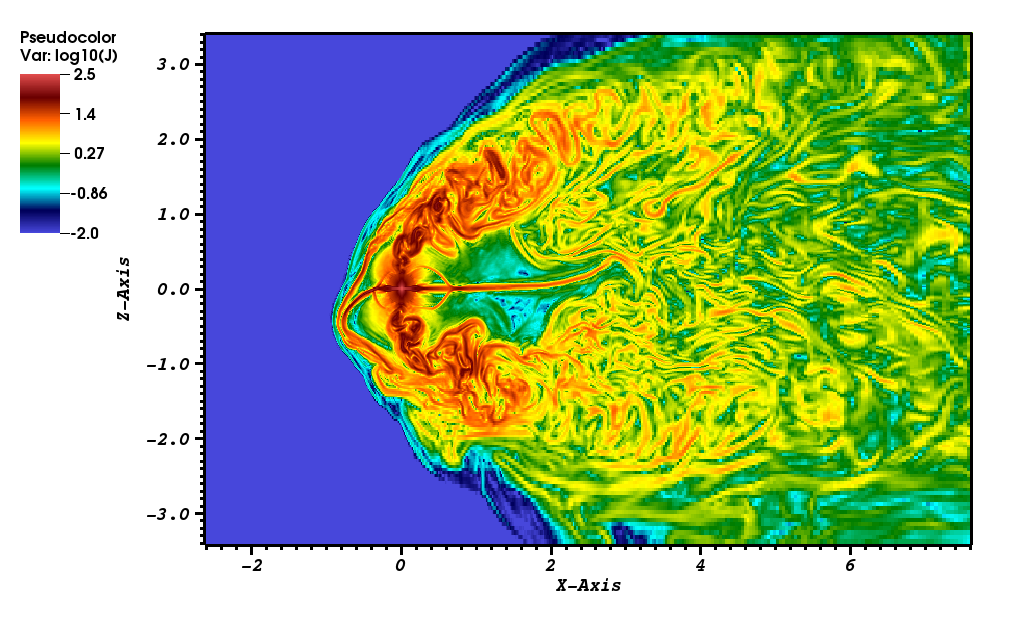}
\includegraphics[width=88mm,angle=-0]{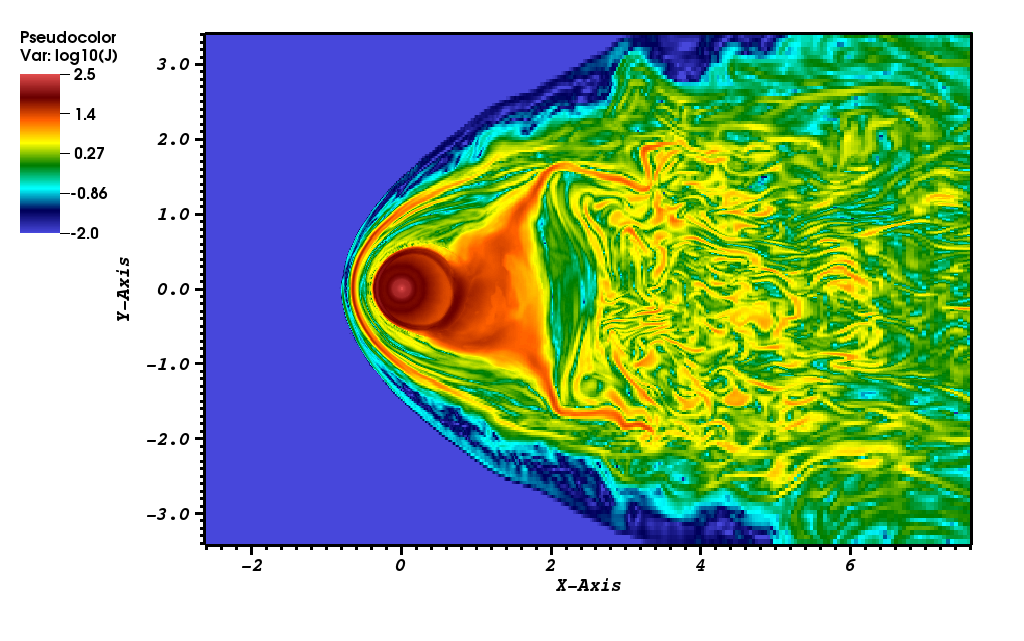}
\caption{Comparison of   \ac{fb} (left column) and \ac{cw} (right column) configurations. 
Top row: pressure (color logarithm) and  velocity field for the model fs1a45, 
middle row:  slice of  current density (color logarithm) for the same model ($\alpha = 45^\circ$),
bottom row: slice of  current density (color logarithm) for the model fs1a10 ($\alpha = 10^\circ$).}
\label{fig:Compare}
\end{figure*}

Results of the modeling of the \ac{fb}/\ac{cw}  configurations are presented in Figs.~\ref{fig:frisbee}, \ref{fig:Compare}. 
The overall structure is not axial-symmetric. The \ac{fb}/\ac{cw} geometry shows the formation of a magnetically confined  plume (jet-like structure) 
initially normal to the pulsar velocity. This plume considerably distorts the shape of the \R, making it to be ``cross-like'', if seen head on. In this case, the equatorial extension
is due to a larger pulsar wind power in the equatorial plane, and the vertical  extension is due to the hoop stresses of the toroidal magnetic field.

%fffffffffffffffffffffffffffffffffffffffffffffffffffffffffffffffffr
%fffffffffffffffffffffffffffffffffffffffffffffffffffffffffffffffff 

In general,  in the \ac{fb}/\ac{cw} configurations the shape of the \TS is more stable as compared to the \ac{rb} case, with no significant oscillations seen in the  Mach disk position (see Fig.~\ref{fig:frisbee}).  
Due to high energy flux in the pulsar wind close to the equatorial plane, Mach disks transform to a  narrow Mach lines in the \ac{fb}/\ac{cw} geometry.

% we should discuss it.
In Fig.~\ref{fig:Compare} we show  electric current distribution for \ac{fb}/\ac{cw} models (fs1a45 and fs1a10). 
In general, all \ac{fb}/\ac{cw} models show  a structure of currents similar to the sketch in  Fig.~\ref{fig:tail-Bfield}. 
We see two outflow currents at the pulsar rotation axis and a return current near the equatorial plane. 
%{\bf It is not clear what should be here....} 
Interestingly, the equatorial current after \TS 
forms a thin layer which is locally stable but after strong perturbation its bends to south or north pulsar's pole region.
The front \TS is quite stable, the back \TS wobbles in a range of $\sim30$\%. This wobbling motion is similar to obtained  
in simulations of the Crab Nebula \citep[\eg,][]{2014MNRAS.438..278P}.   

%{\bf It is not clear what should be here....}
We also point out that the size of the unshocked wind cavity depends on the magnetic inclination angle (compare the middle and bottom rows in Fig.~\ref{fig:Compare}). This is due to the fact that for higher inclination angles the low magnetization equatorial zone occupies larger sector. 
{As a result, this part of the flow has larger compressibility that pushes the shock further out. }.

\subsection{Mixed \acl{rb}~--~\acl{fb} configuration}

\begin{figure*}
\includegraphics[width=88mm,angle=-0]{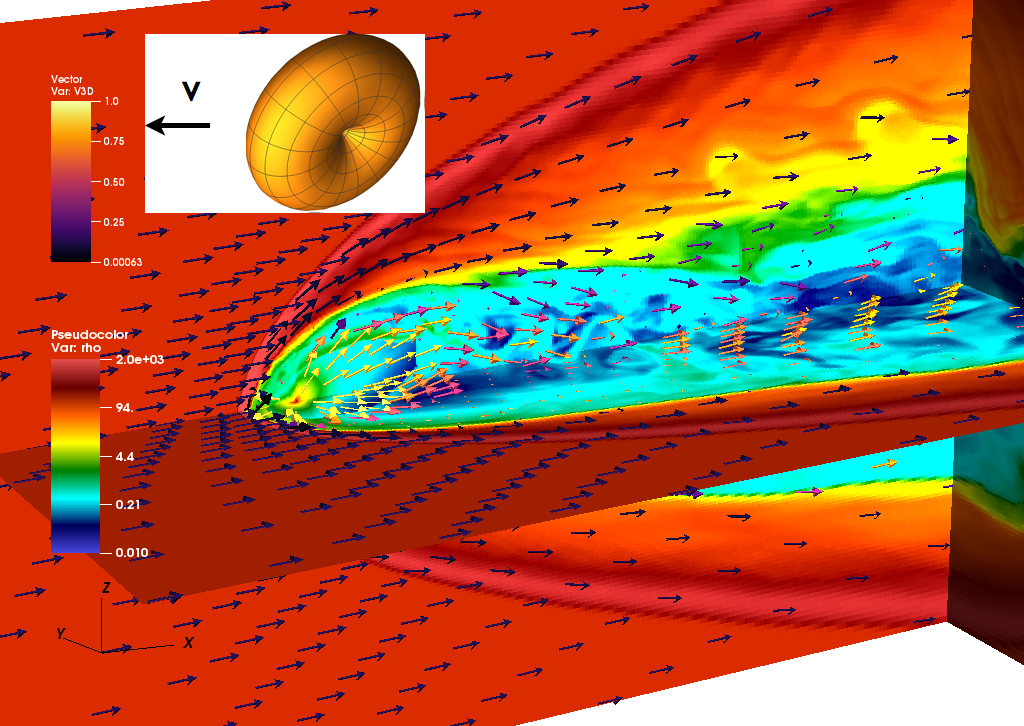}
\includegraphics[width=88mm,angle=-0]{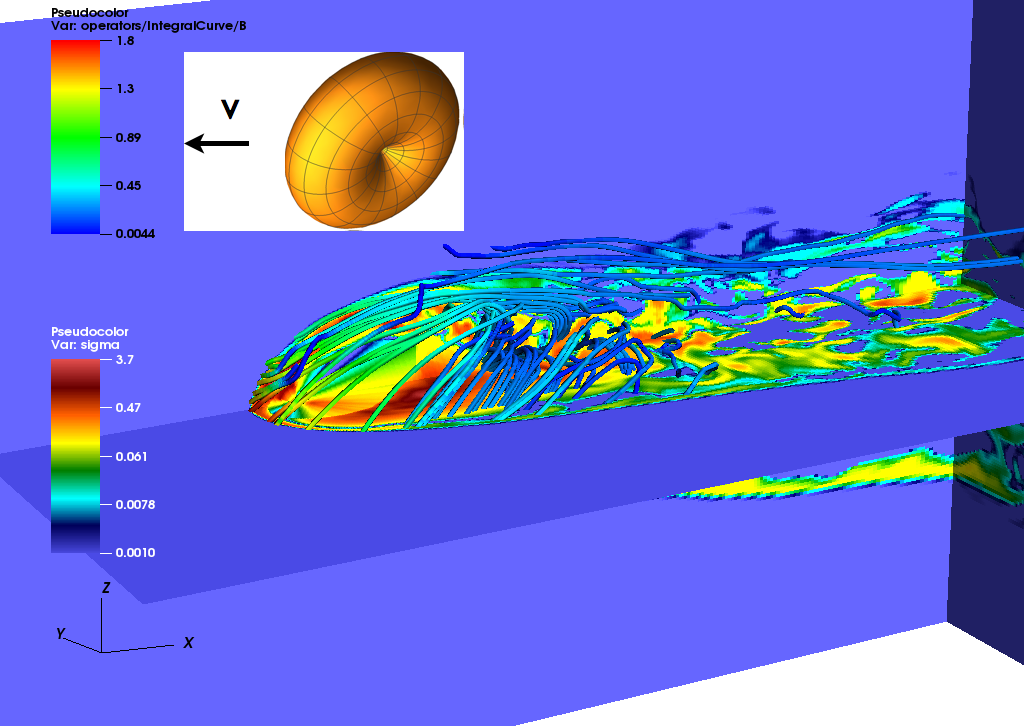}
\includegraphics[width=88mm,angle=-0]{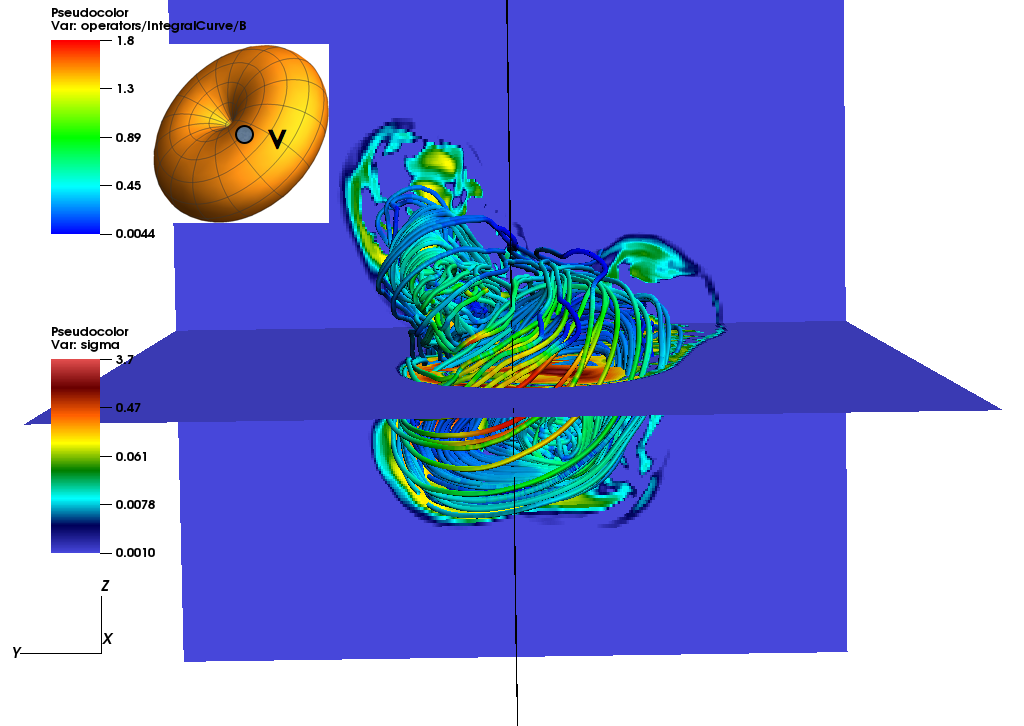}
\includegraphics[width=88mm,angle=-0]{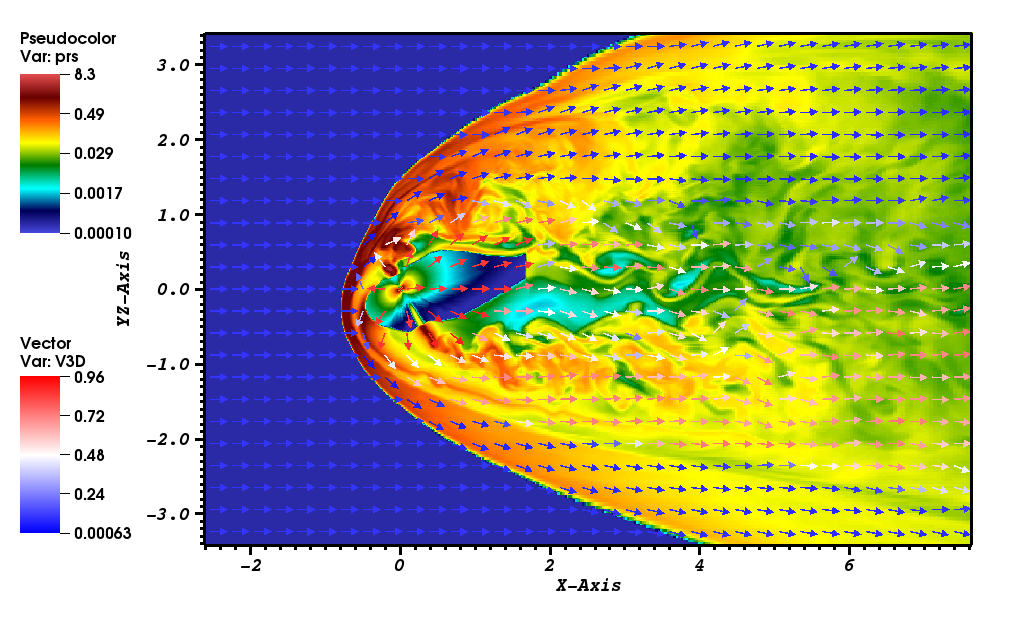}
\caption{Same as Fig.~\ref{fig:frisbee} but in the mixed \ac{rb}~--~\ac{fb} configuration, model fbs1a45 and Pressure/velocity plot at bottom right.  }
\label{fig:mixed}
\end{figure*}

Results of modeling of the mixed \ac{rb}~--~\ac{fb} configuration are shown in Fig.~\ref{fig:mixed}.  The structure of
\R in the \ac{rb}~--~\ac{fb} geometry is a mixture of two discussed above. The most important new effect is
that the whole structures is highly non-symmetric (\ac{fb} and \ac{cw} geometries still have up-down symmetry). This is again due to the effects of magnetic hoop stresses near the rotational axis: in the head part the
wind is slowed down and efficiently confined by the \ISM ram pressure. This allows magnetic stresses to accumulate and
produce a larger distortion than in the tailward part.

The front outflow forms a narrow jet-like structure. 
The tailward outflow is formed by the back ``jet'' and partially by a matter and magnetic field from front ``jet'' which was turned backwards near the head of the bow shock.
Turned back flow has a \Bf directed differently as compared to the back ``jet''. This provides sites suitable for the magnetic field reconnection in the tail. A similar configuration is formed in 
the pure \ac{rb} configuration.

The mixed \ac{rb}~--~\ac{fb} configuration  features a quite stable free wind zone (see Fig.~\ref{fig:mixed}) similarly to  the case of \ac{fb}/\ac{cw} models. 
The shape of pulsar wind \TS  is complicated, similar to the one revealed with \ac{3d} hydrodynamic simulations by \citet{2007MNRAS.374..793V}.

%fffffffffffffffffffffffffffffffffffffffffffffffffffffffffffffffff
%fffffffffffffffffffffffffffffffffffffffffffffffffffffffffffffffff 

%MAX, THIS IS ONE OF THE KEY SECTIONS DESCRIBING YOUR WORK, IT NEEDS TO BE CLARIEFIED/BETTER STRUCTURED, MORE DETAIELD DESCRIPTION  
%OF DIFFERENT PARAMETER REGIMES, HOW SENSITIVE WE ARE TO SOME PARAMETERS ETC. I CANNOT DO IT FOR YOU BECAUSE I JUST DO NOT KNOW THE DETAILS. 
%{\bf BMV: I change he structure of the section, but I need more ideas what to describe. like bullet points.}

%Independently on pulsar spin axis orientation fast mowing \ISM form bow shock structure with an extended cometary tail.
%On another hand the structure of pulsar wind and topology of magnetic field in the tail 
%changes drastically depend on properties of injected pulsar wind orientation.
%In the frame work of \ac{2d} \gls{rmhd} can be simulated only bullet setup.

%WE PERFORMED A NUMBER OF SIMULATIONS IN THE BULLET CONFIGURATION.
% Different magnetizations
% Different inclination angle

\subsection{Overall conclusion: internal \R structure in different configurations}

Summarizing the results of our \ac{3d}  \ac{rmhd} simulations we can formulated the following key findings.
\begin{itemize}
\item Effects of anisotropic wind energy flux and dynamically important \Bfs lead to a very complicated, non-symmetric morphology of bow-shock \Rs.
\item In plasma that originates in the pulsar polar outflows, magnetic pinching generates filamentary regions of high magnetic fields that are prone to the kink instabilities.
\item In the \ac{rb} configuration the location of the Mach disk oscillates with a large amplitude, which is presumably caused  by the developments of the kink instabilities in the tailward region.
\item The internal structure of \R is also sensitive to the magnetic inclination angle: for larger inclinations the unshocked pulsar wind zone occupies a larger volume (compare the middle and bottom rows in Fig.~\ref{fig:Compare}).
\end{itemize}

\section{Emissivity maps}
\label{s:em}

\subsection{Synchrotron emission}
\label{synch}

The observed X-ray emission from \Rs is generated via synchrotron radiation by non-thermal particles, which are presumably accelerated at shocks and/or in 
the reconnection events within the \Rs. Conventional ideal \ac{rmhd} simulations produce only hydrodynamic quantities -- density, thermal pressure, 
velocity, and \Bf. Thus, we have no direct information about energy distribution and density of non-thermal particles. To obtain this additional information one needs
to perform dedicated simulation of evolution of non-thermal particles \citep[see \eg][]{kennel_coroniti_84b,2018ApJ...865..144V}. However, if the particle cooling is dominated by adiabatic losses one can
use a simplified approach and reconstruct the spectrum of non-thermal particles based on \mhd parameters only \citep[see][in Appendix~\ref{ap:syn} we extend this approach to the case relevant here]{bbr18}. 

In bow-shock \Rs the strength of the magnetic field might be quite high, exceeding the field inferred in \Rs around
slow-moving pulsars. Although the structure of the \Bf in \Rs is quite complicated, the characteristic magnetic field
can be obtained from the pressure, \(\sim\rho_\ism V_\ns^2\),  required to support the nebula. Thus, one obtains
\be
B\sim 2\times10^{-4} n^{1/2}_{\ism, 0}V_{\ns,7.5}\rm\,G\,.
\ee
For the magnetic field of strength $B'$ and photon energy \(\epsilon'\) (both in flow co-moving frame) the required \Lf
of the radiating particles is
\be
\gamma_\syn \simeq \sqrt{\frac{2}3\frac{ m_e c \epsilon_{\gamma}' }{e \hbar B'}} = 7
\times 10^6 {\epsilon_{\gamma,\rm 1\, keV}'}^{1/2} {B'}_{-3}^{-1/2}\,,
\label{gammaw}
\ee
which should be easily attainable for non-thermal particles in \Rs  \citep[\eg in Crab Nebula one expects particle acceleration to PeV energies, see][]{kennel_coroniti_84b,1996MNRAS.278..525A}.

The corresponding synchrotron cooling time is 
\be
\begin{split}
  t_{\syn} &\simeq 10^8\, {\epsilon'}_{\gamma,\rm 1\, keV}^{-1/2} {B'}_{-3}^{-3/2}\; {\rm s}\\
  &\simeq10^9{\epsilon'}_{\gamma,\rm 1\, keV}^{-1/2} n^{-3/4}_{\ism, 0}V_{\ns,7.5}^{-3/2}.
\end{split}
% =\left(\frac{3}{2}\right)^{5/4}\left(\frac{ {\hbar } m_e^{5}c^{9}}{ e^{7} {\epsilon_\gamma' }{B'}^{3}}\right)^{1/2}=
%6\times10^7 \, {\epsilon'}_{\gamma,\rm 1\, keV}^{-1/2} {B'}_{-3}^{-3/2}\; {\rm s}.
\label{eq:tsyn}
\ee
This cooling time should be compared to adiabatic cooling time, which can be estimated as the time required for the flow to cross the characteristic hydrodynamic scale, \eg \(r_s\): 
% The travel time of shocked pulsar wind to the stand-off distance is given by
%
\be
t_{\textsc{ad}} \simeq \frac{r_s}{c} \simeq 10^6 n_{\ism, 0}^{-1/2}\, V_{\ns,7.5}^{-1} L^{1/2}_{w,36}\mbox{ s}\,.
\label{eq:tnt}
\ee
% This time $t_{nr}$ can be treated as the typical time for adiabatic loses.
The ratio of synchrotron and adiabatic cooling time is
\be
\frac{t_\syn}{t_\textsc{ad}} \simeq 5\times10^2 {\epsilon'}_{\gamma,\rm 1\, keV}^{-1/2} n_{\ism, 0}^{-1/4}\, V_{\ns,7.5}^{-1/2} L^{-1/2}_{w,36}\,,
 \label{eq:rad_eff}
\ee
which implies that in bow-shock \Rs the cooling of particles, responsible for \xray emission, proceeds predominately due
to adiabatic losses. Thus, one can utilize the simple approach for computing synchrotron radiation (see Appendix
\ref{ap:syn}).

% The synchrotron emissivity is then proportional to the  NT particles' energy density devided by  the local  cooling time.
% \be
% \epsilon_{\rm SY} \propto \frac{u_{\rm cell,NT}}{t_{\rm syn}}.
% \label{eq:symap}
% \ee
% A conventional approach is to scale the properties of the non-thermal population with the   hydrodynamic quantities, 
% effectively assuming that  some fixed  fraction of thermal energy is transferred to non-thermal particles. 

To compute synthetic synchrotron emission maps, we follow a procedure outlined below.
\begin{itemize}
\item Our simulations produce \ac{3d} distribution of pressure, density, velocity and \Bf.
\item Using gas pressure, $p$, and \Bf, \(\B'\), we calculate the synchrotron emissivity according to various prescriptions described in Appendix~\ref{ap:syn}; we also calculate the corresponding \ac{ic} signal, Appendix \ref{ap:IC}.
\item For the obtained local synchrotron emissivity integrate the emissivity,  
 assuming optically thin regime and taking into account the local velocity and the corresponding Doppler factor.
\end{itemize}

We adopt the \R around \ac{j1509} as a prototype \Rs for our simulations. This system is powered by a pulsar with spin
period of \(89\)~ms and period derivative $9.2\times10^{-15}$ (so that its spin-down power is $L_w = 5.1\times10^{35}\ergs$), 
which moves through the \ISM with proper velocity of $V_{\ns}>1.6\times10^7 n^{1/2}_{\ism,0} \mbox{ cm s}^{-1}$
\citep{kkr16}. Adopting \(n_\ism = 1\,\rm cm^{-3}\) and  the lower limit value as the pulsar velocity, we estimate the stand-off distance (the key parameter for
the simulations) to \(r_s = 5.6 \times 10^{16}\rm\,cm\).  That corresponds to the characteristic magnetic field of $0.1$~mG in the \R.

% To calculate the synchrotron emissivity we first find the thermal energy in a given cell in the flow frame (here $dV' = \Gamma  dV$):
% \be
% u_{\rm cell,NT} = \eta_{\rm NT}  u_{\rm cell} = \eta_{\rm NT}  3P  \Gamma dV  , 
% \label{eq:uNT}
% \ee
% where $3P$ is internal energy, $\dif{V}$ is the cell volume in the lab frame, $\Gamma$ is the
% bulk Lorentz factor, and $\eta_{\rm NT}$ is the non-thermal particle fraction parameter \citep[see details in ][]{bbr18}.

% The radiation of a fast moving plasma is affected by Doppler boosting,
% so observer will see 
% \be
% \epsilon_{\rm SY,\delta} \propto \frac{u_{\rm cell,NT}}{t_{\rm syn}} \frac{\delta^3}{\Gamma}.
% \label{eq:symapdelta}
% \ee

To make clearly visible the \xray morphology, we use different quantities to produce the synthetic maps, depending on the orientation of the line-of-sight. Namely, we found that the emission intensity maps are more illustrative in the case if the pulsar moves toward the observer. If the pulsar moves side way, then we plot the square-root of the emission intensity. The latter quantity is somewhat arbitrary, chosen to allow a better highlighting of faint \xray features.

% we produce two kinds of  emission  maps,  scaling the  emissivity with    $\int \epsilon_B \delta_i^3/\Gamma \dif{x_i}$ if  the pulsar moving towards the observer, and  scaling the  emissivity with $\sqrt{\int \epsilon_B \delta_i^3/\Gamma \dif{x_i}}$  
% if pulsar moves side way. The latter, somewhat arbitrary,  choice allows us to  better highlight weaker features

\subsection{Emission maps -- \acl{rb} configuration}

%fffffffffffffffffffffffffffffffffffffffffffffffffffffffffffffffff
\begin{figure*}
\includegraphics[width=70mm,angle=-0]{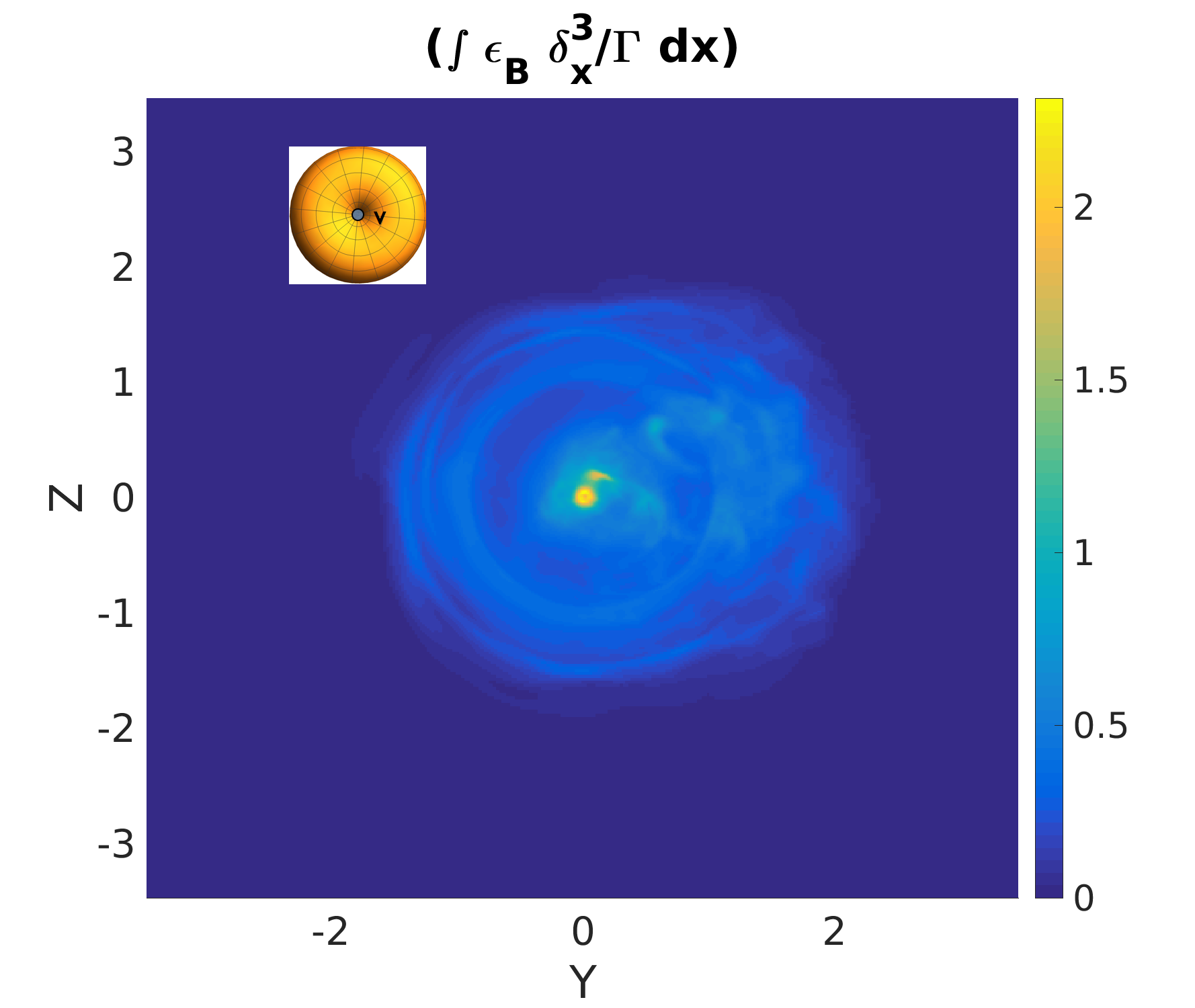}
\includegraphics[width=105mm,angle=-0]{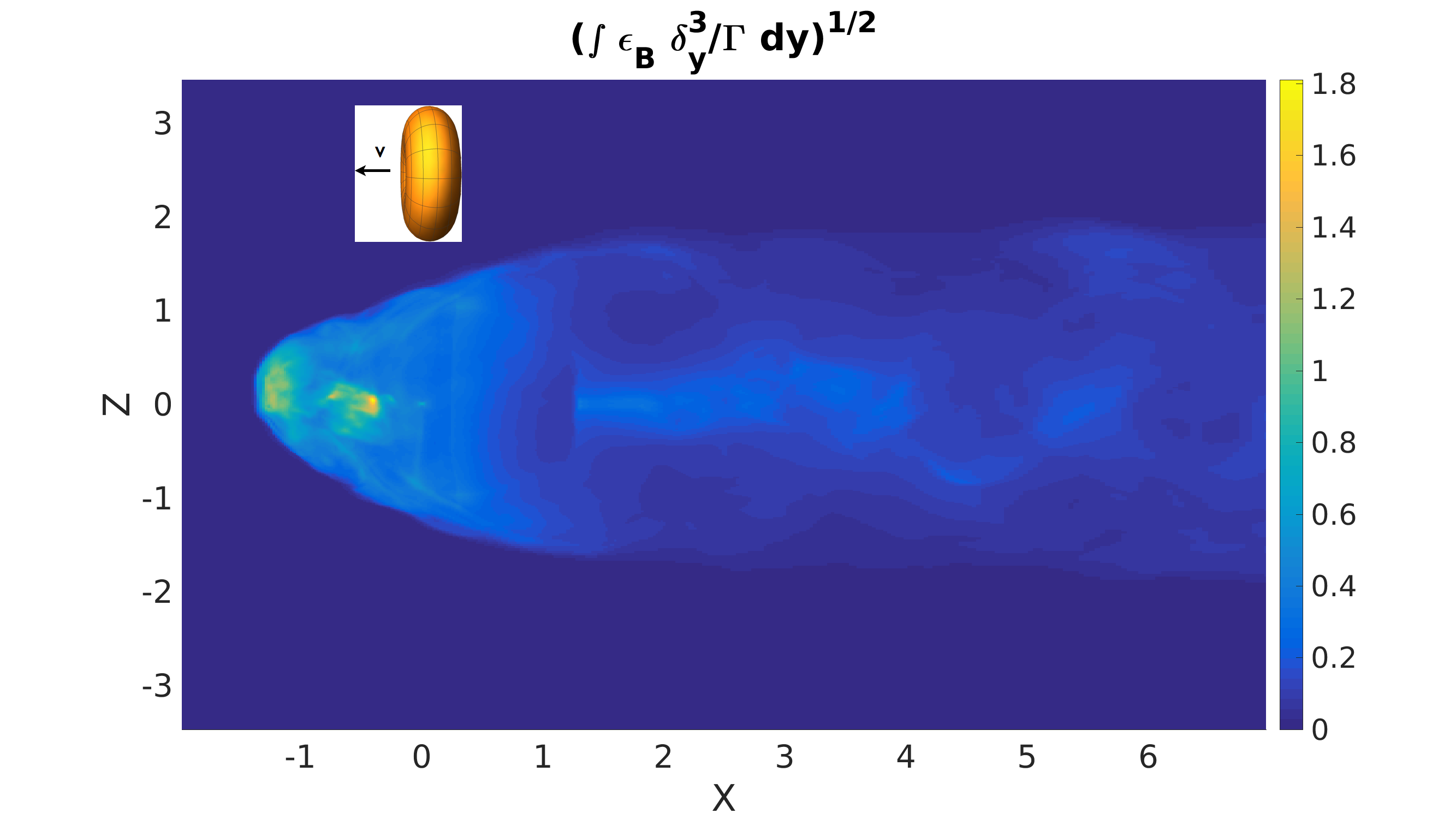}
\includegraphics[width=70mm,angle=-0]{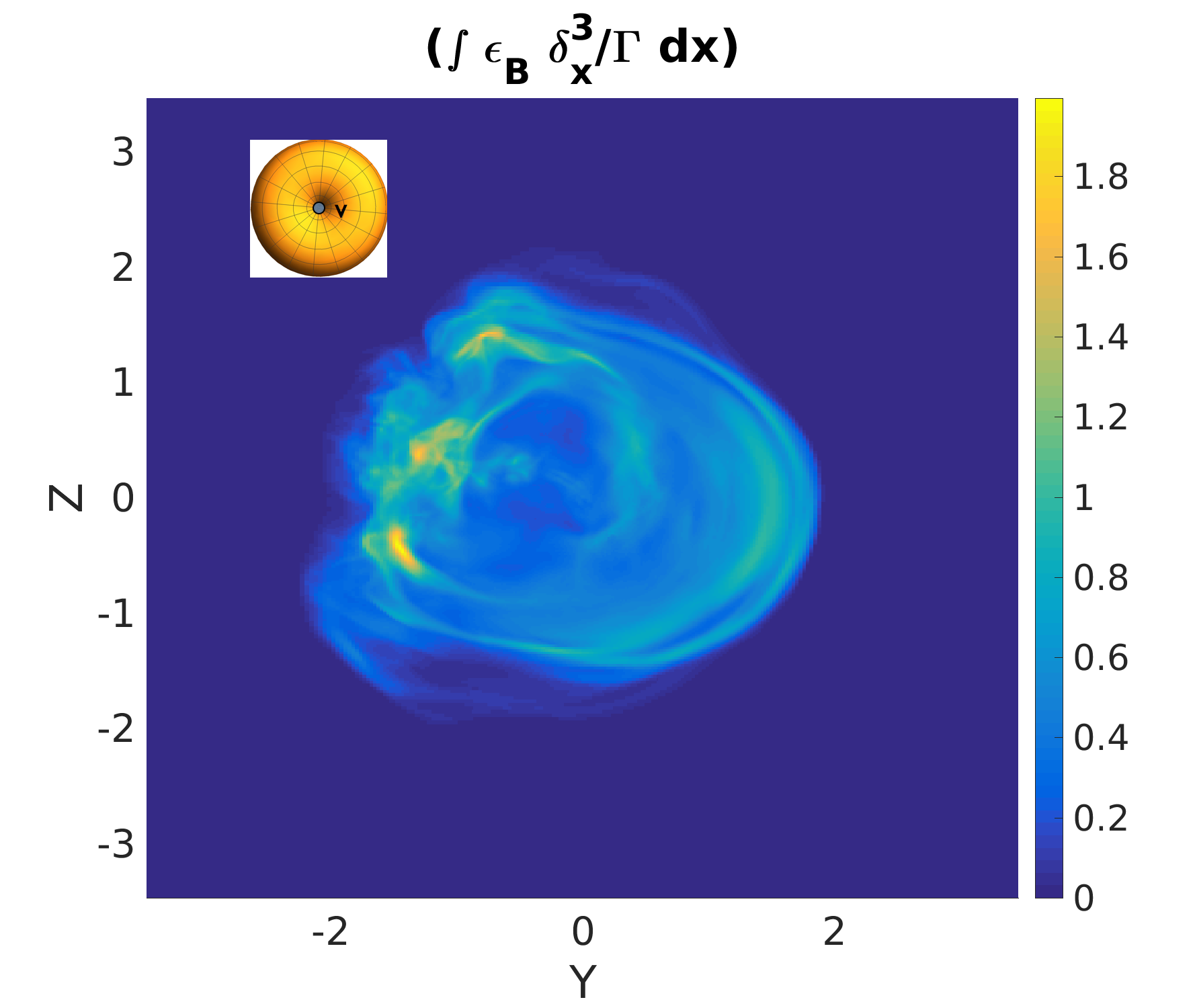}
\includegraphics[width=105mm,angle=-0]{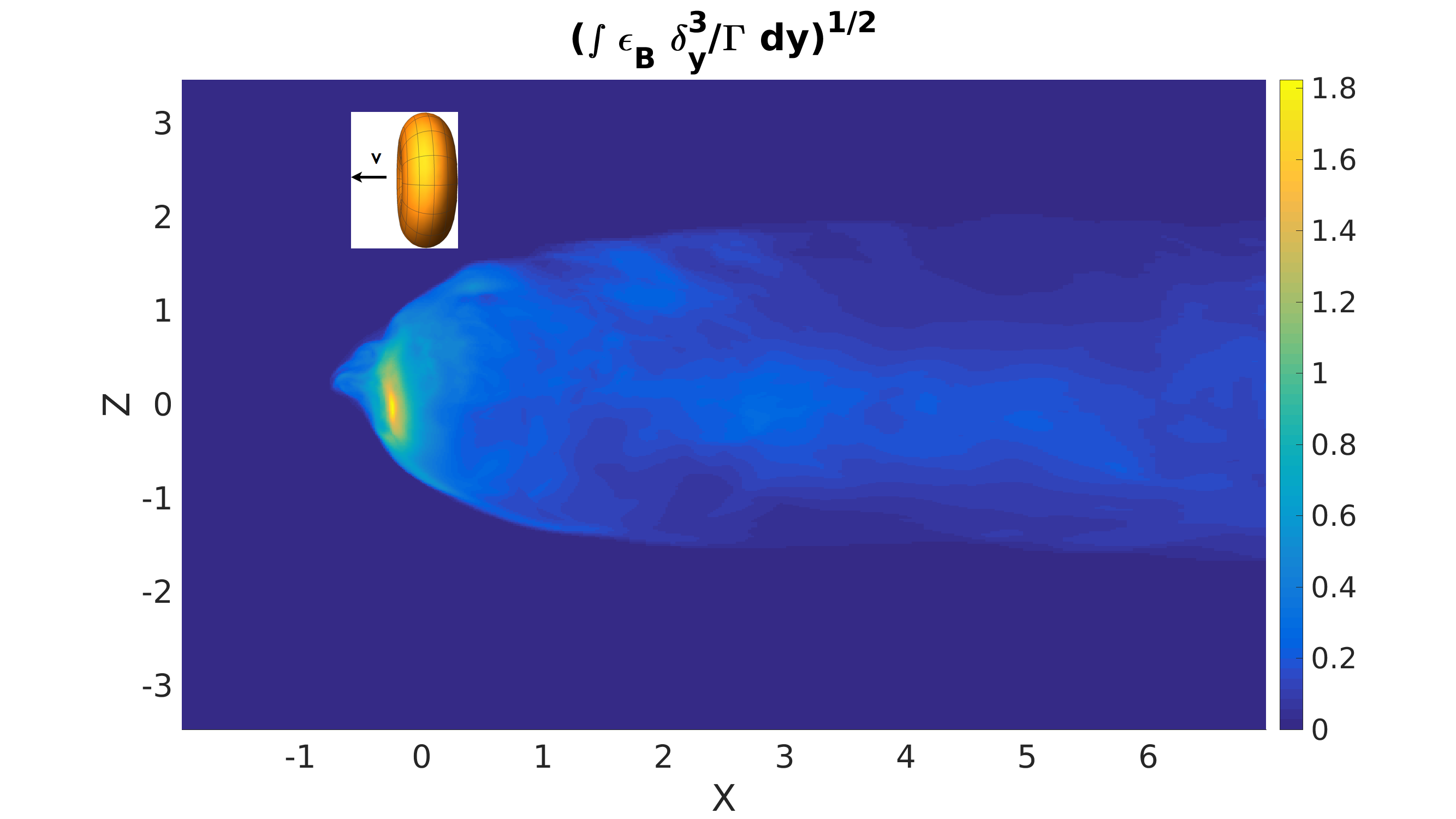}
\caption{Emissivity map   in the \ac{rb} configuration  projected along X (left column)  and Y  (right column) axis for the models bs1a45 ($\sigma = 1$, top) 
bs01a45 ($\sigma = 0.1$, bottom).}
\label{fig:b_Ebd}
\end{figure*}
%fffffffffffffffffffffffffffffffffffffffffffffffffffffffffffffffff 

Synthetic synchrotron maps for models  bs1a45 (\ac{rb} configuration, $\sigma_0 = 1$) and bs01a45 (\ac{rb} configuration, $\sigma_0 = 0.1$) are presented in 
Fig.~\ref{fig:b_Ebd} (top and bottom panels, respectively). The pulsar moves towards the observer (left panels) and to the left (right panels).
In the case of the low wind magnetization (\(\sigma_0=0.1\), bs01a45) the head of the bow shock is bright while the  structures in the tail are barely distinguishable.
This result is similar to the emissivity maps obtained by \citet{bad05}. If the pulsar moves toward the observer, a bright ring-like structure with characteristic radius 
$\approx 1.5 r_s$ is seen.
In the case of high magnetization, \(\sigma_0=1\), \R has a more complicated morphology, similar to a `mushroom' if seen off-axis. The head ``jet'' forms the mushroom cap and the tailward ``jet'' after 
the Mach disk forms the stalk of the mushroom. If the pulsar moves toward the observer, the forward jet emission dominates the morphology, thus the \R appears as a compact (but variable, \(r_s/c\sim\)~month) source.

\subsection{Emission maps --  \acl{fb} configuration}

%fffffffffffffffffffffffffffffffffffffffffffffffffffffffffffffffff
\begin{figure*}
\includegraphics[width=88mm,angle=-0]{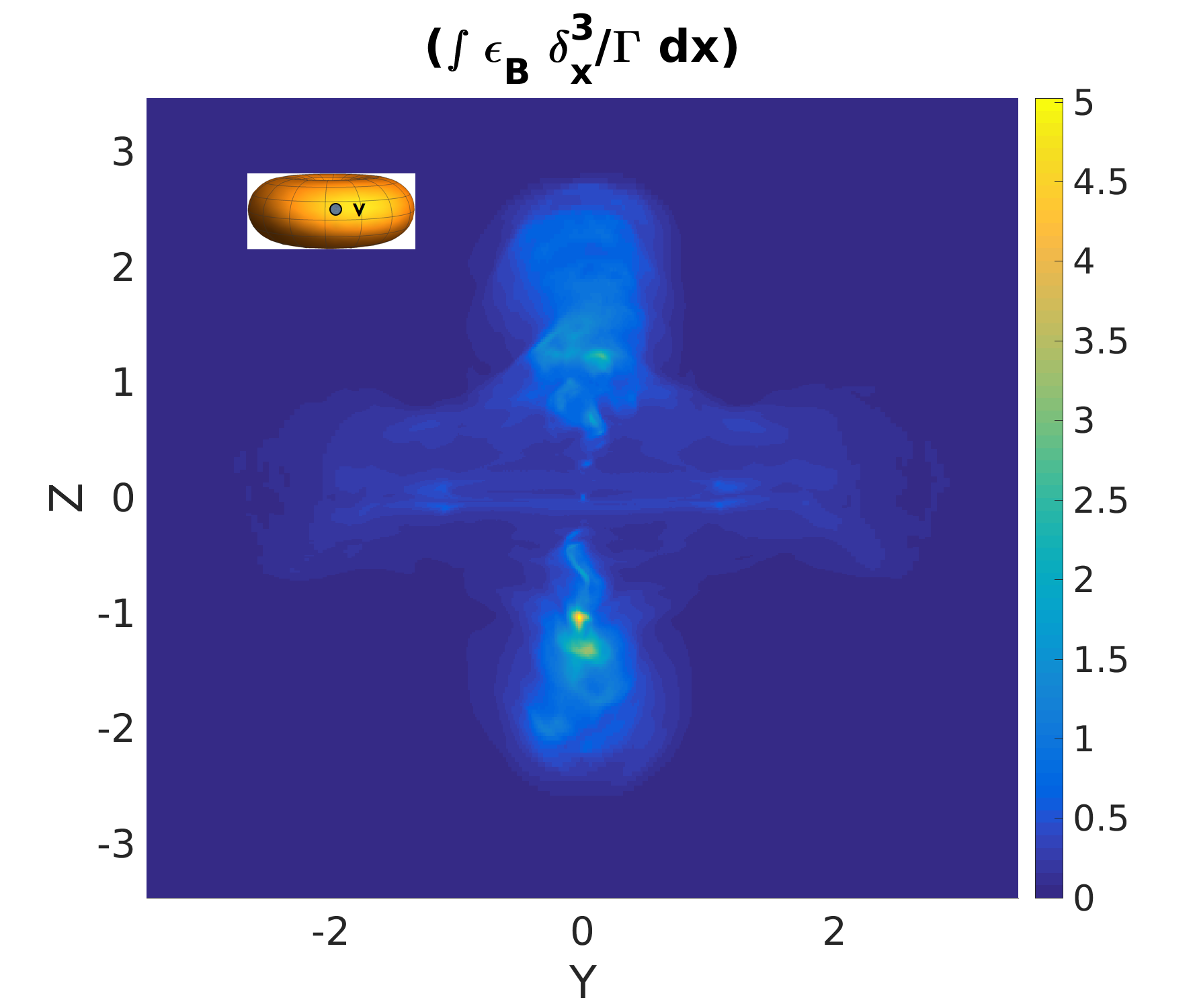}
\includegraphics[width=88mm,angle=-0]{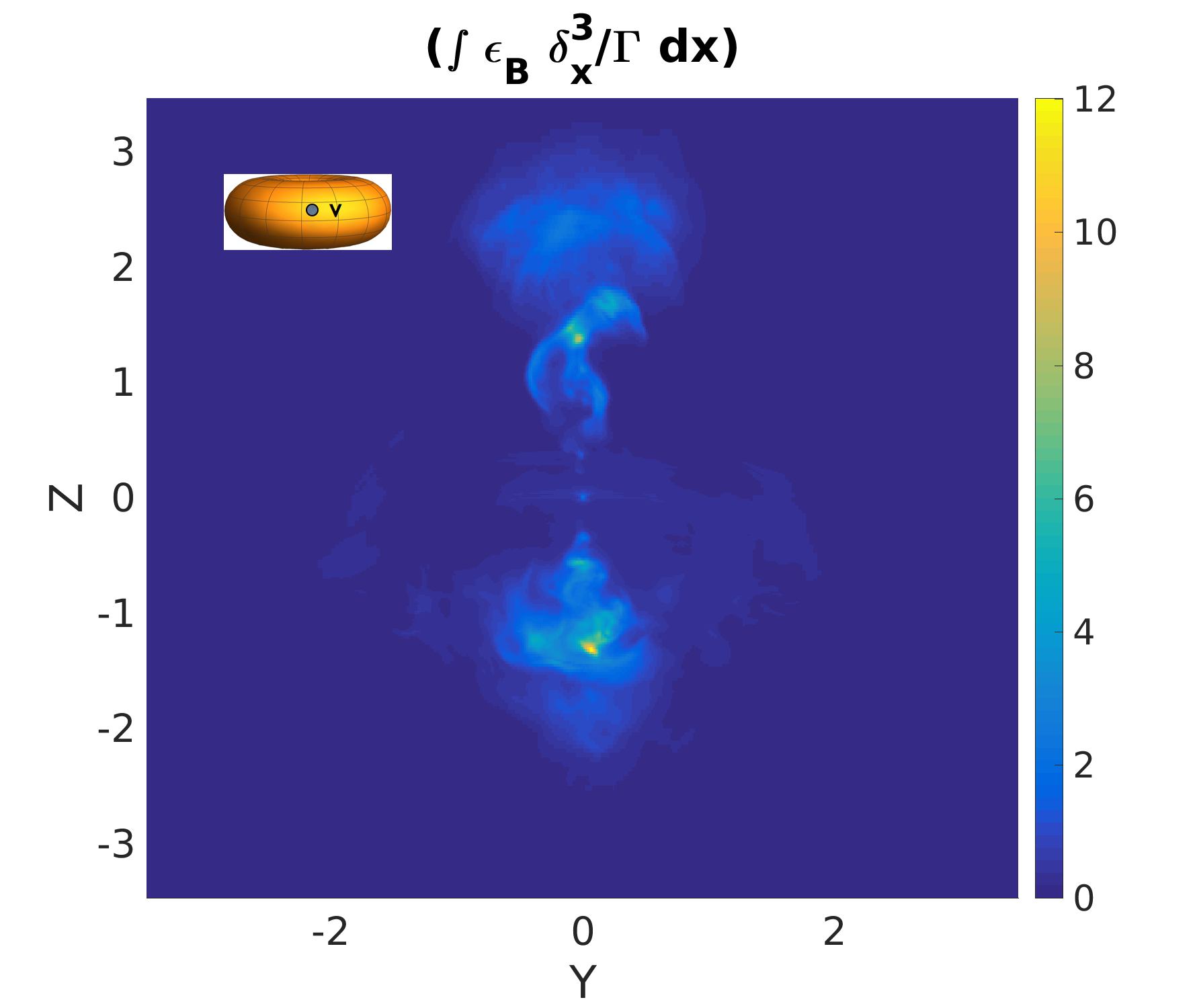}
\includegraphics[width=88mm,angle=-0]{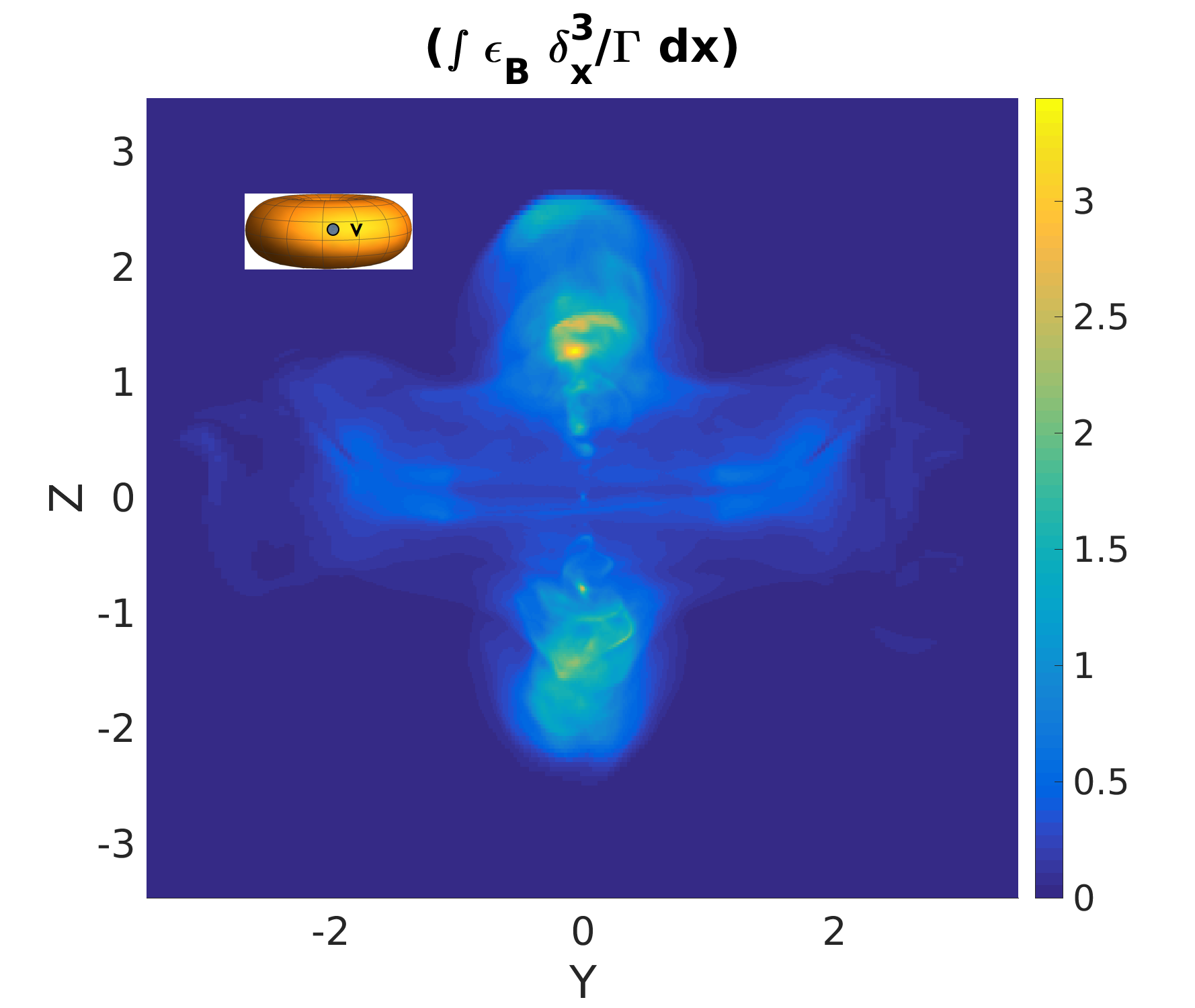}
\includegraphics[width=88mm,angle=-0]{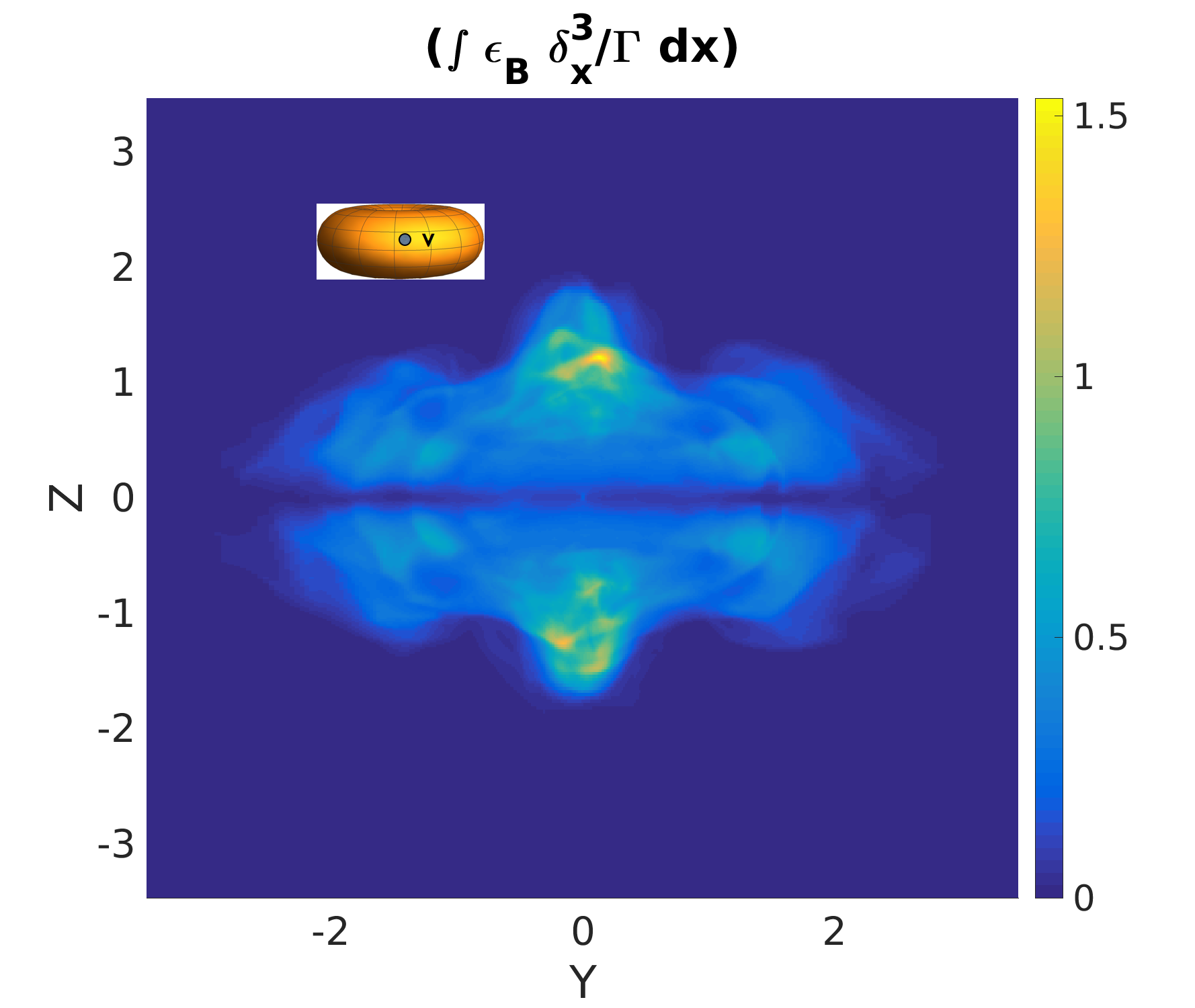}
\caption{Synchrotron   emissivity map  projected along X axis for the models:
top left fs3a45, top right fs1a10, bottom left fs1a45 and bottom right fs01a45. 
%WHAT ARE THE PARAMTERS IN THESE MODELS? AND ALL OTHER MAPS
%1) Please use table 1
%2) first letter - case; second magnetization and fird angle. We use similar upproach with Serguei, no one complains.
}
\label{fig:X_Ebd}
\end{figure*}
%fffffffffffffffffffffffffffffffffffffffffffffffffffffffffffffffff 

%fffffffffffffffffffffffffffffffffffffffffffffffffffffffffffffffff
\begin{figure*}
\includegraphics[width=88mm,angle=-0]{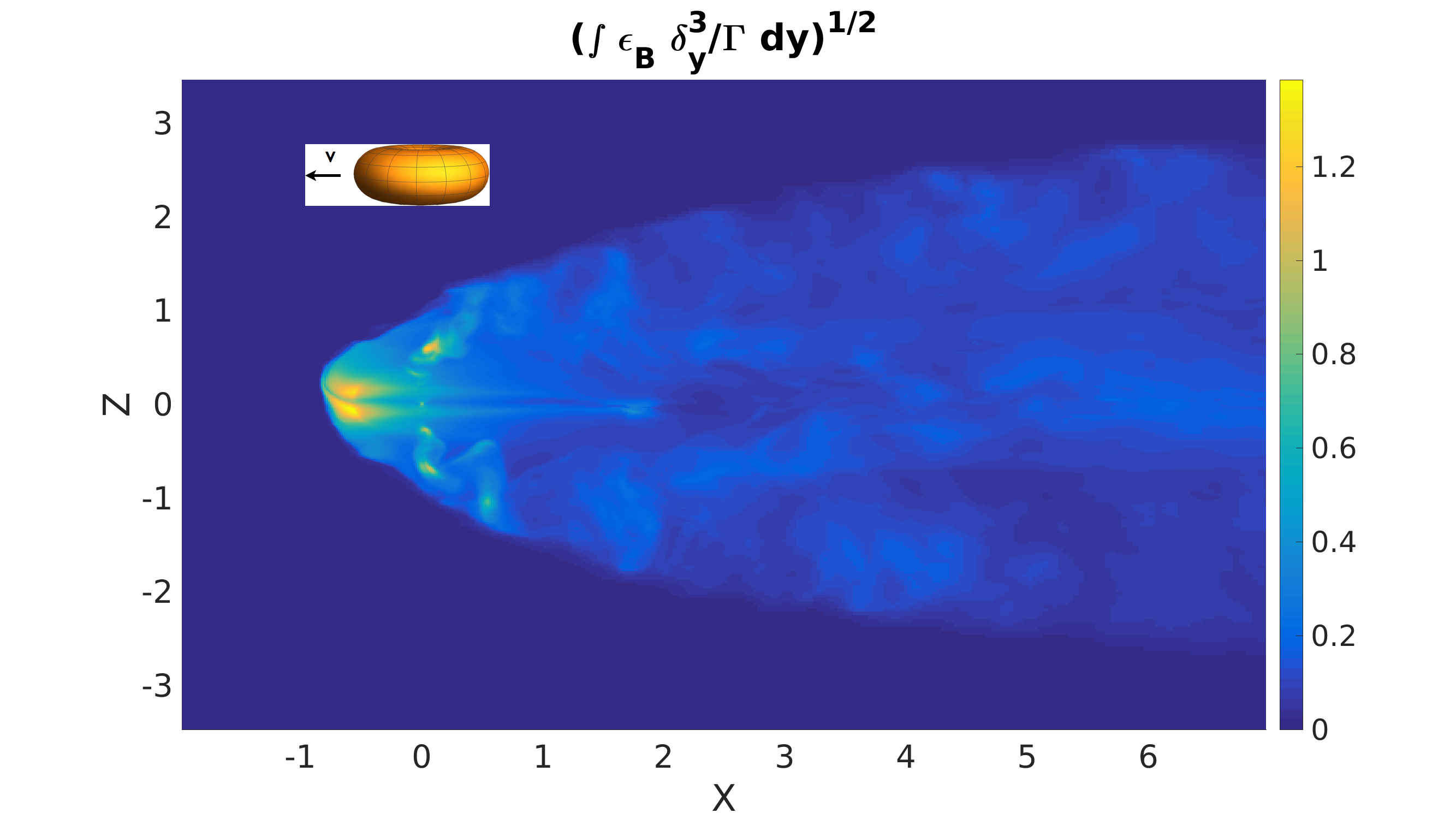}
\includegraphics[width=88mm,angle=-0]{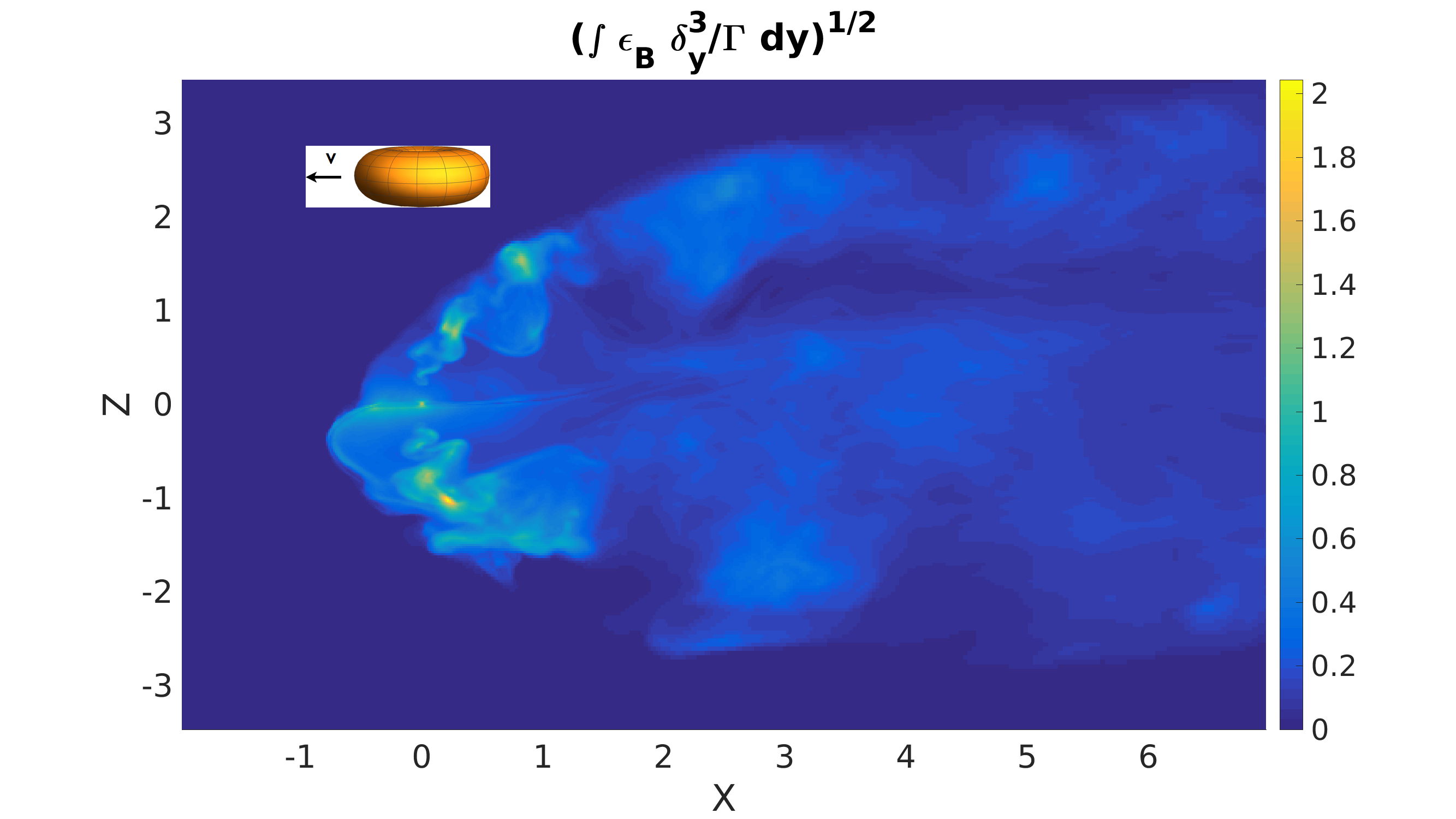}
\includegraphics[width=88mm,angle=-0]{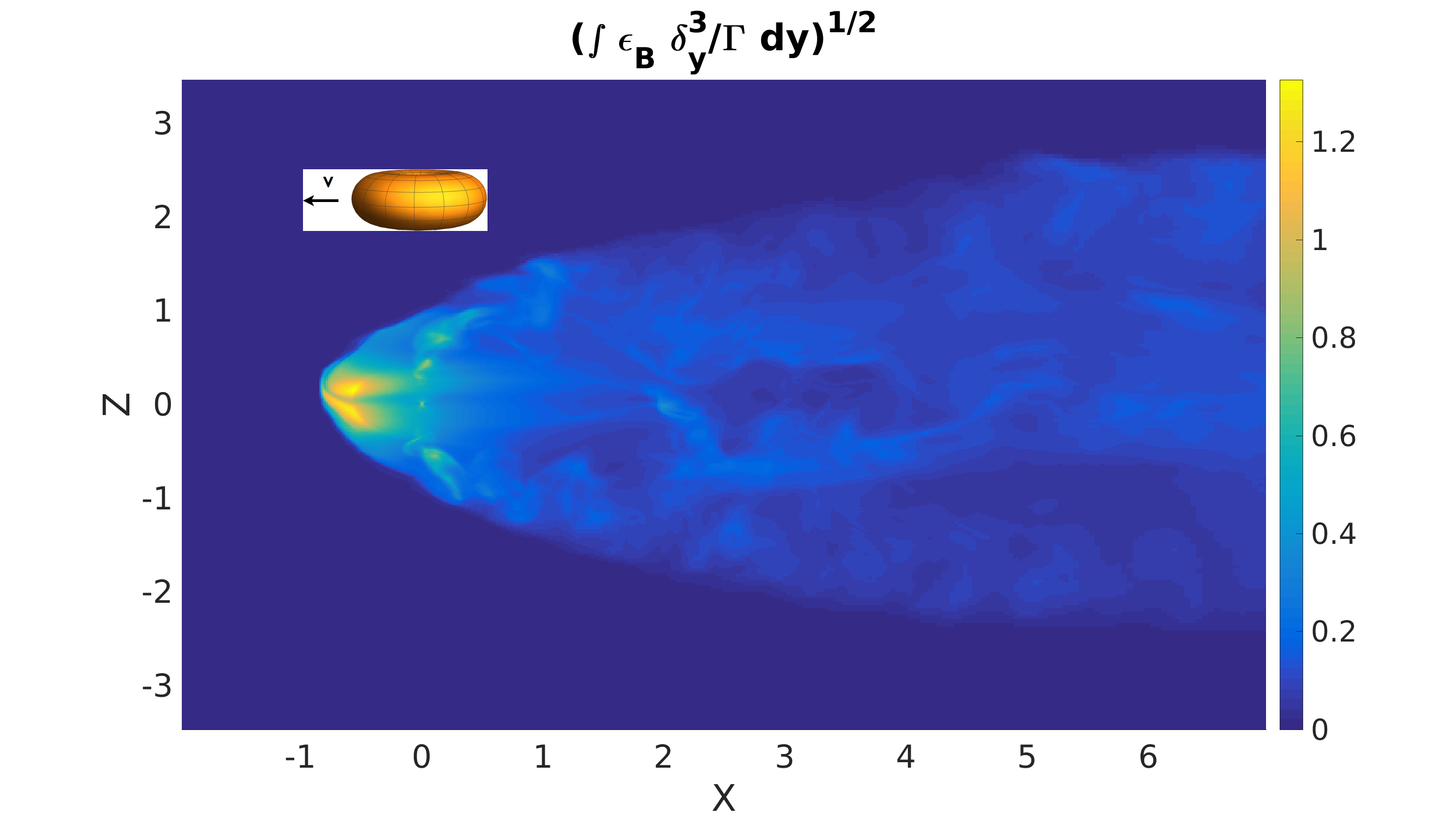}
\includegraphics[width=88mm,angle=-0]{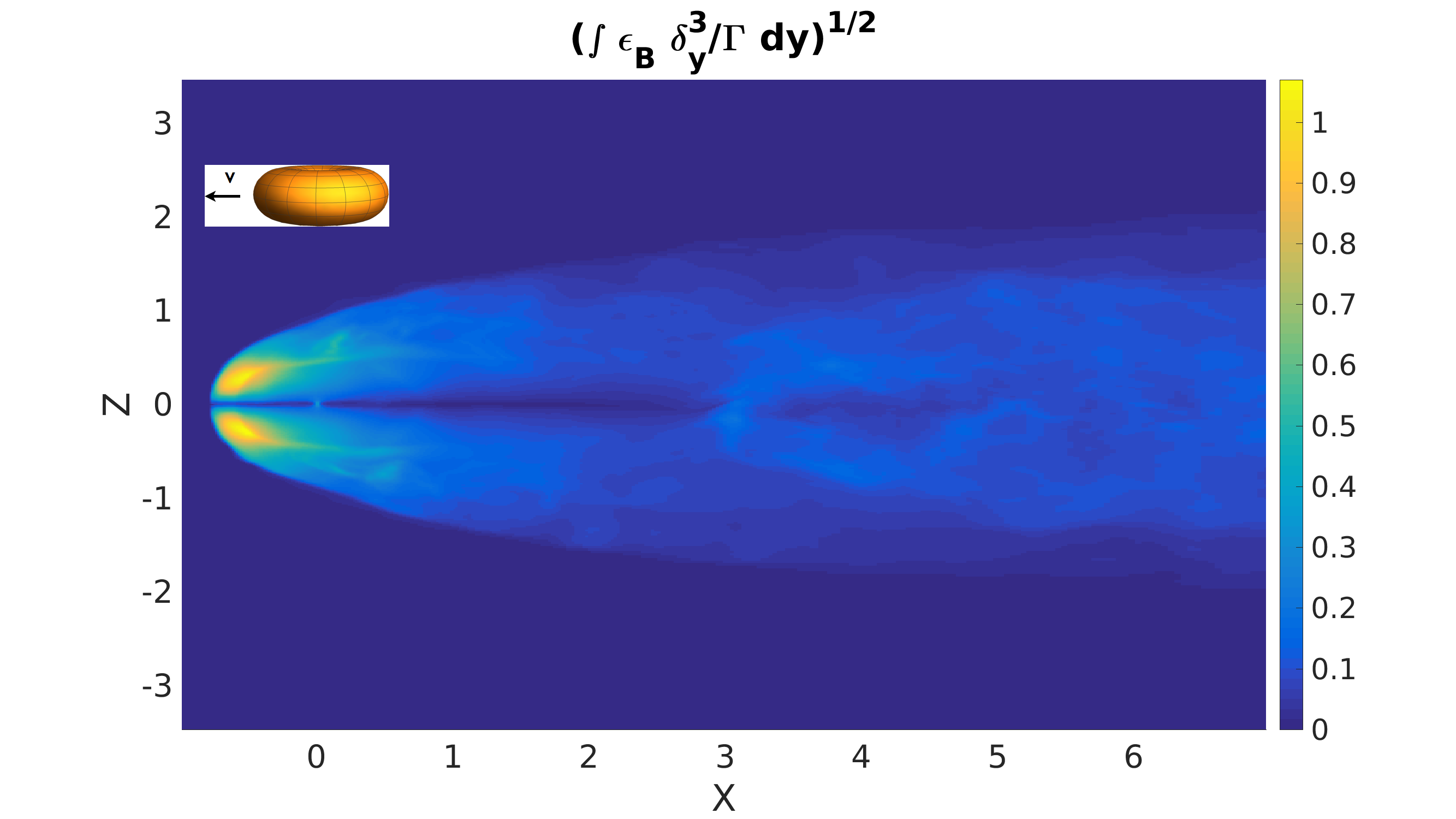}
\caption{Synchrotron   emissivity map   projected along Y axis (\ac{fb}) for the models:
top left fs3a45, top right fs1a10, bottom left fs1a45, and bottom right fs01a45. Here we change magnetization $\sigma_0$ and $\alpha$.}
\label{fig:Y_Ebd}
\end{figure*}
%fffffffffffffffffffffffffffffffffffffffffffffffffffffffffffffffff 

In Fig.~\ref{fig:X_Ebd}  we show synthetic synchrotron maps obtained for the \ac{fb}/\ac{cw} configuration for a pulsar moving toward the observer 
(models fs3a45, fs1a10, fs1a45, and fs01a45). In this case \R appears like an extended source with a typical extension of \(5r_s\). Unless the wind magnetization is small, \(\sigma_0\leq0.1\), the morphology is dominated by two jets). Two equatorial extended features appear visible for \(\sigma_0\leq1\).
For high magnetization $\sigma_0 > 1$ or smaller inclination  angle $\alpha$, jet-like plume structure gets significantly brighter (note the change of the scale in different panels of Fig.~\ref{fig:X_Ebd}).

%Similarly, the equatorial part of the outflow is less pronounced  (\eg, model fs1a10, right upper panel in Fig.~\ref {fig:Y_Ebd}). For  smaller magnetization $\sigma \ll 1$ the equatorial torus structure becomes brighter while  the jet structure  disappears  (\eg, model fs01a45, right low panel in Fig.~\ref {fig:Y_Ebd}). 

In Fig.~\ref{fig:Y_Ebd} we show the synthetic synchrotron maps for the \ac{fb} configuration. If the pulsar inclination is large, \(\alpha=\upi/4\), the bright  head part dominates the morphology. With increasing wind magnetization the jet-like plume becomes more pronounced. For \(\sigma_0\geq1\), a characteristic ``whiskers'' morphology becomes visible: arc-like features trace approximately the shape of the bow shock. If the inclination angle became small, \eg  $\alpha=\upi/18$,  the head part gets significantly fainter and the jet-like plume is clearly seen. 
%For higher magnetization the structure is dominated by extended ``whiskers'' - arc-like features that trace approximately the shape of the bow shock. 
{One of the most prominent  features is a short equatorial tails (top rows and left bottom panel in Fig.~\ref{fig:Y_Ebd}). 
This future is formed due to heating triggered by the dissipation of the \Bf at the equatorial current sheet.}

%Bullet configuration forms mushroom like structure (wide head and narrow tail).

\subsection{Emission maps -- \acl{cw} configuration}

%fffffffffffffffffffffffffffffffffffffffffffffffffffffffffffffffff
\begin{figure*}
\includegraphics[width=88mm,angle=-0]{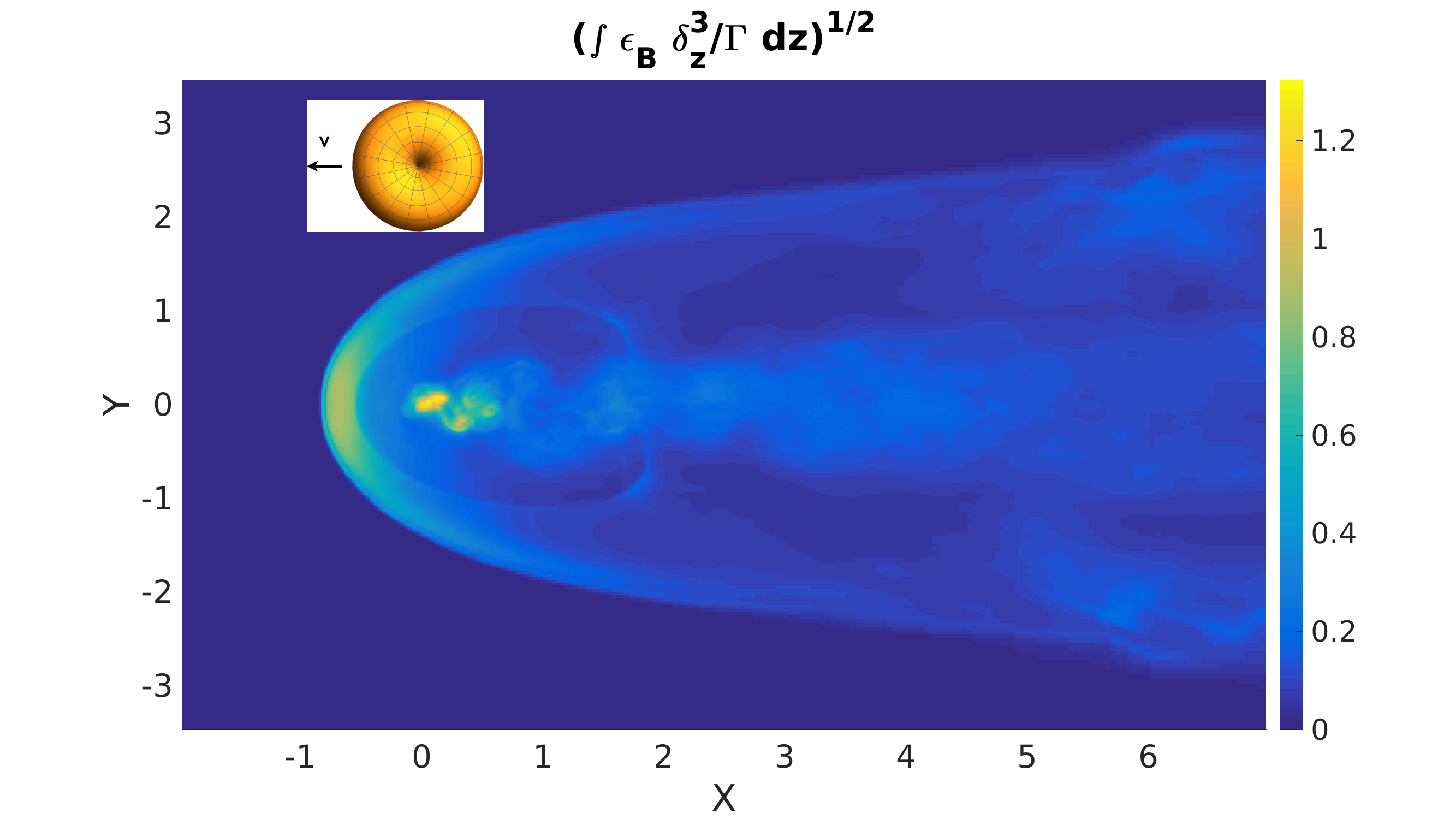}
\includegraphics[width=88mm,angle=-0]{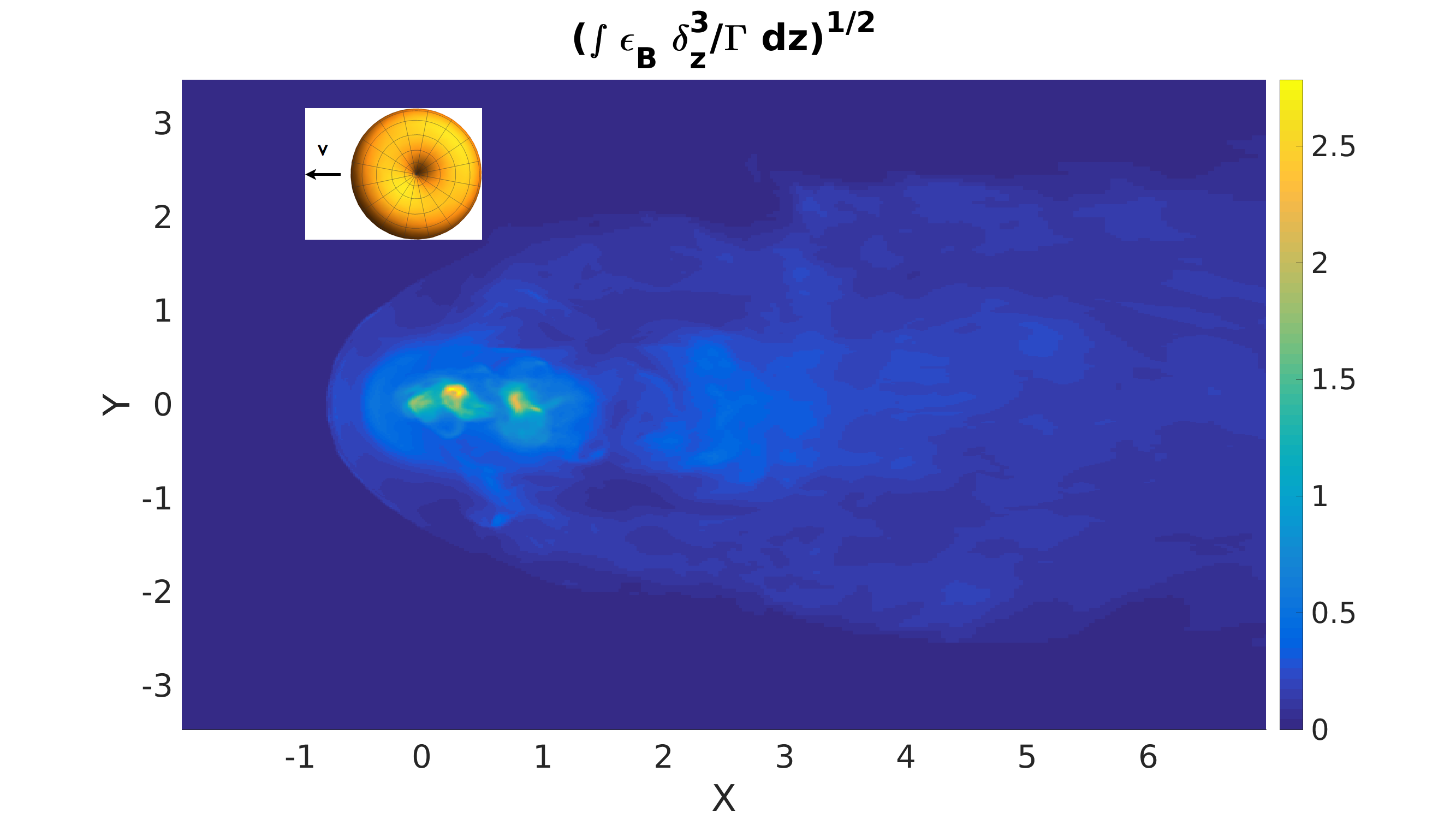}
\includegraphics[width=88mm,angle=-0]{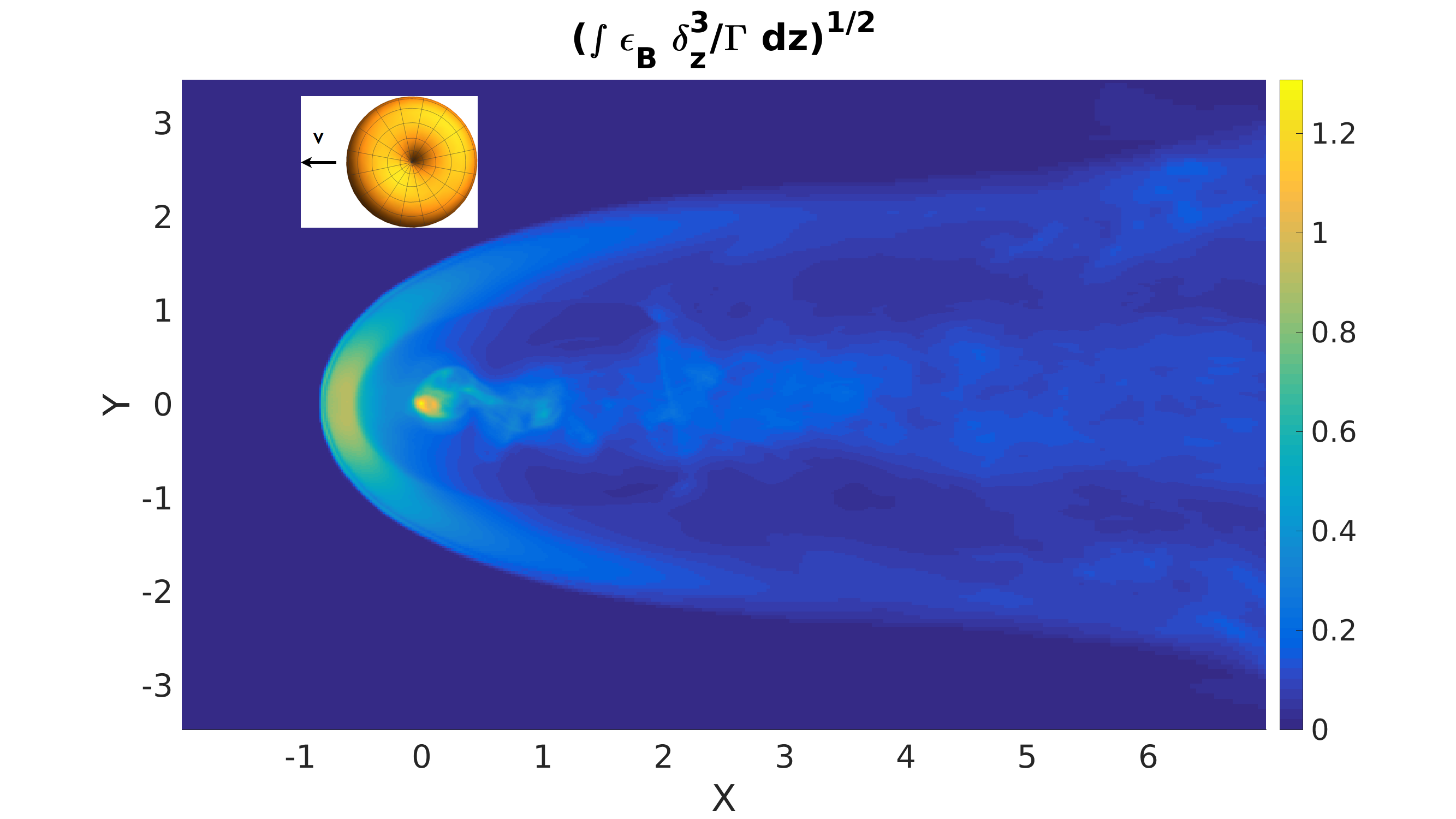}
\includegraphics[width=88mm,angle=-0]{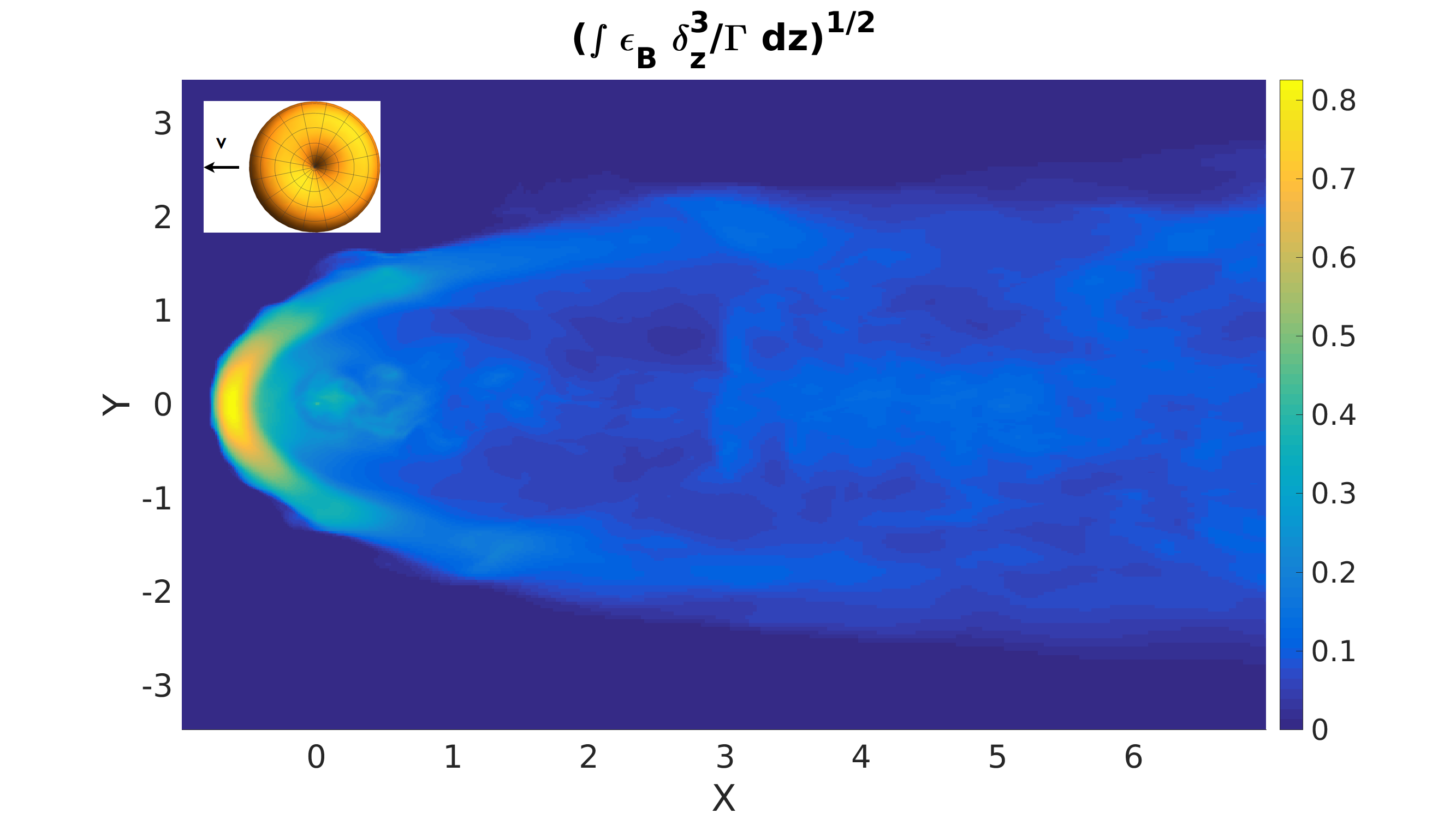}
\caption{Synchrotron   emissivity map   projected along Z axis (\ac{cw}) for the models:
 fs3a45 (top left, $\sigma_0 =3,\; \alpha=\upi/4$),  fs1a10 (top right, $\sigma =1,\; \alpha=\upi/18$),  fs1a45 (bottom left,  $\sigma_0 =1, \; \alpha=\upi/4$),  
 fs01a45 (bottom right,  $\sigma_0 =0.1,\; \alpha=\upi/4$)}
\label{fig:Z_Ebd}
\end{figure*}
%fffffffffffffffffffffffffffffffffffffffffffffffffffffffffffffffff 

In Fig.~\ref{fig:Z_Ebd}  we show synthetic synchrotron emission maps for the \ac{cw} configuration. Similar to the \ac{fb} case, the large magnetic inclination models feature  a  bright head 
 structure (right bottom panel of Fig.~\ref{fig:Z_Ebd}). With increasing magnetization, \(\sigma_0\geq1\), the head part becomes fainter and start wobbling. If the pulsar magnetic inclination is small, \eg \(\alpha=\upi/18\), the front bow-shock structure becomes almost invisible (fs1a10, right top panel of Fig.~\ref{fig:Z_Ebd})). All models with high wind magnetization (top row and left bottom panel in  Fig.~\ref{fig:Z_Ebd}) show prominent narrow tail. 
As comparing with the \ac{fb} orientation, \ac{cw}-like \Rs can have both ``a single tail'' and ``bow shock  plus tail'' features.

\subsection{Emission maps -- mixed \acl{rb}~--~\acl{fb} configuration}

%fffffffffffffffffffffffffffffffffffffffffffffffffffffffffffffffff
\begin{figure*}
\includegraphics[width=70mm,angle=-0]{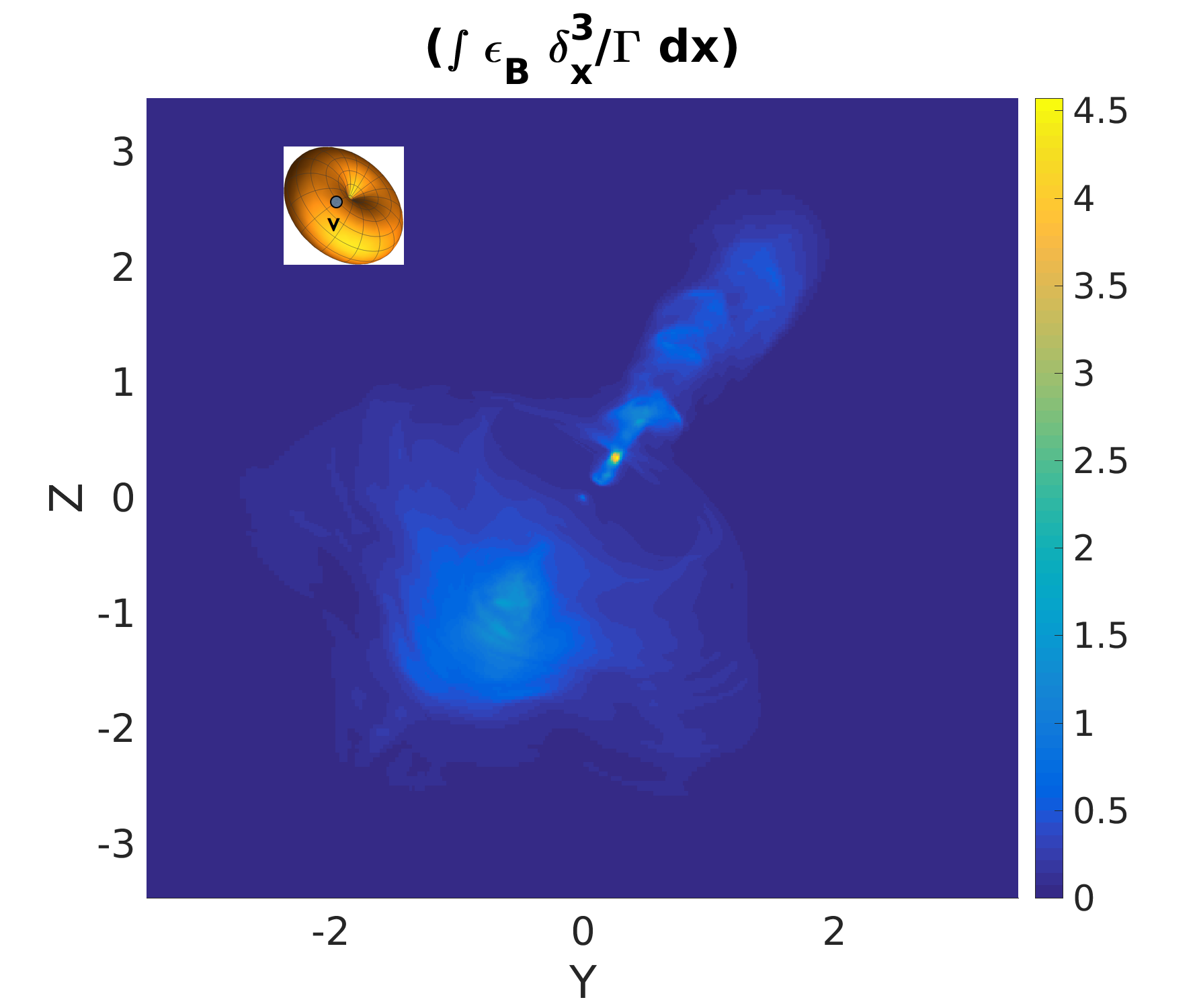}
\includegraphics[width=105mm,angle=-0]{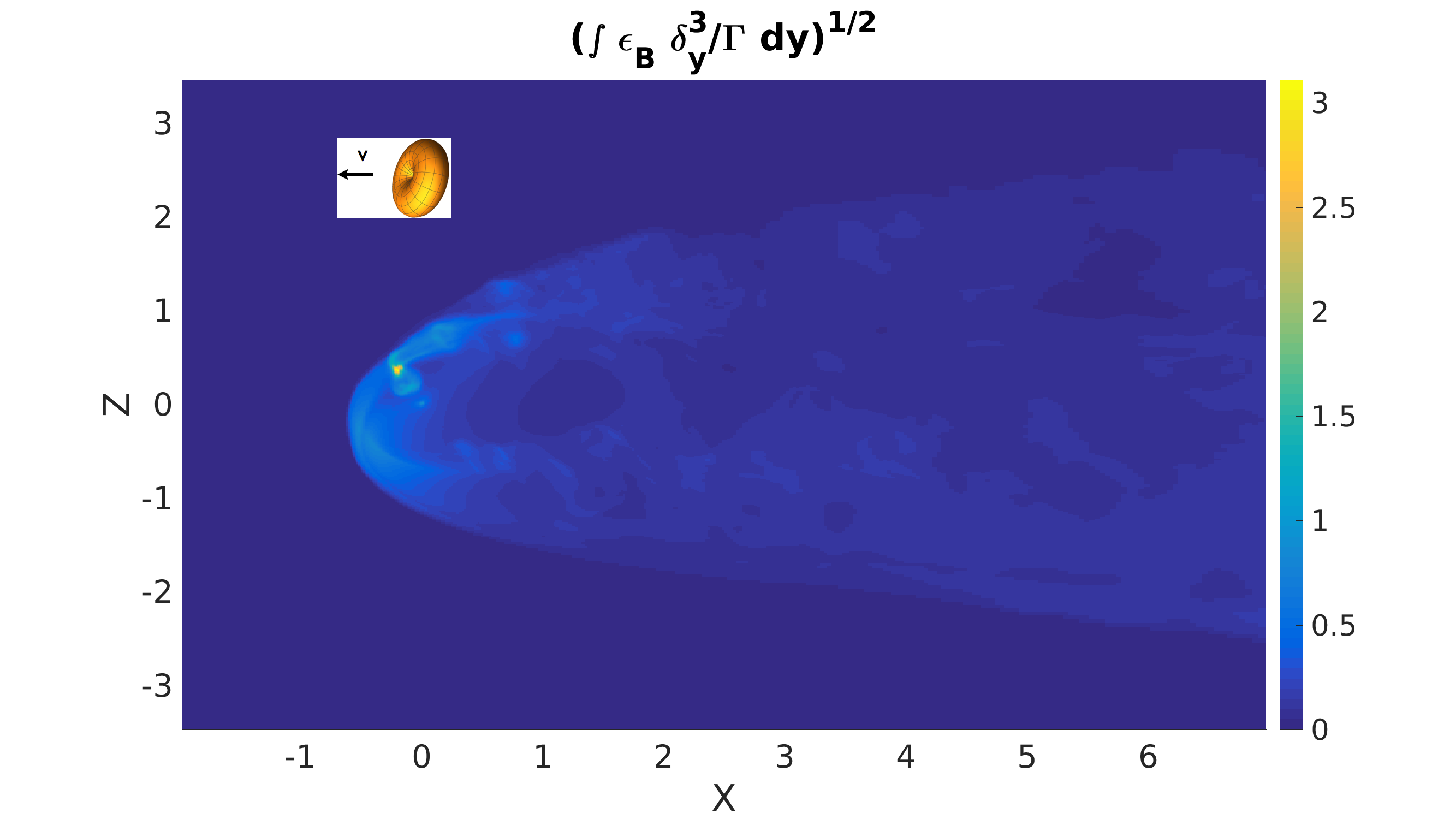}
\caption{Synchrotron   emissivity map  projected along X (top) and Y (bottom) axis for the models fbs1a45.}
\label{fig:fb_Ebd}
\end{figure*}
%fffffffffffffffffffffffffffffffffffffffffffffffffffffffffffffffff 

In Fig.~\ref{fig:fb_Ebd} we show synthetic synchrotron maps for a mixed \ac{rb}~--~\ac{fb} configuration (model
fbs1a45). If seen along the proper velocity, then the morphology is similar to the \ac{fb} case (\cf Fig.~\ref{fig:X_Ebd} and left
panel of Fig.~\ref{fig:fb_Ebd}). If the \R is seen side way, then the \R appears as an asymmetric bow-shock structure
(Fig.~\ref{fig:fb_Ebd}, right panel) - the front/up ``jet'' is brighter and narrower as compared to the down/back
``jet''.

\subsection{Emission maps -- overall conclusion}

Depending on the geometrical configuration and the line of sight we can reproduce many/most of the observed  \xray morphological features seen in the bow-shock \Rs:
\begin{itemize}
\item some \Rs have bow-shock structure (we call them ``whiskers'') that follows the $H\alpha$ bow shocks,
\item some \Rs have tailward extended  features (often called ``jets''),
\footnote{We prefer not to call these features as ``jet'' - the term is usually used for highly supersonic flows in, \eg, YSO, AGNe, {\it etc}. Most of the extended features that we observe in \Rs  are more like plumes - mildly sonic,  spacially  elongated structures}
\item some \Rs have extended  ``filled-in'' morphology,
\item some \Rs have ``mushroom-type'' morphology, 
\item some \Rs asymmetric features (often called ``one-sided jets'').
\end{itemize}

\section{Particular bow-shock \Rs}
\label{particular}

Next we compare the \R morphology observed in \xray band to our synthetic emissivity maps. We define  four general types of \Rs which corresponds to specific orientation of pulsar rotation axis, its proper  velocity, and the line of sight.

\subsection{\acl{rb}: \ac{mouse} and \ac{mushroom}  \Rs}

The wide-head and tail morphology (``mushroom'') can be formed in the \ac{rb} configuration with moderate magnetization $\sigma_0\sim 1$  (\eg model bs1a45 in Fig.~\ref{fig:b_Ebd}). The cap of the ``Mushroom'' is formed by the spread front  ``jet'' and the equatorial outflow; the mushroom stalk is formed by the back ``jet''.

\subsection{\acl{fb}: \ac{gem} and \ac{j1509} \Rs}
\label{sec:gem_1509}

\Rs with ``three-jet'' structure can be naturally interpreted as the \ac{fb} case (see models fs3a45, fs1a10, and fs1a45 in Fig.~\ref{fig:Y_Ebd}). 
\Rs formed by \ac{gem} and \ac{j1509} pulsars are the prototype sources for such systems  \citep{kkr16,2017ApJ...835...66P}.
The jets-like plumes are formed by plasma that originated close to the pulsar polar outflow and the third ``jet'' (middle one) is formed by the equatorial outflow. 
Any quantitative comparison of the synthetic maps and observed morphology needs to be done with images obtained in the same scaling. 
Some \xray features appear clearly seen in images only if one manually adapts the color scale \citep[\eg][]{2017ApJ...835...66P}. Finally, we note a significant dependence of the synthetic maps on the pulsar magnetic inclination angle, \(\alpha\). For example, for \(\alpha=\upi/18\) the \R consists of three nearly equally bright jet-like plumes.
%Case fs1a45 has too bright torus head structure. The most feasible candidate for free jet \Rs is pulsar in \ac{fb} geometry with relatively small $\alpha\sim 10^\circ$  

%and magnetization $\sigma \ge 1$.

The ratio of synchrotron to adiabatic loss rates, Eq.~\eqref{eq:rad_eff}, provides an estimate for the \xray luminosity as a fraction of the spin-down losses. For example, in the case of \ac{j1509} the pulsar spin down luminosity is $\dot{E}_{\rm sd} \approx 5\times 10^{35}\ergs$, and \xray luminosity of the tail
is $L_{\rm \xray} \approx 10^{33}\ergs$. Their ratio, $\dot{E}_{\rm sd}/L_{\rm \xray} \approx 500$, seem to be in a good agreement with Eq.~\eqref{eq:rad_eff}.
%relation of cooling time to adiabatic loses time (see Eq.~\ref{eq:tsyn} and Eq.~\ref{eq:tnt}) $t_{\rm syn}/t_{\rm nr} \approx 1000$.

\subsection{\acl{cw}: \ac{j1741}}

In the \ac{cw} configuration a thin-tail structure can be formed if the jets are banded due to the projection effects and the equatorial bow shock 
is not bright enough to be detected (see Fig.~\ref{fig:Z_Ebd}, models fs1a10). The prototypes of such systems can be \ac{j1741} \citep{asr15}. 
We note that the thin jet like structure is wobbling with time.

\subsection{\acl{fb}~--~\acl{rb}: \ac{j1135}}

Strongly asymmetric distortion of jet like structures can be formed in \ac{fb}~--~\ac{rb} geometry  (see Fig.~\ref{fig:fb_Ebd}, model fbs1a45).
\ac{j1135} can be considered a the prototype of such a configuration \citep[see Fig.~\ref{fig:gem_1509}]{mar12}.

\section{Discussion}
\label{Discussion}

\subsection{Overall \xray efficiency of bow-shock \Rs}

The developed \ac{mhd} model naturally explains the apparent 
low \xray efficiency of the  bow-shock \Rs \citep{2001ApJ...562L.163K,kpkr17}.  
Bow-shock \Rs are very inefficient in converting spin-down luminosity in \xrays, with  efficiencies 
$\sim 10^{-3}-10^{-5}$ \citep{kpkr17}. These values are significantly smaller than for \Rs around slow-moving pulsars, where 
 the conversion efficiency can be as high as tens of percent \citep[\eg][]{2015SSRv..191..391K}.

As we discussed in \Sec\ref{sec:gem_1509}, such low effectiveness can be explained by strong adiabatic loses  
 $L_{\rm \xray} / \dot{E}_{\rm sd} \sim t_{\textsc{ad}}/t_{\textsc{syn}} < 1/1000$  in the head of \Rs.
 In bow-shock \Rs  the crossing time of the relativistic plasma  through the tail
is shorter than  synchrotron life time. As a results particles are able to emit only a small fraction of the  energy
 that they acquired during acceleration at the reverse shock or in reconnection sites - 
adiabatic cooling dominates and
most of the   wind  luminosity is spent on  $p\dif{V}$ work inflating the bubble at the large distance from the pulsar.

This model may also explain the apparent disagreement between the 
estimate of the magnetic field  from equipartition arguments
(even initially  weakly magnetized flow after the shock transition reaches 
approximate equipartition) and the observed length of the \xray tail
\citep{2001ApJ...562L.163K} - particles are quickly adiabatically cooled.

% BMV: the statement bellow contradict previous one.

%The pressure in the tail  may also decrease considerably due to the radiative losses.
%This is likely to be an important effect. The observed tail of the Duck pulsar fades away on a scale $\sim$ 10
%stand-off radii. In the adiabatic limit the 1-D  brightness of the tail should in fact 
%{\it increase } since for a constant density and pressure the particles
%spend  in the tail more time  further from the pulsar.

%Radiative losses will result in a loss of a pressure and
% contraction of the tail.

\subsection{Limitations of the  approach}

We made a significant step forward in understanding of \R formation by fast moving pulsars, but several further important improvements have to be done in future works. 
Our simulation have a good resolution near the pulsar but it quickly decrease with distance. The low resolution triggers fast magnetic energy dissipation in the \R tail, that affects morphology of 
the flow and emissivity maps. Moreover, it is important to follow \R tail evolution on much larger distances.

In the present work we injected  only toroidal magnetic field in pulsar wind. Poloidal component of magnetic field can change significantly flow evolution in the polar regions and affect the ``jet-like'' structure formation.  

We used fairly simple  model for calculation of emissivity maps. We expect that new release of the PLUTO code, which contains module for non-thermal particles evolution calculation \citep{2018ApJ...865..144V}, will allow us to improve the quality 
of emissivity maps.

\subsection{Effects of realistic values of \ISM speed and magnetization}

In our simulation we  focused on intrinsic structure of fast mowing \Rs. To minimize the number of free parameters we fix magnetization of \ISM on very low level ($\sigma_\ism=0.01$).
The effect of the strong,  large scale magnetic field can be significant and  may  change  the shape of \Rs tail, somewhat  similar  to the effects  observed in the heliotail \citep[see ][ and references therein]{2015ApJ...812L...6P,2017SSRv..212..193P}.

Another numerical simplification used is the relatively high \ISM speed ($v_\ism=0.1c$), a realistic value should be at least ten times smaller. Previously, we performed simulation with such a realistic value \citep[see details in ][]{blkb18} and found 
that this effect not very significant. The main difference in \ac{fb}~--~\ac{rb} model is in the separation of front jet-like structure from equatorial/bottom jet-like flow. The shocked \ISM matter moves around front jet-like structure 
and fill the gap between  front jet-like structure and equatorial/bottom jet-like. So, pulsar wind tail becomes separated on several flows. How robust such conclusion it is difficult to say so far. To check this result it is necessary to perform 
simulation with realistic \ISM speed during tenths of dynamical time scales for \ISM. Such a simulation requires a few mega CPUhours.  

\subsection{Connection with \fer observation and pulsar kicks}

One of the  implication of our model is  a possibility to reconstruct information about the relative  orientation of pulsars'  motion, the  line of sight and  magnetic  inclination angle $\alpha$. 
The latter can also independently be deduced from modeling  of  radio and 
 \fer gamma-ray  light curves \citep{1969ApL.....3..225R,2009ApJ...695.1289W,2015A&A...575A...3P}.  In addition, observation of the overall structures and/or proper motion of pulsars often indicate the direction of motion. 
 We leave a comparison of our results with \fer data for future work.
 
The origin of pulsars' high velocities is a long-standing puzzle \citep{1994Natur.369..127L,1996AIPC..366...38B,1998Natur.393..139S}. 
Previously \cite{1996AIPC..366...38B,2001ApJ...549.1111L} argued for the alignment of spin and spacial velocities. Large scale turbulent filaments developing  during the core collapse supernovae explosions
can accelerate protoneutron star up to 1000~\kms \citep[see \eg][]{2013A&A...552A.126W,1998Natur.393..139S}. Depending on the duration of the kick (shorter or longer than the protoneutron star rotational period),  
 the resulting kick can be either directionally  random or along the rotation axis \citep{1998Natur.393..139S}.
   
Our research, in principle may provide additional insight -- 
from morphological features we can distinguish \ac{rb}, \ac{fb}, \ac{cw} or mixed configurations. Since we do not do fits to the parameters, 
we cannot provide statistical analysis (\eg,  the occurrence rates of different configurations) - only  in  the most clear-cut cases we infer particular configuration. 

Out of six systems where  we could clearly claim the structure  we inferred two \ac{rb} and four \ac{fb}/\ac{cw}/mixed cases. Based on these numbers our result do not favor alignment of spin 
and linear velocity (that would produce the \ac{rb} configuration).  

\subsection{Unresolved issues:  the tail structures}

One of the few remaining  unresolved issues, in our view, is the variations in the structure of the tail far downstream. Often, the large-scale morphology shows quasi-periodic variations in the \xray, radio and/or $H\alpha$ intensity. For example, the Guitar nebula shows several  ``closed-in-the-back'' morphological features. In the spirit of our approach - attributing morphological features to intrinsic dynamics, not external density variations - we associate these ubiquitous and  quasi-periodic features with the development of instabilities in the tail flow. Possible instabilities include:
\begin{itemize}
\item  Kelvin-Helmholtz instabilities.  In the tail region the fast, light and strongly magnetized pulsar wind is  moving with respect to the \ISM with  subsonic velocities with respect to the fast magnetosonic velocity in the tail, but supersonically with respect to the sound speed in the \ISM.
\item Current driven instabilities.  The magnetic structure of the axially-symmetric \ac{rb}  configuration  resembles the case of AGN jets, 
where kink instabilities, both global and local, may lead to the magnetic field dissipation and possible disruption of the whole jets 
\citep[e.g.][]{2017MNRAS.467.4647K,2018MNRAS.474.3954K}. 
For the \ac{fb}/\ac{cw} case one might expect the development of the parallel-currents-attract type 
of instability between the two polar currents, see Fig.~\ref{fig:tail-Bfield}.
\item Mass loading instabilities. If a pulsar propagates into partially ionized medium (this is required to produce $H\alpha$ signal), 
the mass loading of the light pulsar wind can be very efficient and can change the wind dynamics \citep{2015MNRAS.454.3886M}. In particular,    
\cite{2015MNRAS.454.3886M} attributed  sudden ``kinks'' in the   $H\alpha$ bow shocks to  the effects of mass-loading of the wind. 
%\item Periodic variations of the tail cross-section, driven, e.g. by the development of the  Kelvin-Helmholtz instabilities or  imposed by the external density structures, can result in the development of parametric instabilities,  resulting in hydraulic jumps and transition to turbulence see Appendix \ref{PeriodicTubeFlow}.
\item The variation of \ISM density along line of pulsar motion also can change cross-section of shocked region 
\citep[][]{2007MNRAS.374..793V,2018arXiv180306240T}.   
\end{itemize}

\section{Conclusion}
\label{sc:conclusion}

In this work we present analytical and numerical \ac{3d} \ac{mhd} calculations of the  interaction of  relativistic  wind produced by fast moving pulsars with the interstellar medium.  We capture both the flow dynamics in the head part of the resulting bow-shock \Rs, as well as the evolution of the flow in the tail part. Our results indicate that magnetic fields play the most important role in shaping the morphology of the bow-shock \Rs. The observed structure depends crucially on the geometrical properties  ---  the relative orientation of the pulsar spin axis and the direction of motion. 

Most importantly, we are able to reproduce both the main observed  morphological features, as well as variations between different systems as arising from internal dynamics of  magnetized pulsar winds. In contrast, external density variation introduces only mild morphological  variations.

%I THINK WE CAN ALSO MAKE PREDICTIONS, Especailly for the head-on cases.

We calculate semi-analytically emission maps for these three cases for various angle-dependent  magnetization parameters of the pulsar winds.  
For low-magnetized, $\sigma < 1$, winds we stress the importance   of  the ``inside magnetic draping''   effect -
 formation of near-equipartition magnetized sheath close to the \CD. This occurs for arbitrary weakly magnetized wind - 
 the thickness of the equipartition region depends on the wind magnetization $\sigma $.

We discuss the interaction of pulsar wind with \ISM analytically and numerically. 
Here we present very first \ac{3d}  \ac{rmhd} simulation, and plot emissivity maps for different geometries.
Our synchrotron emissivity maps can reproduce some  Chandra \xray observation: 1) \Ac{fb} --- \ac{gem} and \ac{j1509} \Rs,
2) \Ac{cw} --- \ac{j1741} \R, 3) \Ac{rb} --- \ac{mouse} and \ac{mushroom} \Rs, 4) \Ac{fb}~--~\ac{rb} --- \ac{j1135} \R. 

We expect that the properties of bow-shock \Rs considered in the present work can be applied to \NS binaries, where the pulsar wind interacts with the stellar wind
 \citep[see eg.][]{2008MNRAS.387...63B,2011A&A...535A..20B,2012A&A...544A..59B,2012MNRAS.419.3426B,2015A&A...577A..89B,2015A&A...581A..27D,
2016MNRAS.456L..64B,2017A&A...598A..13D,2017MNRAS.471L.150B}.

%%%%%%%%%%%%%%%%%%%%%%%%%%%%%%%%%%%%%%%%%%%%%%%%%%%%%%%%%%%%%%%%%
\section{Acknowledgments}
%%%%%%%%%%%%%%%%%%%%%%%%%%%%%%%%%%%%%%%%%%%%%%%%%%%%%%%%%%%%%%%%% 
%We want to thank the referee for a constructive and useful report.
We would like to thank Anatoliy Spitkovsky, Joseph Gelfand, Oleg Kargaltsev, Victoria Kaspi, Andrey Bykov, and Mallory
Roberts for numerous enlightening discussions.

The calculations were carried out in the CFCA cluster of \acl{naoj}.
We thank the {\it PLUTO} team for the possibility to use the {\it PLUTO} code and for technical support. 
The visualization of the results performed in the VisIt package \citep{HPV:VisIt}. 
This work had been supported by   NSF  grant AST-1306672, DoE grant DE-SC0016369, NASA grant 80NSSC17K0757, JSPS KAKENHI Grant Numbers JP18H03722, JP24105007, JP16H02170.

% 
% %\clearpage
% \bibliographystyle{mn2e}
% %\bibliography{ismpw}  
% \bibliography{ismpw,BibTex} 

\begin{thebibliography}{}

\bibitem[\protect\citeauthoryear{{Atoyan} \& {Aharonian}}{{Atoyan} \&
  {Aharonian}}{1996}]{1996MNRAS.278..525A}
{Atoyan} A.~M.,  {Aharonian} F.~A.,  1996, \mnras, 278, 525

\bibitem[\protect\citeauthoryear{{Auchettl}, {Slane}, {Romani}, {Posselt},
  {Pavlov}, {Kargaltsev}, {Ng}, {Temim}, {Weisskopf}, {Bykov} \&
  {Swartz}}{{Auchettl} et~al.}{2015}]{asr15}
{Auchettl} K.,  {Slane} P.,  {Romani} R.~W.,  {Posselt} B.,  {Pavlov} G.~G.,
  {Kargaltsev} O.,  {Ng} C.-Y.,  {Temim} T.,  {Weisskopf} M.~C.,  {Bykov} A.,
   {Swartz} D.~A.,  2015, \apj, 802, 68

\bibitem[\protect\citeauthoryear{{Bandiera}}{{Bandiera}}{2008}]{2008A&A...490L...3B}
{Bandiera} R.,  2008, \aap, 490, L3

\bibitem[\protect\citeauthoryear{{Baranov}, {Krasnobaev} \&
  {Kilikovskii}}{{Baranov} et~al.}{1971}]{bkk71}
{Baranov} V.~B.,  {Krasnobaev} K.~V.,    {Kilikovskii} A.~G.,  1971, Soviet
  Physics - Doklady, 15, 791

\bibitem[\protect\citeauthoryear{{Barkov} \& {Bosch-Ramon}}{{Barkov} \&
  {Bosch-Ramon}}{2018}]{bbr18}
{Barkov} M.,  {Bosch-Ramon} V.,  2018, in preparation

\bibitem[\protect\citeauthoryear{{Barkov}, {Lyutikov}, {Klingler} \&
  {Bordas}}{{Barkov} et~al.}{2018}]{blkb18}
{Barkov} M.,  {Lyutikov} M.,  {Klingler} N.,    {Bordas} P.,  2018, in
  preparation

\bibitem[\protect\citeauthoryear{{Barkov} \& {Bosch-Ramon}}{{Barkov} \&
  {Bosch-Ramon}}{2016}]{2016MNRAS.456L..64B}
{Barkov} M.~V.,  {Bosch-Ramon} V.,  2016, \mnras, 456, L64

\bibitem[\protect\citeauthoryear{{Bisnovatyi-Kogan}}{{Bisnovatyi-Kogan}}{1996}]{1996AIPC..366...38B}
{Bisnovatyi-Kogan} G.~S.,  1996, in {Rothschild} R.~E.,  {Lingenfelter} R.~E.,
  eds, High Velocity Neutron Stars Vol.~366 of American Institute of Physics
  Conference Series, {High velocity neutron stars as a result of asymmetric
  neutrino emission}.
pp 38--42

\bibitem[\protect\citeauthoryear{{Bogovalov}}{{Bogovalov}}{1999}]{bogovalov_99}
{Bogovalov} S.~V.,  1999, \aap, 349, 1017

\bibitem[\protect\citeauthoryear{{Bogovalov}, {Chechetkin}, {Koldoba} \&
  {Ustyugova}}{{Bogovalov} et~al.}{2005}]{2005MNRAS.358..705B}
{Bogovalov} S.~V.,  {Chechetkin} V.~M.,  {Koldoba} A.~V.,    {Ustyugova} G.~V.,
   2005, \mnras, 358, 705

\bibitem[\protect\citeauthoryear{{Bogovalov} \& {Khangoulian}}{{Bogovalov} \&
  {Khangoulian}}{2002}]{2002MNRAS.336L..53B}
{Bogovalov} S.~V.,  {Khangoulian} D.~V.,  2002, \mnras, 336, L53

\bibitem[\protect\citeauthoryear{{Bogovalov} \& {Khangoulyan}}{{Bogovalov} \&
  {Khangoulyan}}{2002}]{2002AstL...28..373B}
{Bogovalov} S.~V.,  {Khangoulyan} D.~V.,  2002, Astronomy Letters, 28, 373

\bibitem[\protect\citeauthoryear{{Bogovalov}, {Khangulyan}, {Koldoba},
  {Ustyugova} \& {Aharonian}}{{Bogovalov} et~al.}{2012}]{2012MNRAS.419.3426B}
{Bogovalov} S.~V.,  {Khangulyan} D.,  {Koldoba} A.~V.,  {Ustyugova} G.~V.,
  {Aharonian} F.~A.,  2012, \mnras, 419, 3426

\bibitem[\protect\citeauthoryear{{Bogovalov}, {Khangulyan}, {Koldoba},
  {Ustyugova} \& {Aharonian}}{{Bogovalov} et~al.}{2008}]{2008MNRAS.387...63B}
{Bogovalov} S.~V.,  {Khangulyan} D.~V.,  {Koldoba} A.~V.,  {Ustyugova} G.~V.,
   {Aharonian} F.~A.,  2008, \mnras, 387, 63

\bibitem[\protect\citeauthoryear{{Bosch-Ramon} \& {Barkov}}{{Bosch-Ramon} \&
  {Barkov}}{2011}]{2011A&A...535A..20B}
{Bosch-Ramon} V.,  {Barkov} M.~V.,  2011, \aap, 535, A20

\bibitem[\protect\citeauthoryear{{Bosch-Ramon}, {Barkov}, {Khangulyan} \&
  {Perucho}}{{Bosch-Ramon} et~al.}{2012}]{2012A&A...544A..59B}
{Bosch-Ramon} V.,  {Barkov} M.~V.,  {Khangulyan} D.,    {Perucho} M.,  2012,
  \aap, 544, A59

\bibitem[\protect\citeauthoryear{{Bosch-Ramon}, {Barkov}, {Mignone} \&
  {Bordas}}{{Bosch-Ramon} et~al.}{2017}]{2017MNRAS.471L.150B}
{Bosch-Ramon} V.,  {Barkov} M.~V.,  {Mignone} A.,    {Bordas} P.,  2017,
  \mnras, 471, L150

\bibitem[\protect\citeauthoryear{{Bosch-Ramon}, {Barkov} \&
  {Perucho}}{{Bosch-Ramon} et~al.}{2015}]{2015A&A...577A..89B}
{Bosch-Ramon} V.,  {Barkov} M.~V.,    {Perucho} M.,  2015, \aap, 577, A89

\bibitem[\protect\citeauthoryear{{Bucciantini}}{{Bucciantini}}{2002}]{Bucciantini02a}
{Bucciantini} N.,  2002, \aap, 387, 1066

\bibitem[\protect\citeauthoryear{{Bucciantini}, {Amato} \& {Del
  Zanna}}{{Bucciantini} et~al.}{2005a}]{bad05}
{Bucciantini} N.,  {Amato} E.,    {Del Zanna} L.,  2005a, \aap, 434, 189

\bibitem[\protect\citeauthoryear{{Bucciantini}, {Amato} \& {Del
  Zanna}}{{Bucciantini} et~al.}{2005b}]{2005A&A...434..189B}
{Bucciantini} N.,  {Amato} E.,    {Del Zanna} L.,  2005b, \aap, 434, 189

\bibitem[\protect\citeauthoryear{{Bykov}, {Amato}, {Petrov}, {Krassilchtchikov}
  \& {Levenfish}}{{Bykov} et~al.}{2017}]{2017SSRv..207..235B}
{Bykov} A.~M.,  {Amato} E.,  {Petrov} A.~E.,  {Krassilchtchikov} A.~M.,
  {Levenfish} K.~P.,  2017, \ssr, 207, 235

\bibitem[\protect\citeauthoryear{{Coroniti}}{{Coroniti}}{1990}]{Coroniti90}
{Coroniti} F.~V.,  1990, \apj, 349, 538

\bibitem[\protect\citeauthoryear{{Cranfill}}{{Cranfill}}{1974}]{cran71}
{Cranfill} C.~W.,  1974, Ph.D.~Thesis

\bibitem[\protect\citeauthoryear{{de la Cita}, {Bosch-Ramon},
  {Paredes-Fortuny}, {Khangulyan} \& {Perucho}}{{de la Cita}
  et~al.}{2017}]{2017A&A...598A..13D}
{de la Cita} V.~M.,  {Bosch-Ramon} V.,  {Paredes-Fortuny} X.,  {Khangulyan} D.,
     {Perucho} M.,  2017, \aap, 598, A13

\bibitem[\protect\citeauthoryear{{Dubus}, {Lamberts} \& {Fromang}}{{Dubus}
  et~al.}{2015}]{2015A&A...581A..27D}
{Dubus} G.,  {Lamberts} A.,    {Fromang} S.,  2015, \aap, 581, A27

\bibitem[\protect\citeauthoryear{{Dyson}}{{Dyson}}{1975}]{1975Ap&SS..35..299D}
{Dyson} J.~E.,  1975, \apss, 35, 299

\bibitem[\protect\citeauthoryear{{Gaensler} \& {Slane}}{{Gaensler} \&
  {Slane}}{2006}]{2006ARA&A..44...17G}
{Gaensler} B.~M.,  {Slane} P.~O.,  2006, \araa, 44, 17

\bibitem[\protect\citeauthoryear{{Goldreich} \& {Julian}}{{Goldreich} \&
  {Julian}}{1969}]{GoldreichJulian}
{Goldreich} P.,  {Julian} W.~H.,  1969, \apj, 157, 869

\bibitem[\protect\citeauthoryear{Hank~{Childs}, {Brugger}, {Whitlock} \& {et
  al.}}{Hank~{Childs} et~al.}{2012}]{HPV:VisIt}
Hank~{Childs} H.,  {Brugger} E.,  {Whitlock} B.,    {et al.} 2012, in , {High
  Performance Visualization--Enabling Extreme-Scale Scientific Insight}.
pp 357--372

\bibitem[\protect\citeauthoryear{Harten}{Harten}{1983}]{HLL83}
Harten A.,  1983, Journal of Computational Physics, 49, 357–393

\bibitem[\protect\citeauthoryear{{Kargaltsev}, {Cerutti}, {Lyubarsky} \&
  {Striani}}{{Kargaltsev} et~al.}{2015}]{2015SSRv..191..391K}
{Kargaltsev} O.,  {Cerutti} B.,  {Lyubarsky} Y.,    {Striani} E.,  2015, \ssr,
  191, 391

\bibitem[\protect\citeauthoryear{{Kargaltsev} \& {Pavlov}}{{Kargaltsev} \&
  {Pavlov}}{2008}]{2008AIPC..983..171K}
{Kargaltsev} O.,  {Pavlov} G.~G.,  2008, in {Bassa} C.,  {Wang} Z.,  {Cumming}
  A.,   {Kaspi} V.~M.,  eds, 40 Years of Pulsars: Millisecond Pulsars,
  Magnetars and More Vol.~983 of American Institute of Physics Conference
  Series, {Pulsar Wind Nebulae in the Chandra Era}.
pp 171--185

\bibitem[\protect\citeauthoryear{{Kargaltsev}, {Pavlov}, {Klingler} \&
  {Rangelov}}{{Kargaltsev} et~al.}{2017}]{kpkr17}
{Kargaltsev} O.,  {Pavlov} G.~G.,  {Klingler} N.,    {Rangelov} B.,  2017,
  Journal of Plasma Physics, 83, 635830501

\bibitem[\protect\citeauthoryear{{Kaspi}, {Gotthelf}, {Gaensler} \&
  {Lyutikov}}{{Kaspi} et~al.}{2001}]{2001ApJ...562L.163K}
{Kaspi} V.~M.,  {Gotthelf} E.~V.,  {Gaensler} B.~M.,    {Lyutikov} M.,  2001,
  \apjl, 562, L163

\bibitem[\protect\citeauthoryear{{Kennel} \& {Coroniti}}{{Kennel} \&
  {Coroniti}}{1984a}]{kc84}
{Kennel} C.~F.,  {Coroniti} F.~V.,  1984a, \apj, 283, 694

\bibitem[\protect\citeauthoryear{{Kennel} \& {Coroniti}}{{Kennel} \&
  {Coroniti}}{1984b}]{kennel_coroniti_84b}
{Kennel} C.~F.,  {Coroniti} F.~V.,  1984b, \apj, 283, 710

\bibitem[\protect\citeauthoryear{{Khangoulian} \& {Bogovalov}}{{Khangoulian} \&
  {Bogovalov}}{2003}]{2003AstL...29..495K}
{Khangoulian} D.~V.,  {Bogovalov} S.~V.,  2003, Astronomy Letters, 29, 495

\bibitem[\protect\citeauthoryear{{Khangulyan}, {Aharonian} \&
  {Kelner}}{{Khangulyan} et~al.}{2014}]{2014ApJ...783..100K}
{Khangulyan} D.,  {Aharonian} F.~A.,    {Kelner} S.~R.,  2014, \apj, 783, 100

\bibitem[\protect\citeauthoryear{{Khangulyan}, {Bosch-Ramon} \&
  {Uchiyama}}{{Khangulyan} et~al.}{2018}]{2018MNRAS.481.1455K}
{Khangulyan} D.,  {Bosch-Ramon} V.,    {Uchiyama} Y.,  2018, \mnras, 481, 1455

\bibitem[\protect\citeauthoryear{{Kim}, {Balsara}, {Lyutikov} \&
  {Komissarov}}{{Kim} et~al.}{2017}]{2017MNRAS.467.4647K}
{Kim} J.,  {Balsara} D.~S.,  {Lyutikov} M.,    {Komissarov} S.~S.,  2017,
  \mnras, 467, 4647

\bibitem[\protect\citeauthoryear{{Kim}, {Balsara}, {Lyutikov} \&
  {Komissarov}}{{Kim} et~al.}{2018}]{2018MNRAS.474.3954K}
{Kim} J.,  {Balsara} D.~S.,  {Lyutikov} M.,    {Komissarov} S.~S.,  2018,
  \mnras, 474, 3954

\bibitem[\protect\citeauthoryear{{Klingler}, {Kargaltsev}, {Rangelov},
  {Pavlov}, {Posselt} \& {Ng}}{{Klingler} et~al.}{2016}]{kkr16}
{Klingler} N.,  {Kargaltsev} O.,  {Rangelov} B.,  {Pavlov} G.~G.,  {Posselt}
  B.,    {Ng} C.-Y.,  2016, \apj, 828, 70

\bibitem[\protect\citeauthoryear{{Klingler}, {Rangelov}, {Kargaltsev},
  {Pavlov}, {Romani}, {Posselt}, {Slane}, {Temim}, {Ng}, {Bucciantini},
  {Bykov}, {Swartz} \& {Buehler}}{{Klingler} et~al.}{2016}]{krkp16}
{Klingler} N.,  {Rangelov} B.,  {Kargaltsev} O.,  {Pavlov} G.~G.,  {Romani}
  R.~W.,  {Posselt} B.,  {Slane} P.,  {Temim} T.,  {Ng} C.-Y.,  {Bucciantini}
  N.,  {Bykov} A.,  {Swartz} D.~A.,    {Buehler} R.,  2016, \apj, 833, 253

\bibitem[\protect\citeauthoryear{{Komissarov}}{{Komissarov}}{2013}]{2013MNRAS.428.2459K}
{Komissarov} S.~S.,  2013, \mnras, 428, 2459

\bibitem[\protect\citeauthoryear{{Komissarov} \& {Lyubarsky}}{{Komissarov} \&
  {Lyubarsky}}{2004a}]{KomissarovLyubarsky}
{Komissarov} S.~S.,  {Lyubarsky} Y.~E.,  2004a, MNRAS, 349, 779

\bibitem[\protect\citeauthoryear{{Komissarov} \& {Lyubarsky}}{{Komissarov} \&
  {Lyubarsky}}{2004b}]{komissarov_04}
{Komissarov} S.~S.,  {Lyubarsky} Y.~E.,  2004b, \mnras, 349, 779

\bibitem[\protect\citeauthoryear{{Kompaneets}}{{Kompaneets}}{1960}]{Komp}
{Kompaneets} A.~S.,  1960, Soviet Physics Doklady, 5, 46

\bibitem[\protect\citeauthoryear{{Lai}, {Chernoff} \& {Cordes}}{{Lai}
  et~al.}{2001}]{2001ApJ...549.1111L}
{Lai} D.,  {Chernoff} D.~F.,    {Cordes} J.~M.,  2001, \apj, 549, 1111

\bibitem[\protect\citeauthoryear{{Landau} \& {Lifshitz}}{{Landau} \&
  {Lifshitz}}{1959}]{LLVI}
{Landau} L.~D.,  {Lifshitz} E.~M.,  1959, {Fluid mechanics}.
Course of theoretical physics, Oxford: Pergamon Press, 1959

\bibitem[\protect\citeauthoryear{{Lyne} \& {Lorimer}}{{Lyne} \&
  {Lorimer}}{1994}]{1994Natur.369..127L}
{Lyne} A.~G.,  {Lorimer} D.~R.,  1994, \nat, 369, 127

\bibitem[\protect\citeauthoryear{{Lyubarsky}}{{Lyubarsky}}{2002}]{2002MNRAS.329L..34L}
{Lyubarsky} Y.~E.,  2002, MNRAS, 329, L34

\bibitem[\protect\citeauthoryear{{Lyubarsky}}{{Lyubarsky}}{2003}]{lub03}
{Lyubarsky} Y.~E.,  2003, \mnras, 345, 153

\bibitem[\protect\citeauthoryear{{Lyutikov}}{{Lyutikov}}{2002}]{2002PhFl...14..963L}
{Lyutikov} M.,  2002, Physics of Fluids, 14, 963

\bibitem[\protect\citeauthoryear{{Lyutikov}}{{Lyutikov}}{2006}]{Lyutikovdraping}
{Lyutikov} M.,  2006, MNRAS, 373, 73

\bibitem[\protect\citeauthoryear{{Lyutikov}, {Komissarov} \&
  {Porth}}{{Lyutikov} et~al.}{2016}]{2016MNRAS.456..286L}
{Lyutikov} M.,  {Komissarov} S.~S.,    {Porth} O.,  2016, \mnras, 456, 286

\bibitem[\protect\citeauthoryear{{Lyutikov}, {Pariev} \&
  {Blandford}}{{Lyutikov} et~al.}{2003}]{2003ApJ...597..998L}
{Lyutikov} M.,  {Pariev} V.~I.,    {Blandford} R.~D.,  2003, \apj, 597, 998

\bibitem[\protect\citeauthoryear{{Lyutikov}, {Sironi}, {Komissarov} \&
  {Porth}}{{Lyutikov} et~al.}{2017}]{2017JPlPh..83f6302L}
{Lyutikov} M.,  {Sironi} L.,  {Komissarov} S.~S.,    {Porth} O.,  2017, Journal
  of Plasma Physics, 83, 635830602

\bibitem[\protect\citeauthoryear{{Marelli}}{{Marelli}}{2012}]{mar12}
{Marelli} M.,  2012, ArXiv e-prints

\bibitem[\protect\citeauthoryear{{Michel}}{{Michel}}{1969}]{1969ApJ...158..727M}
{Michel} F.~C.,  1969, \apj, 158, 727

\bibitem[\protect\citeauthoryear{{Michel}}{{Michel}}{1973}]{Michel73}
{Michel} F.~C.,  1973, \apj, 180, 207

\bibitem[\protect\citeauthoryear{{Mignone}, {Bodo}, {Massaglia}, {Matsakos},
  {Tesileanu}, {Zanni} \& {Ferrari}}{{Mignone} et~al.}{2007}]{mbm07}
{Mignone} A.,  {Bodo} G.,  {Massaglia} S.,  {Matsakos} T.,  {Tesileanu} O.,
  {Zanni} C.,    {Ferrari} A.,  2007, \apjs, 170, 228

\bibitem[\protect\citeauthoryear{{Morlino}, {Lyutikov} \& {Vorster}}{{Morlino}
  et~al.}{2015}]{2015MNRAS.454.3886M}
{Morlino} G.,  {Lyutikov} M.,    {Vorster} M.,  2015, \mnras, 454, 3886

\bibitem[\protect\citeauthoryear{{Pierbattista}, {Harding}, {Grenier},
  {Johnson}, {Caraveo}, {Kerr} \& {Gonthier}}{{Pierbattista}
  et~al.}{2015}]{2015A&A...575A...3P}
{Pierbattista} M.,  {Harding} A.~K.,  {Grenier} I.~A.,  {Johnson} T.~J.,
  {Caraveo} P.~A.,  {Kerr} M.,    {Gonthier} P.~L.,  2015, \aap, 575, A3

\bibitem[\protect\citeauthoryear{{Pogorelov}, {Borovikov}, {Heerikhuisen} \&
  {Zhang}}{{Pogorelov} et~al.}{2015}]{2015ApJ...812L...6P}
{Pogorelov} N.~V.,  {Borovikov} S.~N.,  {Heerikhuisen} J.,    {Zhang} M.,
  2015, \apjl, 812, L6

\bibitem[\protect\citeauthoryear{{Pogorelov}, {Fichtner}, {Czechowski},
  {Lazarian}, {Lembege}, {le Roux}, {Potgieter}, {Scherer}, {Stone}, {Strauss},
  {Wiengarten}, {Wurz}, {Zank} \& {Zhang}}{{Pogorelov}
  et~al.}{2017}]{2017SSRv..212..193P}
{Pogorelov} N.~V.,  {Fichtner} H.,  {Czechowski} A.,  {Lazarian} A.,  {Lembege}
  B.,  {le Roux} J.~A.,  {Potgieter} M.~S.,  {Scherer} K.,  {Stone} E.~C.,
  {Strauss} R.~D.,  {Wiengarten} T.,  {Wurz} P.,  {Zank} G.~P.,    {Zhang} M.,
  2017, \ssr, 212, 193

\bibitem[\protect\citeauthoryear{{Porth}, {Komissarov} \& {Keppens}}{{Porth}
  et~al.}{2014}]{2014MNRAS.438..278P}
{Porth} O.,  {Komissarov} S.~S.,    {Keppens} R.,  2014, \mnras, 438, 278

\bibitem[\protect\citeauthoryear{{Posselt}, {Pavlov}, {Slane}, {Romani},
  {Bucciantini}, {Bykov}, {Kargaltsev}, {Weisskopf} \& {Ng}}{{Posselt}
  et~al.}{2017}]{2017ApJ...835...66P}
{Posselt} B.,  {Pavlov} G.~G.,  {Slane} P.~O.,  {Romani} R.,  {Bucciantini} N.,
   {Bykov} A.~M.,  {Kargaltsev} O.,  {Weisskopf} M.~C.,    {Ng} C.-Y.,  2017,
  \apj, 835, 66

\bibitem[\protect\citeauthoryear{{Radhakrishnan} \& {Cooke}}{{Radhakrishnan} \&
  {Cooke}}{1969}]{1969ApL.....3..225R}
{Radhakrishnan} V.,  {Cooke} D.~J.,  1969, \aplett, 3, 225

\bibitem[\protect\citeauthoryear{{Rees} \& {Gunn}}{{Rees} \&
  {Gunn}}{1974}]{reesgunn}
{Rees} M.~J.,  {Gunn} J.~E.,  1974, MNRAS, 167, 1

\bibitem[\protect\citeauthoryear{{Reynolds}, {Pavlov}, {Kargaltsev},
  {Klingler}, {Renaud} \& {Mereghetti}}{{Reynolds}
  et~al.}{2017}]{2017SSRv..207..175R}
{Reynolds} S.~P.,  {Pavlov} G.~G.,  {Kargaltsev} O.,  {Klingler} N.,  {Renaud}
  M.,    {Mereghetti} S.,  2017, \ssr, 207, 175

\bibitem[\protect\citeauthoryear{{Romani}, {Cordes} \& {Yadigaroglu}}{{Romani}
  et~al.}{1997}]{Romani97}
{Romani} R.~W.,  {Cordes} J.~M.,    {Yadigaroglu} I.-A.,  1997, \apjl, 484,
  L137

\bibitem[\protect\citeauthoryear{{Romani}, {Slane} \& {Green}}{{Romani}
  et~al.}{2017}]{2017ApJ...851...61R}
{Romani} R.~W.,  {Slane} P.,    {Green} A.~W.,  2017, \apj, 851, 61

\bibitem[\protect\citeauthoryear{{Spitkovsky}}{{Spitkovsky}}{2006}]{Spitkovsky06}
{Spitkovsky} A.,  2006, \apjl, 648, L51

\bibitem[\protect\citeauthoryear{{Spruit} \& {Phinney}}{{Spruit} \&
  {Phinney}}{1998}]{1998Natur.393..139S}
{Spruit} H.,  {Phinney} E.~S.,  1998, \nat, 393, 139

\bibitem[\protect\citeauthoryear{{Tchekhovskoy}, {Philippov} \&
  {Spitkovsky}}{{Tchekhovskoy} et~al.}{2016}]{2016MNRAS.457.3384T}
{Tchekhovskoy} A.,  {Philippov} A.,    {Spitkovsky} A.,  2016, \mnras, 457,
  3384

\bibitem[\protect\citeauthoryear{{Tchekhovskoy}, {Spitkovsky} \&
  {Li}}{{Tchekhovskoy} et~al.}{2013}]{2013MNRAS.435L...1T}
{Tchekhovskoy} A.,  {Spitkovsky} A.,    {Li} J.~G.,  2013, \mnras, 435, L1

\bibitem[\protect\citeauthoryear{Thompson}{Thompson}{1971}]{thom71}
Thompson P.~A.,  1971, Compressible-fluid dynamics.
Advanced engineering series, McGraw-Hill, New York

\bibitem[\protect\citeauthoryear{{Toropina}, {Romanova} \&
  {Lovelace}}{{Toropina} et~al.}{2018}]{2018arXiv180306240T}
{Toropina} O.~D.,  {Romanova} M.~M.,    {Lovelace} R.~V.~E.,  2018, ArXiv
  e-prints

\bibitem[\protect\citeauthoryear{{Vaidya}, {Mignone}, {Bodo}, {Rossi} \&
  {Massaglia}}{{Vaidya} et~al.}{2018}]{2018ApJ...865..144V}
{Vaidya} B.,  {Mignone} A.,  {Bodo} G.,  {Rossi} P.,    {Massaglia} S.,  2018,
  \apj, 865, 144

\bibitem[\protect\citeauthoryear{{Vigelius}, {Melatos}, {Chatterjee},
  {Gaensler} \& {Ghavamian}}{{Vigelius} et~al.}{2007}]{2007MNRAS.374..793V}
{Vigelius} M.,  {Melatos} A.,  {Chatterjee} S.,  {Gaensler} B.~M.,
  {Ghavamian} P.,  2007, \mnras, 374, 793

\bibitem[\protect\citeauthoryear{{Watters}, {Romani}, {Weltevrede} \&
  {Johnston}}{{Watters} et~al.}{2009}]{2009ApJ...695.1289W}
{Watters} K.~P.,  {Romani} R.~W.,  {Weltevrede} P.,    {Johnston} S.,  2009,
  \apj, 695, 1289

\bibitem[\protect\citeauthoryear{{Wilkin}}{{Wilkin}}{1996}]{Wilkin96}
{Wilkin} F.~P.,  1996, \apjl, 459, L31

\bibitem[\protect\citeauthoryear{{Wongwathanarat}, {Janka} \&
  {M{\"u}ller}}{{Wongwathanarat} et~al.}{2013}]{2013A&A...552A.126W}
{Wongwathanarat} A.,  {Janka} H.-T.,    {M{\"u}ller} E.,  2013, \aap, 552, A126

\bibitem[\protect\citeauthoryear{{Zank}}{{Zank}}{1999}]{zank99}
{Zank} G.~P.,  1999, Space Science Reviews, 89, 413

\end{thebibliography}

\appendix

\section{Synchrotron emission}
\label{ap:syn}
Here we generalize,  for the case of relativistically moving plasma, a simple approach for computing non-thermal emission from \ac{mhd} outflow \citep{bbr18}. We focus primarily on the synchrotron morphology thus we aim obtaining  the synchrotron  specific intensity
\be\label{ap:I}
I_\nu=\dif[t,\Omega,\nu,S]{E}= \int j_\nu \dif{\ell}\,,%=\int\limits\left(\frac{\nu}{\nu'}\right)^2 j_{\nu'}' \dif{\ell} \,,
\ee
where \(j_{\nu}\) is the \MEC and \(\ell\) is a length element along the line of sight. The integral is computed across
the volume \(V_0\) occupied by the outflow and should be taken in the laboratory frame, \(K\). Since the synchrotron
emission is typically computed in the plasma co-moving frame, \(K'\), where the electric field vanishes, it is
convenient to express the \MEC in \(K\) through the \MEC in \(K'\):
\be
j_{\nu} = \left(\frac{\nu}{\nu'}\right)^2 j_{\nu'}'\,.
\ee
Here primed and not-primed quantities correspond to the fluid co-moving and laboratory frames, respectively. The two photon frequencies are related as \(\nu'=\nu / \Db\), where the Doppler factor, \(\Db=1/\G (1-\bb\ort{r}_\obs)\), is determined by the flow bulk velocity, \(\bb\); its Lorentz factor, \(\G=1/\sqrt{1-\bb^2}\); and the direction toward the observer, \(\ort{r}_\obs\).

If the non-thermal particles are distributed isotropically in the plasma co-moving frame, then they can be described
with energy distribution: \(\dif{N} = n'\dif{\ve'}\).  The synchrotron \MEC is
\be 
j'_{\nu'}= \int \dif{\ve'} {\cal K}_\syn(\nu',\ve',B'(\br')\cdot \sin\theta'_\syn)\, \frac{n'(\br',\ve')}{4\upi}
\,, 
\ee 
where \(B'\) is co-moving frame magnetic field; \({\cal K}_\syn\) is synchrotron single particle \MEC; and
\(\theta'_\syn\) is the angle between the local magnetic field and the direction toward the observer in the co-moving
frame, \(\ort{r}_\obs'\). If the particle distribution is a power-law, \(n'=A\ve'^{-\alpha}\), then setting formally the energy range
\(0\leq \ve'<\infty\) allows an analytical convolution for \(\alpha > \nicefrac13\)
\be
j'_{\textsc{pl}\,\nu'}=  \frac{\sqrt{3}}{4\upi}\frac{A {e^3  B'\sin\theta_\syn'}}{(m_ec^2)^{\alpha}}\Big(\frac{2\upi m_ec\nu'}{3eB'\sin\theta_\syn'}\Big)^{-\nicefrac{(\alpha-1)}2}{\cal F}(\alpha)\,, %\frac{\sqrt{3}}{4\upi}
\ee
where the auxiliary function \(\cal F\) is
\[
{\cal F}(\alpha)=\frac{\Gamma_f\Big(\nicefrac\alpha4+\nicefrac{19}{12}\Big)\Gamma_f\Big(\nicefrac\alpha4-\nicefrac{1}{12}\Big)}{(\alpha+1)}\,,
\]
and \(\Gamma_f\) is the gamma function.

If one assumes that the non-thermal particles substitute a fixed fraction of the internal energy, \(\epsilon\), which is fulfilled for the electron energy range where adiabatic losses dominate, then the normalization coefficient can be obtained as 
\be
  \epsilon\eta_\nt =  3p \eta_\nt =  A \int\limits_{\ve'_{\textsc{min}}}^{\ve'_{\textsc{max}}}d\ve'\ve'^{-\alpha +1}=AA_0\,,
\ee
where constant \(\eta_\nt\) determines the contribution of the non-thermal particles to the internal energy, and \(A_0\) is a factor that depends on the non-thermal slope and energy range. If the power-law index is close to \(\alpha\simeq2\), then the factor \(A_0\) does not change along stream lines, thus one can simply redefine the phenomenological \(\eta_\nt\)-parameter: \(\eta_\nt / A_0 \rightarrow \eta_\nt\ll1\). Thus one obtains \(A=3p\eta_\nt\) and for \(\alpha=2\) the \MEC is
\be
j'_{\textsc{pl}\,\nu'}=  \frac{0.3\eta_\nt p{e^3  B'\sin\theta_\syn'}}{(m_ec^2)^2}\Big(\frac{2\upi m_ec\nu'}{3eB'\sin\theta_\syn'}\Big)^{-\nicefrac{1}2}\,.
\ee
\be
j'_{\textsc{pl}\,\nu'}\simeq  \frac{ 0.2\eta_\nt p  (B'\sin\theta_\syn')^{\nicefrac32}e^{\nicefrac72}c^{\nicefrac12}}{(m_ec^2)^{\nicefrac52}\nu'^{\nicefrac12}}\,.
\ee

The pitch angle in the co-moving frame can be obtained as the cross product of the corresponding vectors:
\be
B'\sin\theta_\syn'=|\B'\times \ort{r}_\obs'|\,.
\ee
The used \ac{rmhd} code provides us with the \Bf vector. To obtain the \Bf in the co-moving frame one needs to use the froze-in condition and apply the Lorentz transformation  \citep[see \eg][]{2003ApJ...597..998L}:
\be
\B'=\frac1{\G} \left(\B+\frac{\G^2}{\G+1}(\B\bb)\bb\right)\,.
\ee
The direction toward the observer transforms as \citep[see \eg][]{komissarov_04}
\be
\ort{r}'_\obs= \Db\left(\ort{r}_\obs + \G\bb\left(\frac{\G}{1+\G}(\ort{r}_\obs\bb)-1\right)\right)\,. %\Db\left(\ort{r}_\obs + (\G-1)\frac{(\ort{r}_\obs\bb)\bb}{\bb^2}-\G\bb\right)\,.
\ee
The above equations allow us to obtain the fluid element co-moving magnetic field corrected for the change of the pitch
angle through the quantities measured in the laboratory frame.

\section{Inverse Compton  emission}
\label{ap:IC}

%fffffffffffffffffffffffffffffffffffffffffffffffffffffffffffffffff
\begin{figure*}
\includegraphics[width=88mm,angle=-0]{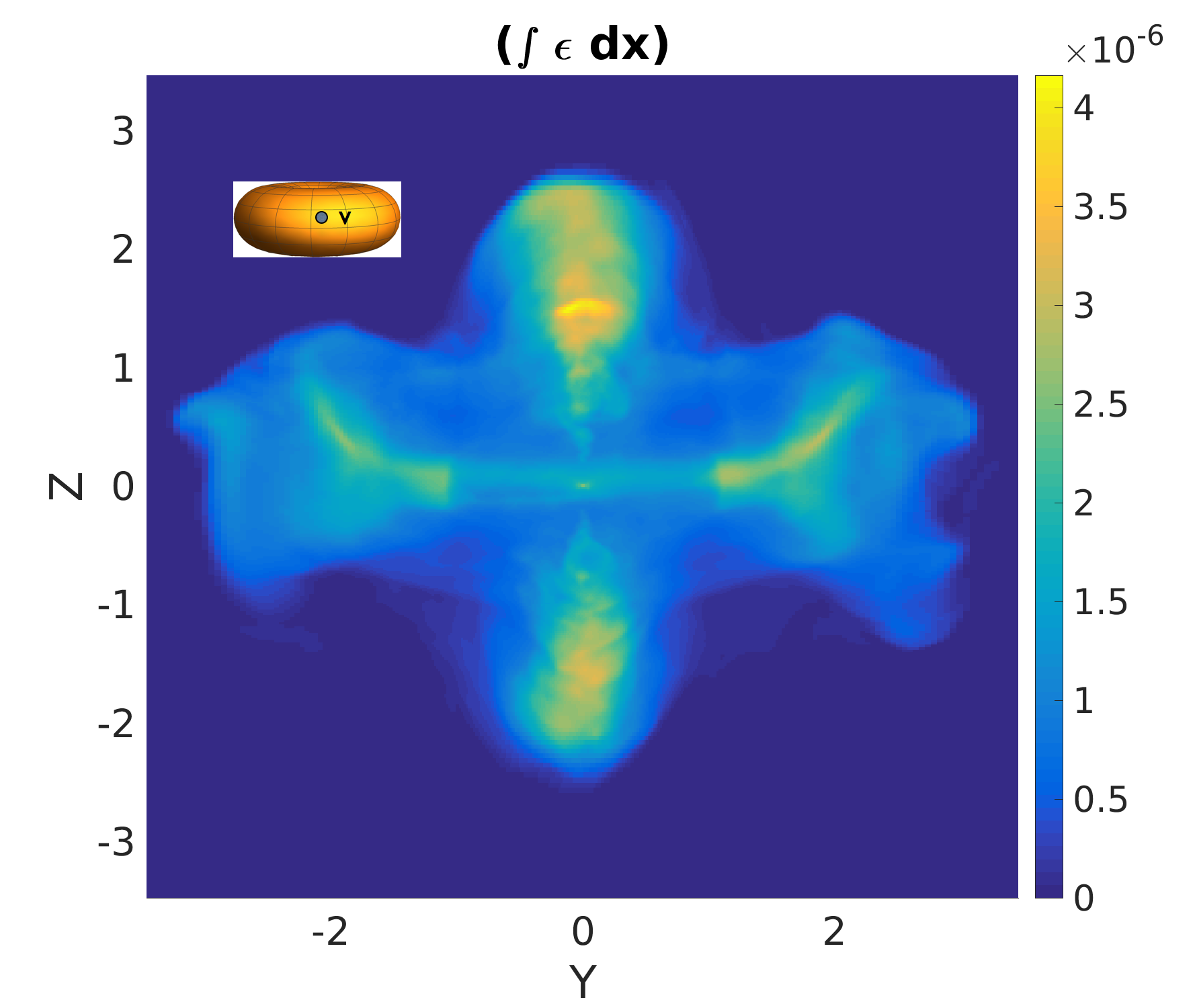}
\includegraphics[width=88mm,angle=-0]{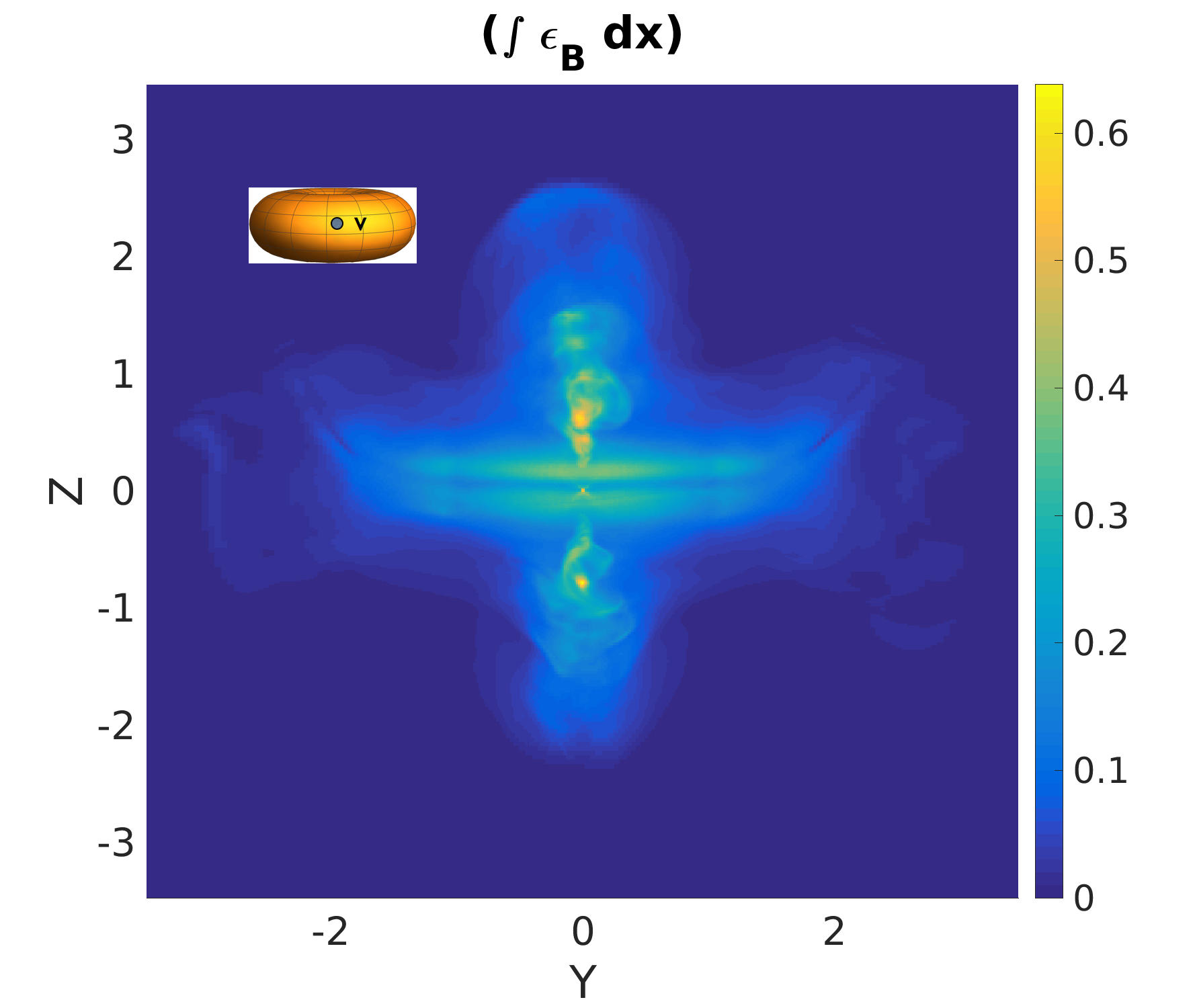}
\includegraphics[width=88mm,angle=-0]{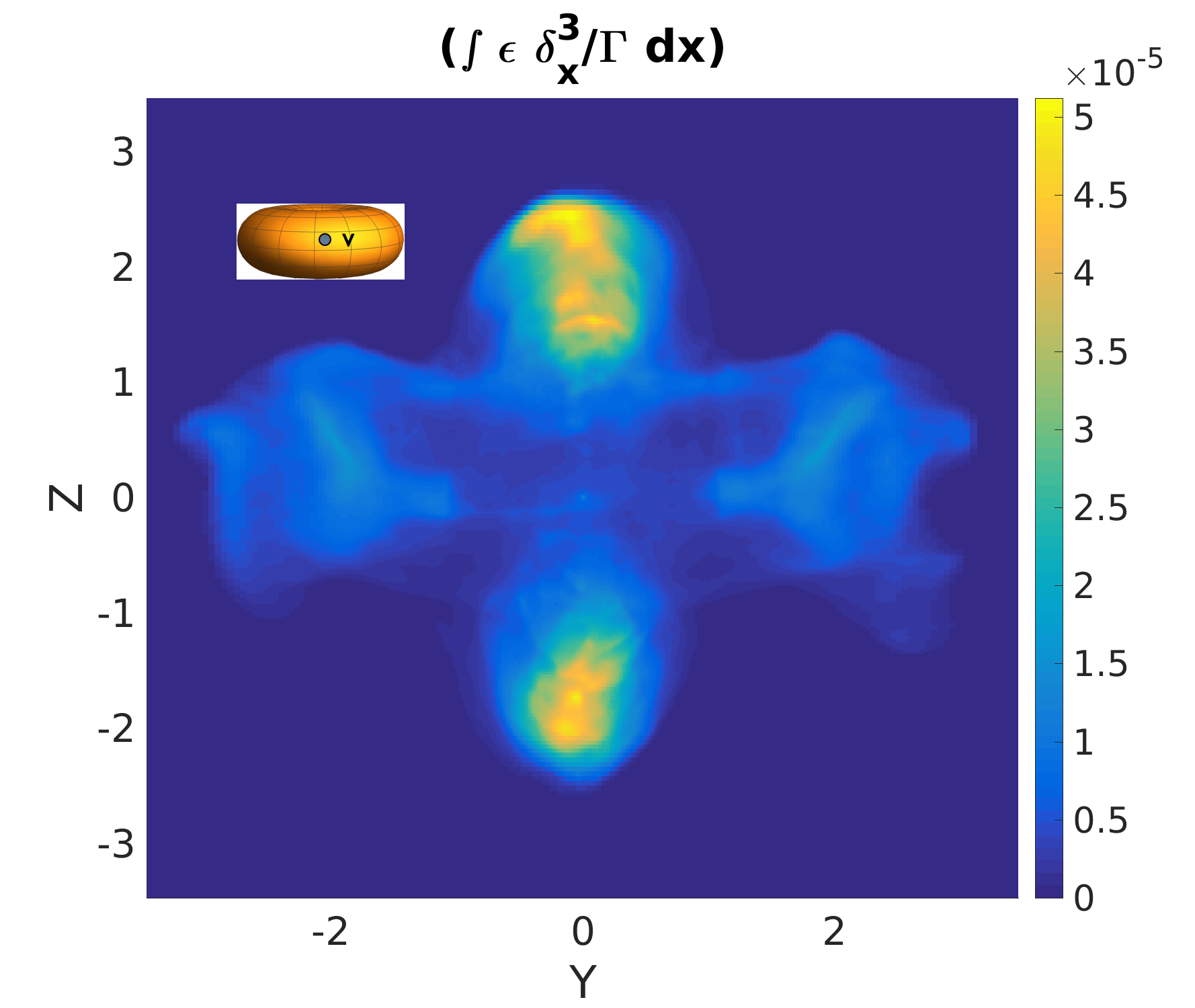}
\includegraphics[width=88mm,angle=-0]{figures/f_m_s1_eint_surf_X_Ebd_0018.png}
\caption{Emissivity maps projected along X axis for the model fs1a45:
top left IC (no Doppler boosting), top right synchrotron (no Doppler boosting), bottom left IC and bottom right synchrotron. The morphology of the \Rs is similar in all the cases.
The synchrotron maps shows a clear cross structure. Doppler boosting (bottom vs top panels) amplifies the jet-like structures and equatorial torus becomes less visible.}
\label{fig:X_f}
\end{figure*}
%fffffffffffffffffffffffffffffffffffffffffffffffffffffffffffffffff 

%fffffffffffffffffffffffffffffffffffffffffffffffffffffffffffffffff
\begin{figure*}
\includegraphics[width=88mm,angle=-0]{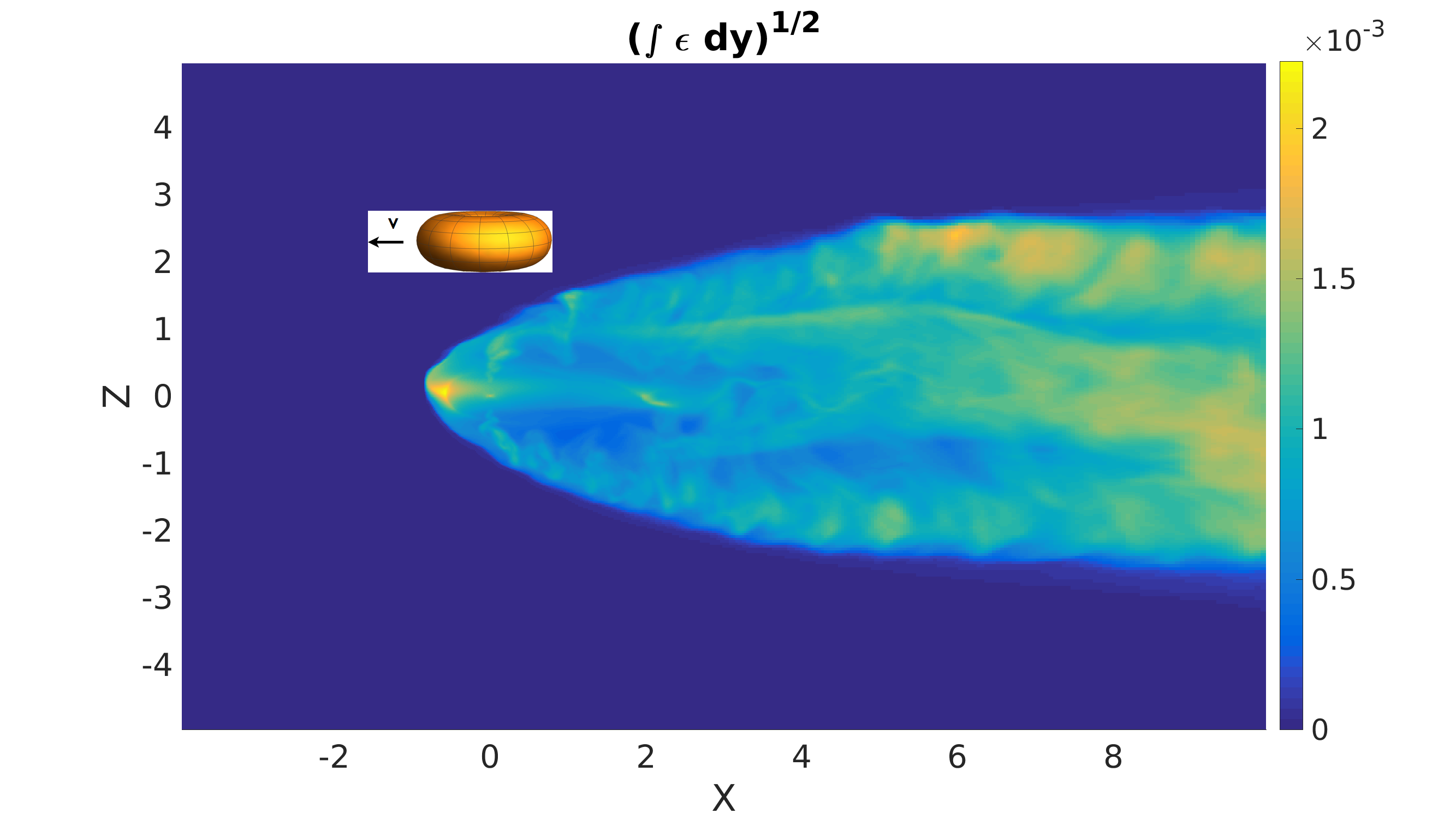}
\includegraphics[width=88mm,angle=-0]{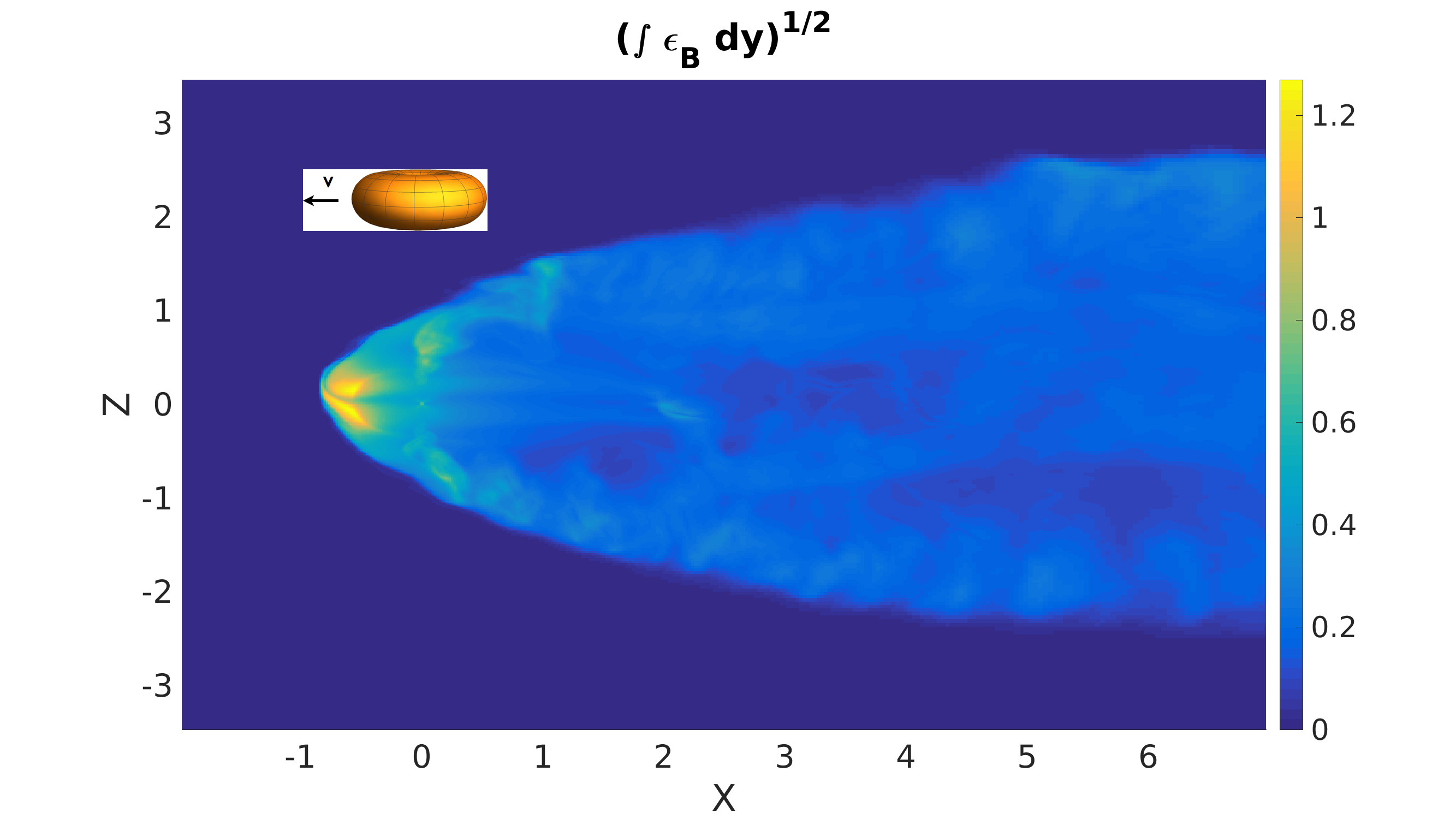}
\includegraphics[width=88mm,angle=-0]{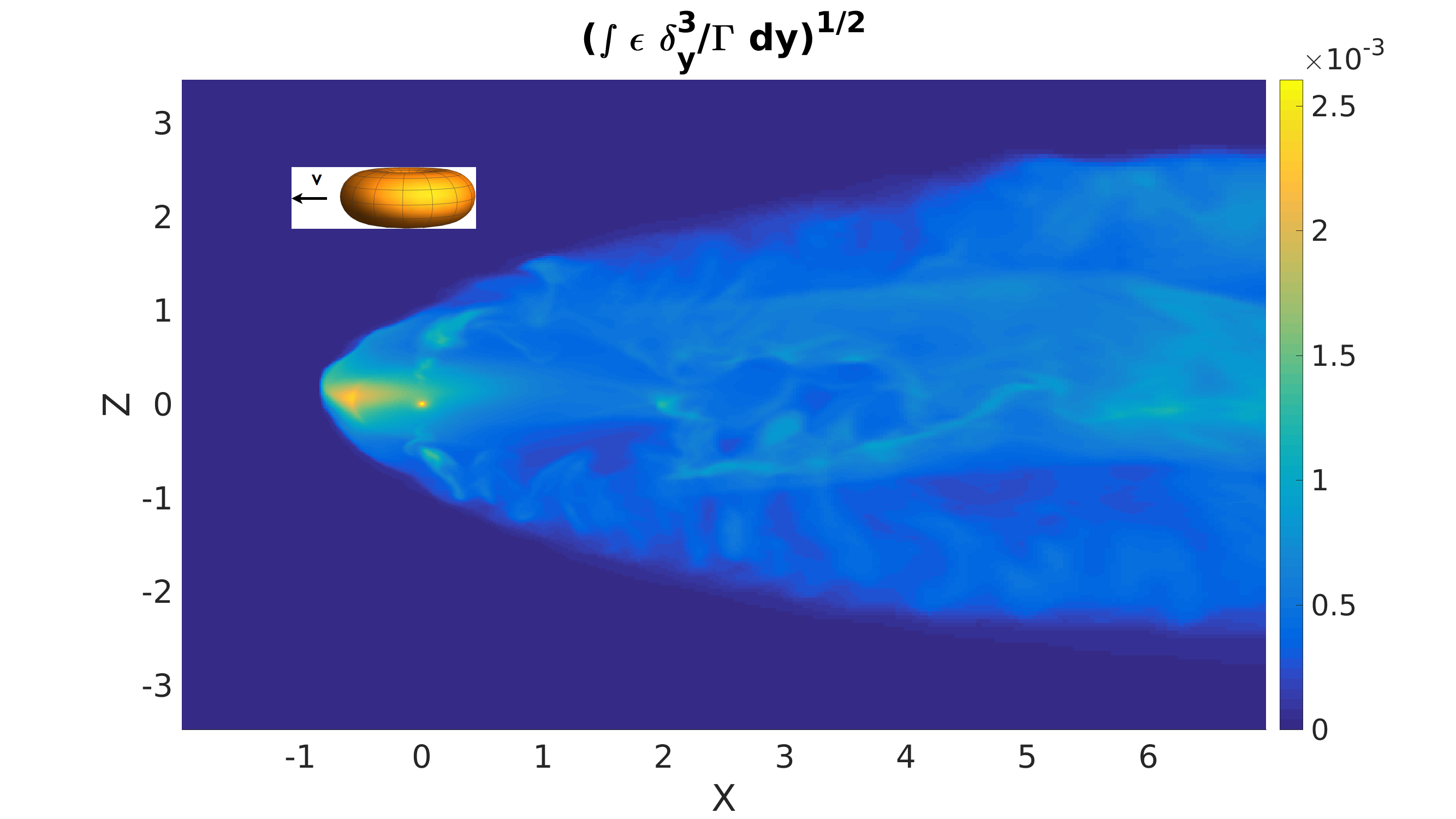}
\includegraphics[width=88mm,angle=-0]{figures/f_m_s1_eint_surf_Y_Ebd_0018.png}
\caption{Emissivity maps projected along Y axis for the model fs1a45 (\ac{fb}):
top left IC (no Doppler boosting), top right synchrotron (no Doppler boosting), bottom left IC and bottom right synchrotron.
As in the Fig.~\ref{fig:X_f} the morphology of the \R is similar in all the cases. The equatorial torus forms a bright head part and 
brightened structure. The jet-like structures are perpendicular to the direction of the pulsar motion. 
The synchrotron maps shows more explicit jets structure. Doppler boosting (bottom vs top panels) amplifies equatorial torus near the bow-shock head.}
\label{fig:Y_f}
\end{figure*}
%fffffffffffffffffffffffffffffffffffffffffffffffffffffffffffffffff

%fffffffffffffffffffffffffffffffffffffffffffffffffffffffffffffffff
\begin{figure*}
\includegraphics[width=88mm,angle=-0]{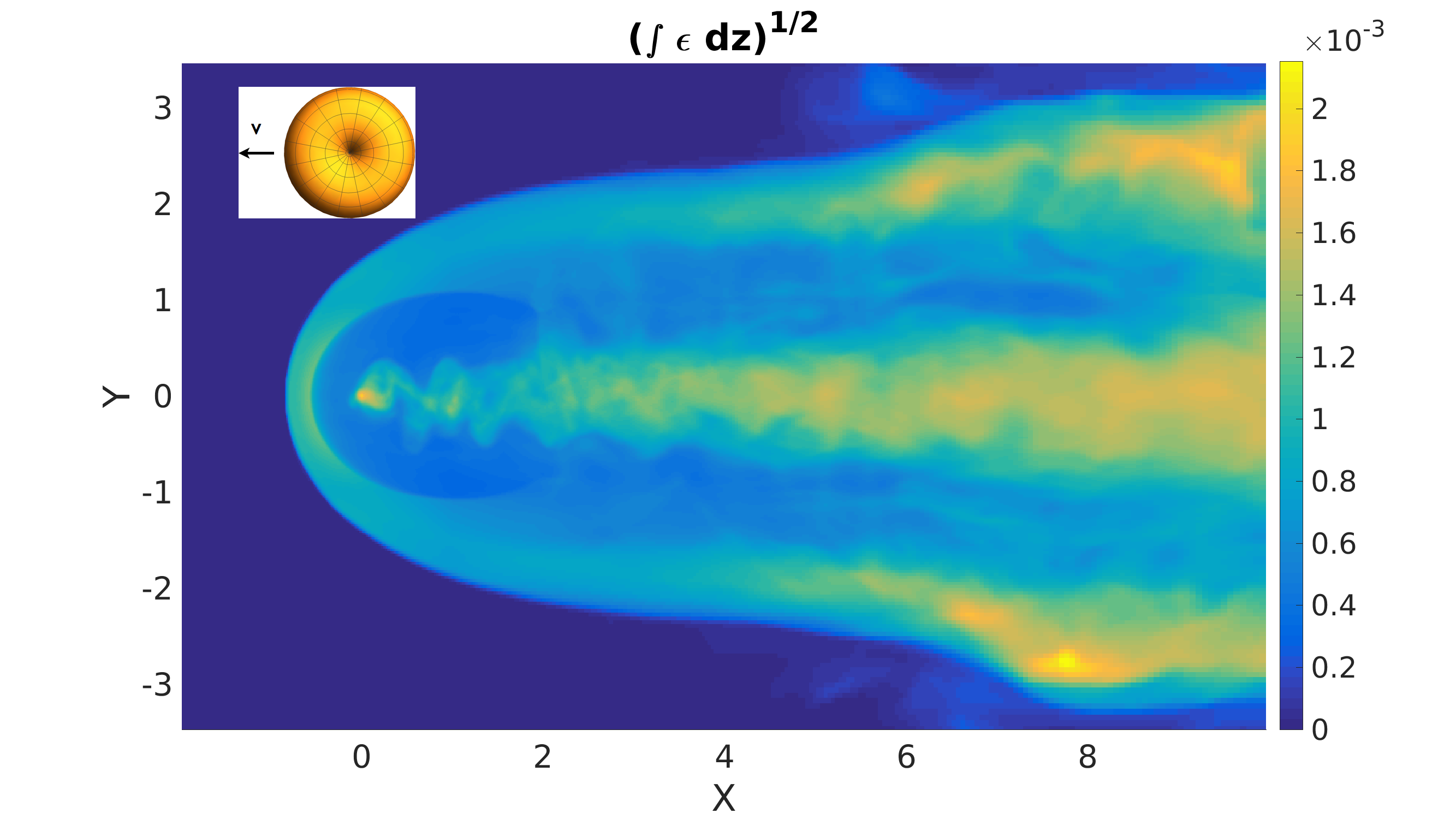}
\includegraphics[width=88mm,angle=-0]{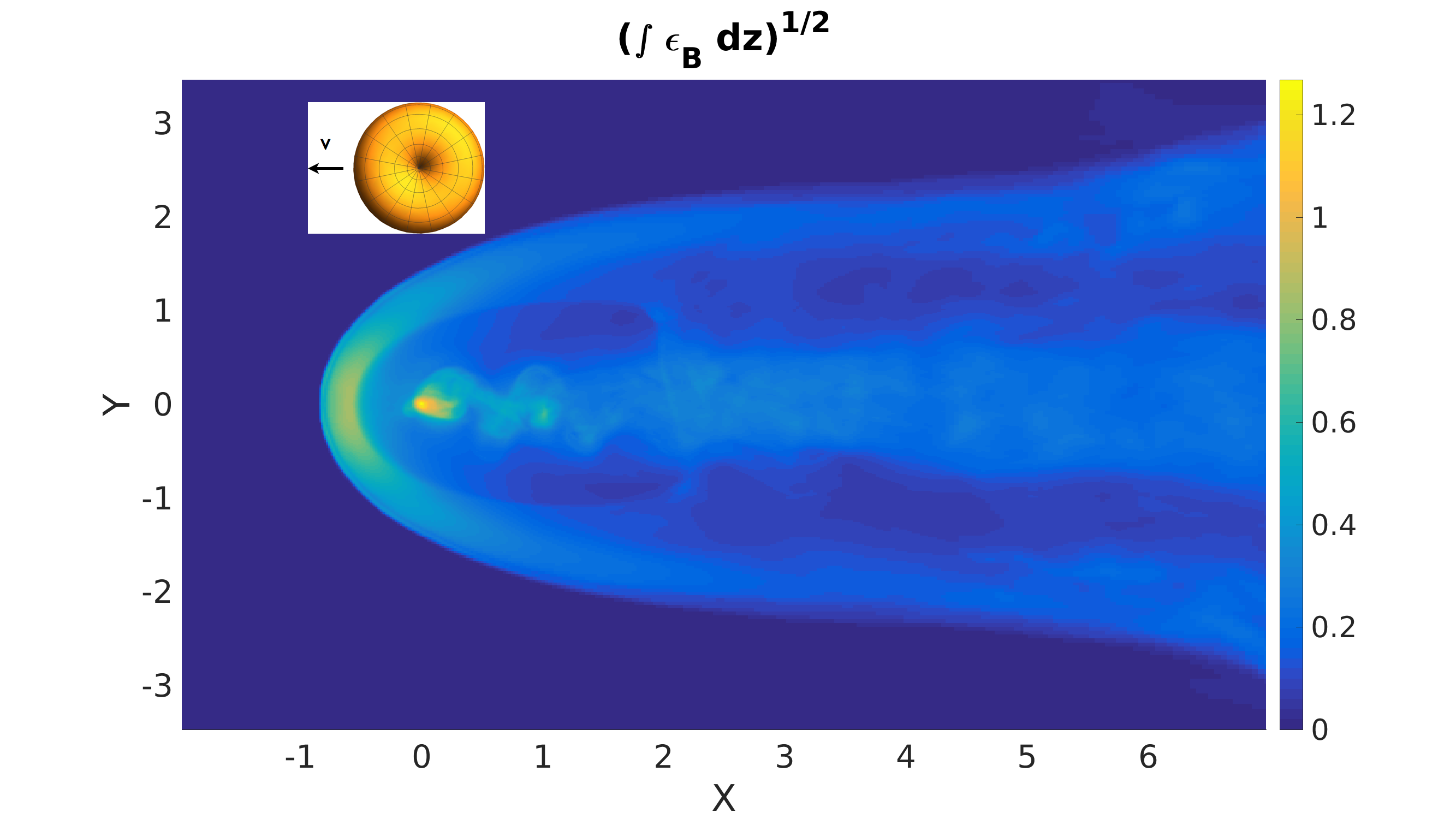}
\includegraphics[width=88mm,angle=-0]{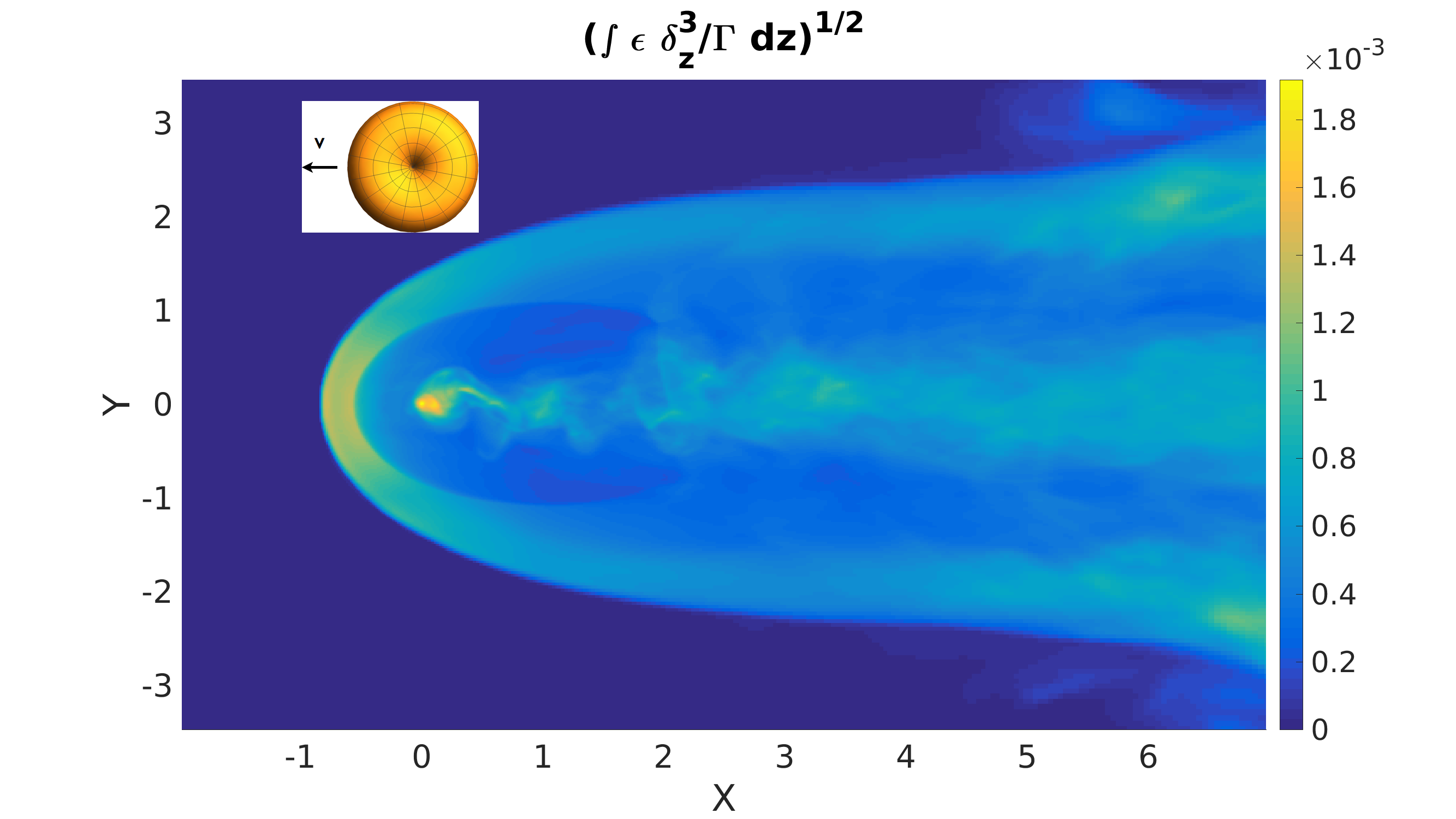}
\includegraphics[width=88mm,angle=-0]{figures/f_m_s1_eint_surf_Z_Ebd_0018.png}
\caption{Emissivity maps projected along Z axis for the model fs1a45 (\ac{cw}):
top left IC (no Doppler boosting), top right synchrotron (no Doppler boosting), bottom left IC and bottom right synchrotron.
As in the Fig.~\ref{fig:X_f} the morphology of the \R is similar in all the cases. The pulsar wind zone and its \TS are clearly visible, later they form wide cometary tail. Also jet-like structure originates from the pulsar and form relatively narrow tail.  
The synchrotron maps show a more explicit jet structure. Doppler boosting (bottom vs top panels) amplifies narrow tail 
and suppresses the head and wide tails.}
\label{fig:Z_f}
\end{figure*}
%fffffffffffffffffffffffffffffffffffffffffffffffffffffffffffffffff

The most efficient radiation channel for production of \xray emission in \Rs is synchrotron radiation, thus synchrotron
emissivity maps, computed accounting for the Doppler boosting effect, should be compared to observations. However, for
sake of completeness, we also provide emissivity maps for synchrotron without Doppler boosting, and \ac{ic} maps with
and without Doppler boosting.

The \ac{ic} cooling time can be estimated  as
\be 
t_{\textsc{ic}} =\frac{3 m_e c}{4 \sigma_T u_{\rm CMB} \gamma_e}=
7\times10^{12} \, \gamma_{\rm w,7}\; {\rm s},
\label{eq:tic}
\ee
here $u_{\rm CMB}$ is \ac{cmb} photon energy density. The energy of \ac{ic} photons for electrons with Lorentz factor $10^7$ is $\sim100$~GeV \citep[see, e.g.,][]{2014ApJ...783..100K}.
 \ac{ic} photons with energy about 1~keV are produced by electrons with Lorentz factor $\sim 1000$, which have cooling time of $\sim 10^{17}$~s.
Using the approach outlined in Appendix \ref{ap:syn}, we calculate the  local synchrotron emissivity. For calculation of \ac{ic} emission we follow a similar procedure using \ac{ic} cooling time Eq.~\eqref{eq:tic} instead of synchrotron one Eq.~\eqref{eq:tsyn}  \citep[we note, however, that there could be minor differences due to different Doppler boosting patterns for the synchrotron and \ac{ic} emission, e.g.,][]{2018MNRAS.481.1455K}.
The $t_{\textsc{ic}}$ does not depends on flow properties (except a small enhancement due to the Doppler boosting), so in the frame of our model the \ac{ic} maps for different energy band are the same 
with the only difference in the normalization factor. 

The comparison of the four emissivity maps for the case fs1a45 projected on X axis (pulsar moves towards us) are 
presented in Fig.~\ref{fig:X_f}, projected on Y axis (pulsar moves to the left) are 
presented in Fig.~\ref{fig:Y_f}, and projected on Z axis (pulsar moves to the left) are 
presented in Fig.~\ref{fig:Z_f}. 
As we can see for the same viewing angle, in general, the morphology is similar in all maps, but we see significant difference in details.
The synchrotron maps feature a brighter jet-like structure as compared to the \ac{ic} case.
The effect of the Doppler boosting reduces the brightness of the equatorial torus and pulsar tail (except head part of the equatorial torus).
Due to fast dissipation (numerical effect) of magnetic energy in the pulsar wind tail the intensity of synchrotron radiation on the synchrotron maps drops down 
significantly faster as compared to the \ac{ic} maps. Magnetic field dissipation not only decreases the \Bf strength but also pumps energy
to  particles as well, that makes tale in the \ac{ic} maps to be brighter.

In conclusion, the synchrotron losses in the bow-shock \Rs dominate over \ac{ic} losses due to (i) 
 the \Bf in the head part of the fast moving pulsars is considerably higher than for stationary ones - this is due to the fact that large ram pressure ($M \gg 1$) leads to smaller  scales of the \TS; (ii) high energy  particles are quickly advected out into the tail - this leads to a quick depletion of \ac{ic} scatters. In contrast,  for stationary pulsars  non-radiatively-cooling lower energy particles are stored in the \Rs producing intense \ac{ic} signal on the synchrotron target.

\section{Study of anisotropy of the pulsar wind} 
\label{ap:sin4}

\begin{figure*}
\includegraphics[width=88mm,angle=-0]{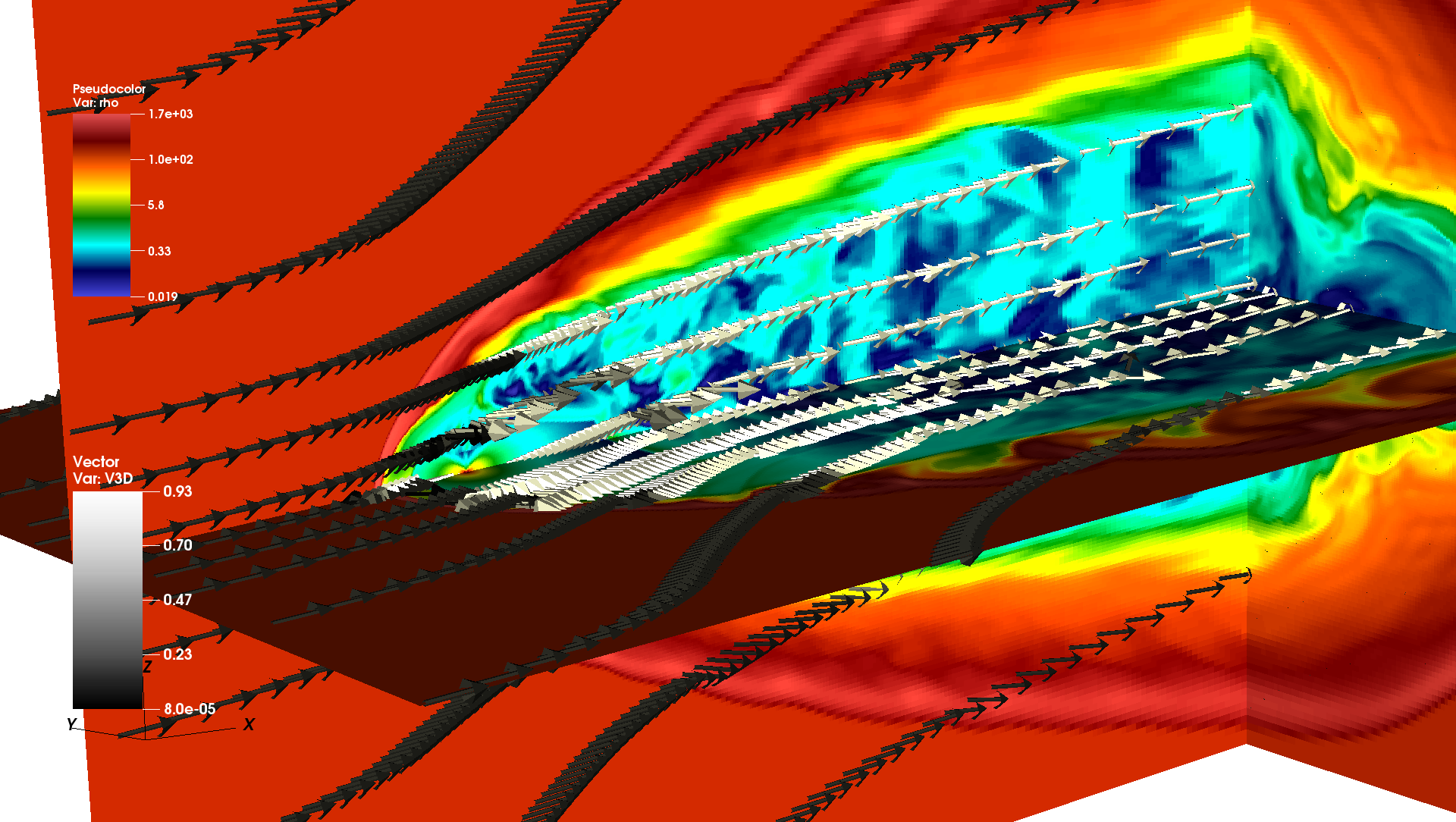}
\includegraphics[width=88mm,angle=-0]{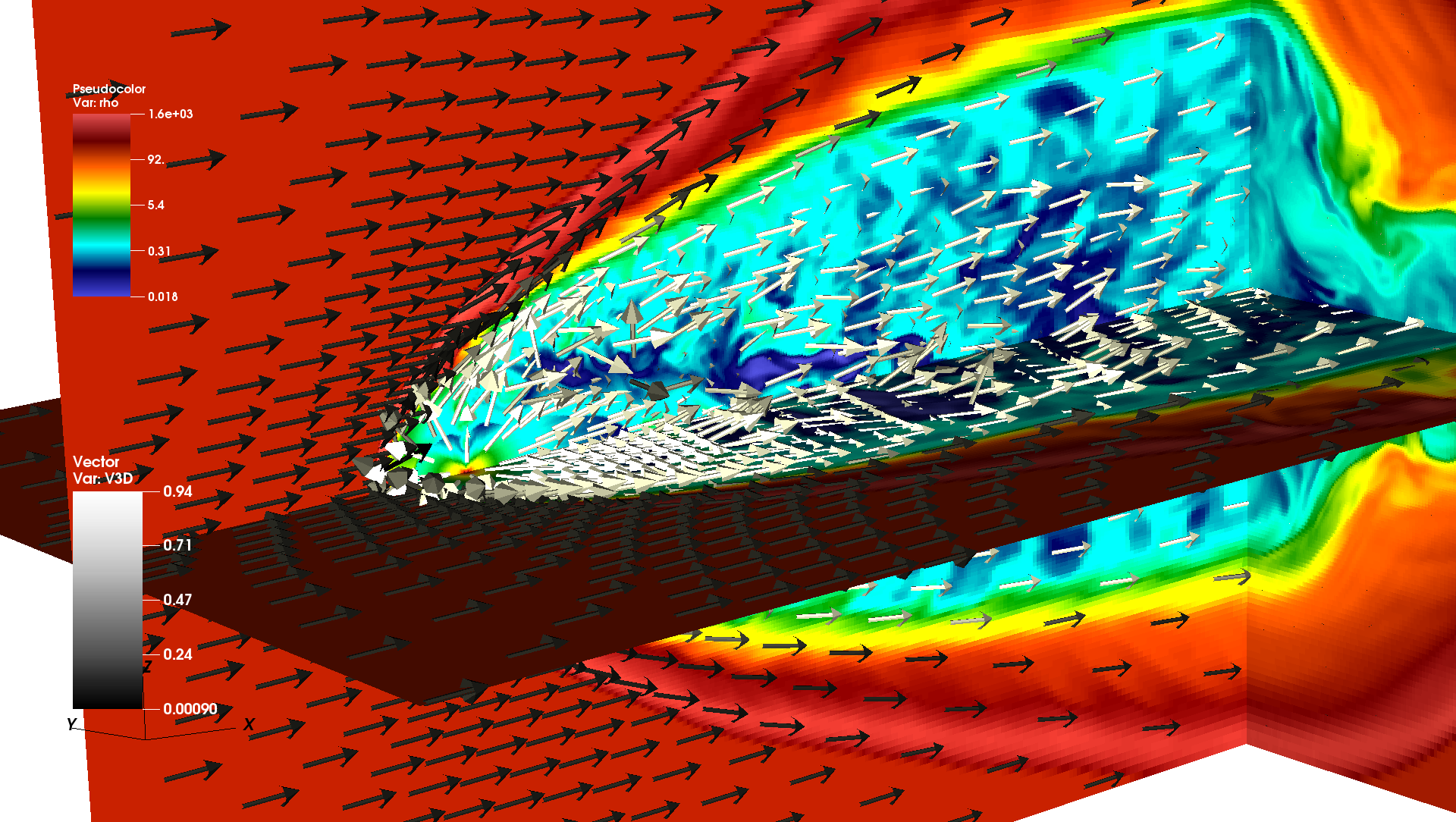}
\caption{\Acl{fb}/\Acl{cw} configuration, models fs1a45 with energy density distribution proportional to $\sin^4\theta_p$ (left panel) and $\sin^2\theta_p$ (right panel). 
\ac{3d} rendering of plasma density  logarithm by color and  velocity field by arrows.}
\label{fig:sin4rho}
\end{figure*}

The recent studies of the formation of pulsar winds favor different polar angle dependence of energy flux distribution in the wind 
\citep[see \eg][]{2013MNRAS.435L...1T,2016MNRAS.457.3384T}. While usually one arguers for a \(\propto\sin^2\theta_p\) 
dependence of the energy flux \citep{bogovalov_99,2002MNRAS.336L..53B}, a significantly sharper dependence, $\propto sin^4\theta_p$, 
cannot be excluded \citep[see \eg][]{2016MNRAS.457.3384T}. In this appendix we present the results obtained adopting a pulsar wind with this sharp dependence 
of the energy flux. As a base case we take a \ac{fb}/\ac{cw} configuration (fs1a45) and change the power inEq.~\ref{eq:wpow} from 2 to 4. 
%{\bf (do you also change the power \(L_0\) by 1.25 to?)  BMV: NO, equatorial pressure was the same, TS stay at the same radius.}. 
The simulation result is shown in Fig.~\ref{fig:sin4rho}. As one can see, the dependence of the energy flux on the polar angle has a weak impact 
on the morphology in general. However,  the jet-like structure is significantly less pronounce in the case with the $\sin^4\theta_p$ dependence. 
From the \mhd point of view the flow, obtained for the $\sin^4\theta_p$ dependence of the energy flux and wind magnetization of \(\sigma_0=1\), 
appears to be in between of two considered cases for the $\sin^2\theta_p$ dependence of the energy flux: fs1a45 and fs01a45, which differ 
by the wind magnetization (\(\sigma_0=1\) and \(0.1\), respectively).

\begin{figure*}
\includegraphics[width=58mm,angle=-0]{figures/f_m_s01_eint_surf_X_Ebd_0076.png}
\includegraphics[width=58mm,angle=-0]{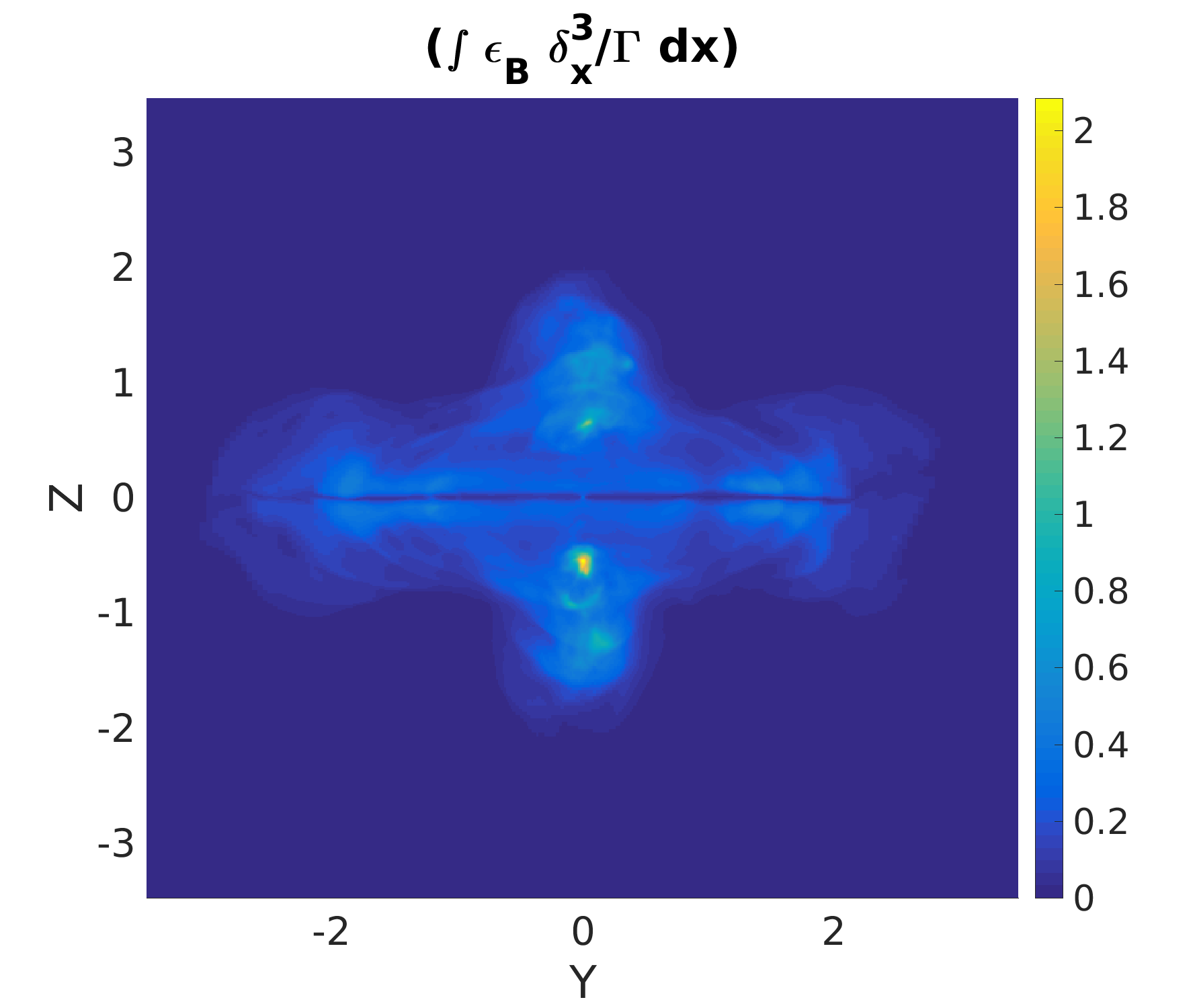}
\includegraphics[width=58mm,angle=-0]{figures/f_m_s1_eint_surf_X_Ebd_0018.png}
\includegraphics[width=58mm,angle=-0]{figures/f_m_s01_eint_surf_Y_Ebd_0076.png}
\includegraphics[width=58mm,angle=-0]{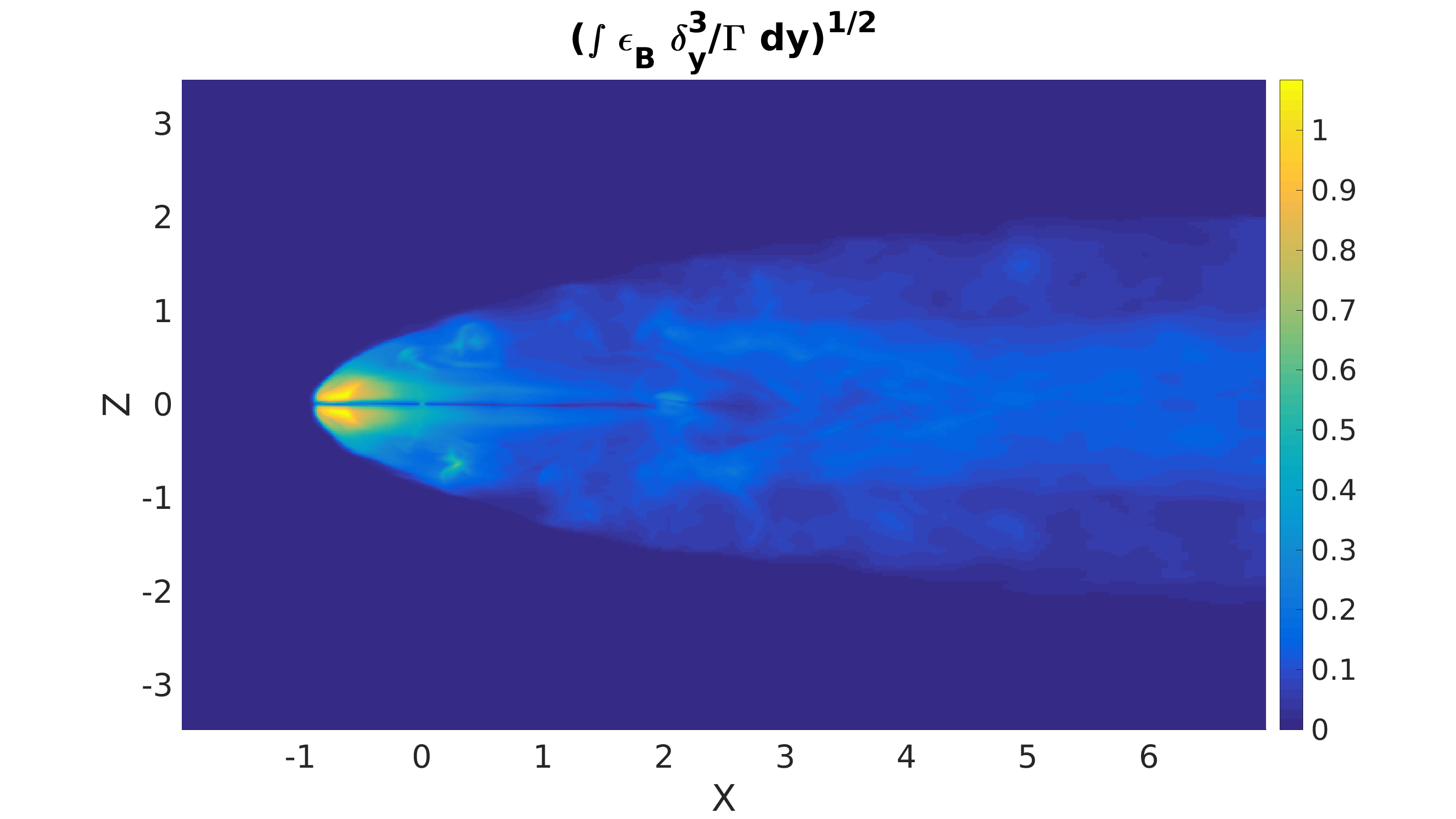}
\includegraphics[width=58mm,angle=-0]{figures/f_m_s1_eint_surf_Y_Ebd_0018.png}
\includegraphics[width=58mm,angle=-0]{figures/f_m_s01_eint_surf_Z_Ebd_0076.png}
\includegraphics[width=58mm,angle=-0]{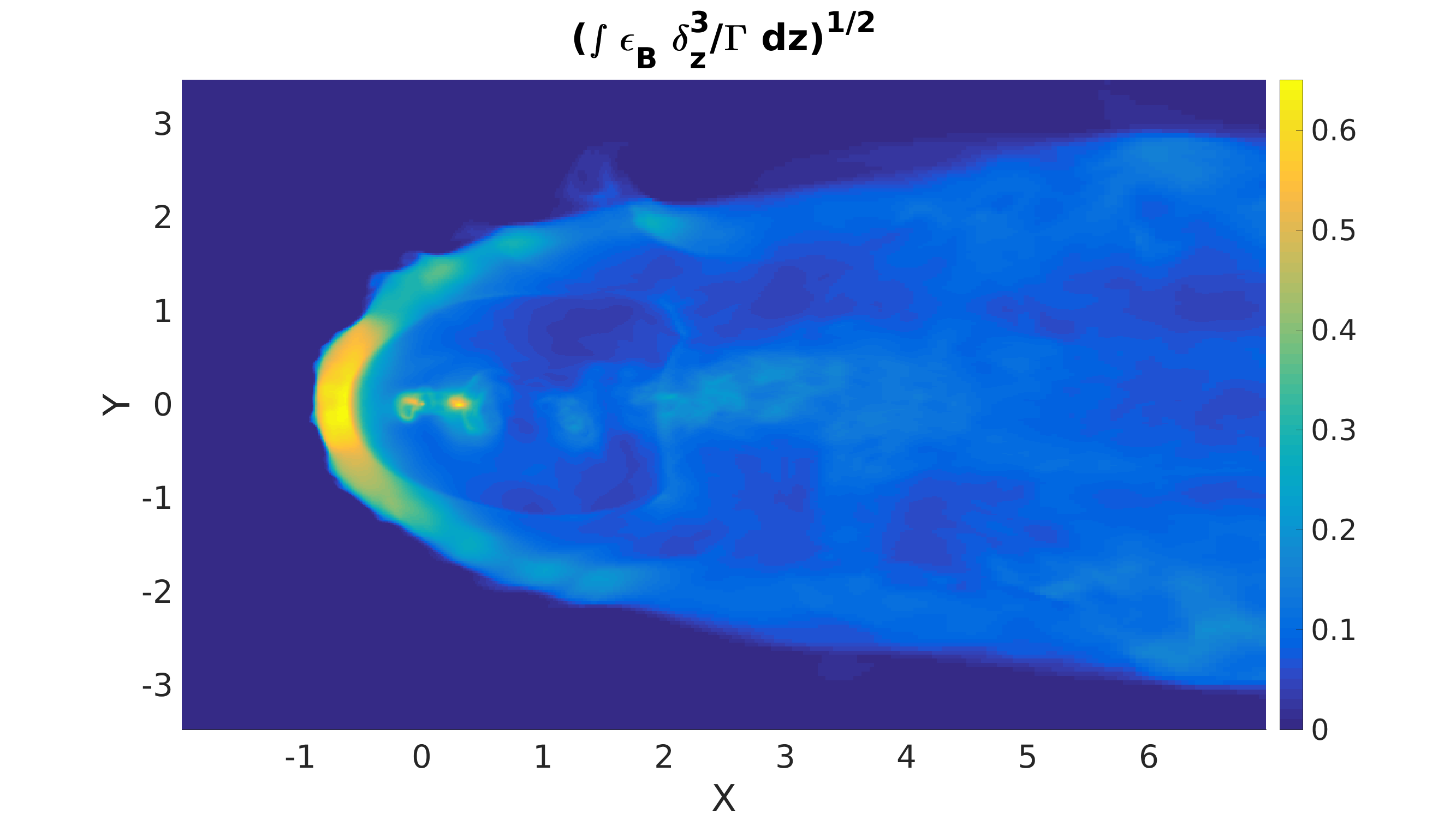}
\includegraphics[width=58mm,angle=-0]{figures/f_m_s1_eint_surf_Z_Ebd_0018.png}
\caption{\ac{fb}/\ac{cw} configuration, models fs01a45 (left panels), fs1a45 with energy density distribution proportional to $sin^4\theta_p$ (central panels) and $sin^2\theta_p$ (right panels). 
Synchrotron   emissivity map   projected along X axis (top raw), Y axis (middle raw), and  Z axis (bottom raw).}
\label{fig:sin4em}
\end{figure*}

A comparison of synthetic synchrotron emissivity maps for the case of the sharp dependence of the energy flux with two
benchmark cases is shown in Fig.~\ref{fig:sin4em}. As we have already inferred from the similarity of the \mhd
structures, models show similar morphology, with the most remarkable change in the plum: for the same wind
magnetization, the jet-like structure is less prominent in the case of the $\sin^4\theta_p$ dependence. A comparison of
the emissivity maps suggests that a change of the wind magnetization has a similar impact on the synchrotron morphology
as a change of the energy flux dependence. Thus, given that values of these parameters are highly uncertain from the
theoretical point of view, we do not perform simulations with the $\sin^4\theta_p$ dependence for other considered
models. We also note that a smaller inclination of the pulsar magnetic moment, \(\alpha\), may have a similar influence
on the morphology.

%fffffffffffffffffffffffffffffffffffffffffffffffffffffffffffffffff
%fffffffffffffffffffffffffffffffffffffffffffffffffffffffffffffffff

%fffffffffffffffffffffffffffffffffffffffffffffffffffffffffffffffff
%fffffffffffffffffffffffffffffffffffffffffffffffffffffffffffffffff

\section{Magnetic field in the head part of the ``Riffle Bullet'' configuration -  inside magnetic draping}
\label{Bfhead}

Above, in \Sec\ref{Kompaneets}, we treated the shape of the \CD under the thin shell approximation. In fact,
  the thickness of the shocked   pulsar
wind  can be a large fraction of the stand-off distance:   the post \TS velocity $v_s$ is mildly  relativistic,
$v_s =c/3$, and the flow has to expand  considerably before  it can be matched
to any non-relativistically  moving (or stationary in the pulsar frame)
\CD.  Thus, the
 pulsarsheath has a large
width.
Inside the sheath the pulsar wind is  slowed down and diverted ``sideways''
by pressure forces.

Let us consider analytically the structure of the magnetic field in the head part of pulsar wind when 
the pulsar is in the  \ac{rb} configuration - the flow is then axially symmetric.  We will calculate the structure of the \Bf\ for small magnetization, $\sigma \ll 1$. 
In this approximation the magnetic field can be treated kinetically, neglecting its influence on the flow parameters. We will demonstrate that however small the \Bf\ is, there is also a highly magnetized layer near the contact discontinuity - inside magnetic draping.

\begin{figure}
\includegraphics[width=0.97\linewidth]{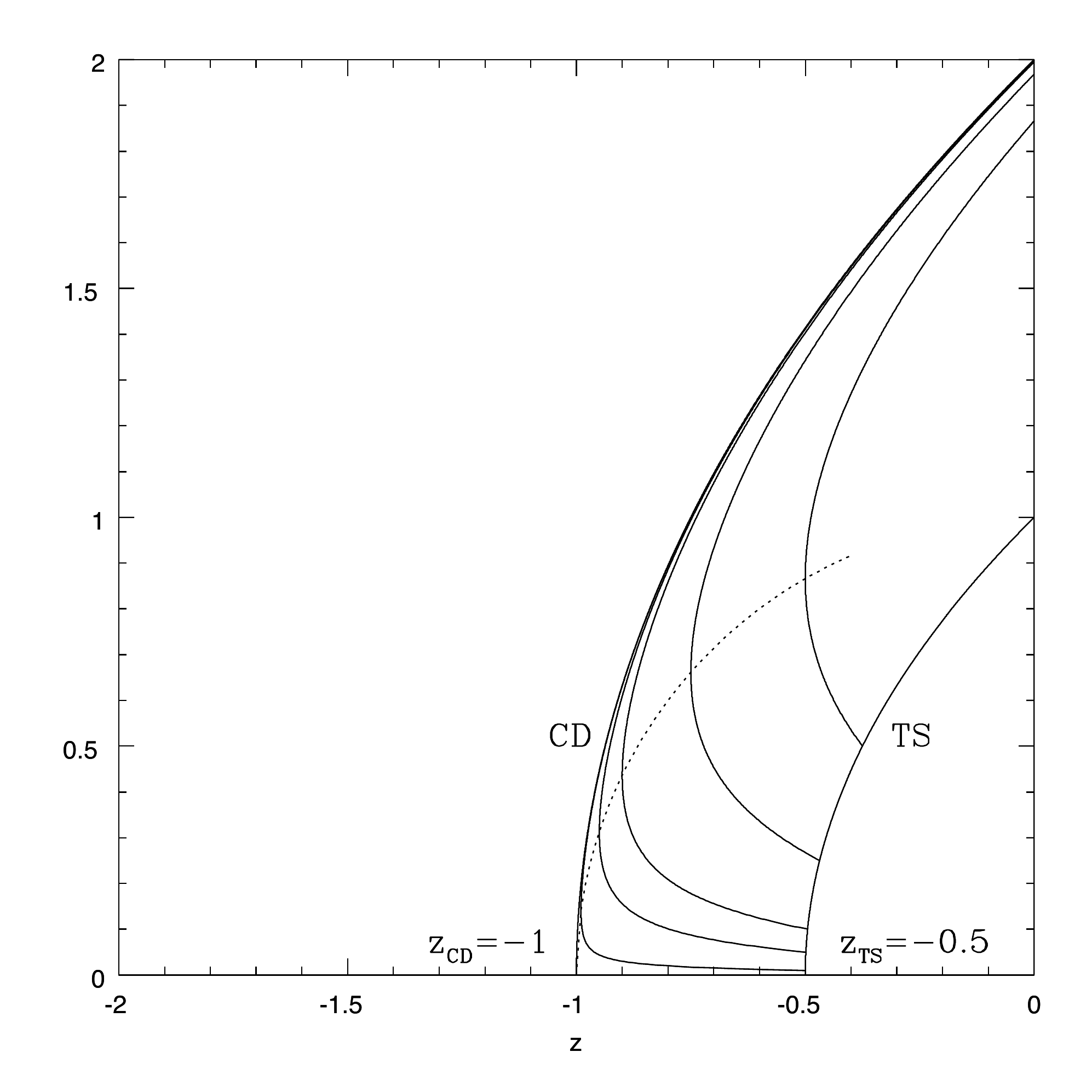}
\caption{Self-similar hyperbolic
 flow in the head part of bow-shock \Rs. Apex of the \CD is at \mbox{$z=-z_\cd=-1$},
apex of the \TS is chosen tom be at \mbox{$z=-z_\tsh = -0.5$}. The curves are parameterized
by \mbox{$\nu_s =0.01$, $0.05$, $0.1$, $0.25$, and $0.5$} (from bottom to top). Self-similar model
provides a good description for the unmagnetized  flow close to the axis. 
Dashed line marks    the inside draping region  beyond which dynamical effects of  \Bf cannot be neglected.}
\label{head}
\end{figure}

Let us illustrate the magnetized  dynamics of the pulsar wind in the head part by the following 
kinematic model.
We assume that the shapes of the \CD and \TS
are parabolic with the focus  on the pulsar and the vertex  located
at $z=-z_\cd$ and $z=-z_\tsh$, respectively (see in Fig.~\ref{head}). We first  construct  a
model of the flow of incompressible irrotational fluid between 
the \CD and \TS, and then add kinematically (neglecting its dynamical influence) a toroidal magnetic field. Since magnetic field is compressed in the subsonic
flow, we estimate  a distance at which
magnetic field becomes dynamically  important and our kinematic approximation
becomes invalid. This procedure is expected to be valid close to the symmetry axis
of the flow, where wind magnetization is low.

Let us introduce parabolic coordinates $\nu, \mu, \phi$ defined as
%u = \mu
%v = \nu
\be 
x= \mu  \nu \cos \phi, \hskip .3 truein
y=\mu  \nu  \sin \phi, \hskip .3 truein
z= \frac{\nu^2 - \mu^2 }{ 2}
\ee
In this coordinates the \CD is located
at $\mu_\cd = \sqrt{2 z_\cd} $ and the \TS is at $\mu_\tsh = \sqrt{2 z_\tsh}$, 
Fig.~\ref{head}.
Assuming that the flow is incompressible and  irrotational,
\be
{\rm div} \v =0 
, \hskip .3 truein
\curl \v =0,
\ee
and the flow is axisymmetric,
  one can introduce flow surfaces $P(\mu,\nu)=$constant, so that velocity is given by
  \be
\v = \frac{ \nabla P \ts \bm{e}_\phi }{ \sqrt{g}}
\label{P}
\ee
where $g = \mu^2 \nu^2 (\mu^2 + \nu^2)  $ is the determinant of the 
metric tensor. 
Explicitly,
\be
v_\mu = \frac{ \partial_\nu P }{ \mu \nu \sqrt{\mu^2 + \nu^2} }
, \hskip .3 truein
v_\nu =-  \frac{ \partial_\mu P }{ \mu \nu \sqrt{\mu^2 + \nu^2} }
\ee
The condition of incompressibility  is  then satisfied automatically, while
 the condition of  irrotational flow gives
\be 
\mu 
\partial_\mu \left( \frac{ \partial_\mu P }{ \mu} \right) 
+ 
\nu
\partial_\nu \left( \frac{ \partial_\nu P }{ \nu} \right) =0
\ee
The boundary conditions require that the component of the
velocity normal to the \CD be zero, and that on the axis the 
velocity is along $\mu$ direction:
\be
\partial_\nu P
\big|_{\mu=\mu_\cd} =0
, \hskip .3 truein
\partial_\mu\big|_{\nu =0}=0
\ee
In addition, the velocity on the \TS should be found from the 
oblique shock conditions for relativistic pulsar wind.

Looking for self-similar solutions
$P=U(\mu) V(\nu)$ we find
that general solutions can be represented as a sum over Bessel
functions
$
U, V
\propto J_1
$. For a given form of the \CD and the assumed radial pulsar wind
we can then find the velocity at the \TS. Expansion of this velocity
 in terms of functions $U$ and $V$ will then give a complete solution to the problem.

Instead deriving  a complete solution according to the above-described procedure,  we will make a simplifying assumption that the  post-shock
velocity is some  given function (not found from the
shock polar)  and illustrate the flow pattern and magnetic field evolution
in this case. 
As a simplest case we  chose
\be
\partial_\mu \left( \frac{ \partial_\mu P }{ \mu} \right) =
\partial_\nu \left( \frac{ \partial_\nu P }{ \nu} \right) =0
\ee
Then, if at the apex of the \TS the post-shock flow velocity is $v_s$, we find
\be
P = v_s \nu^2 \mu_\tsh \frac{ 1 - \mu^2 / \mu_\cd^2 }{ 2( 1 - \mu_\tsh^2 / \mu_\cd^2)}
\ee
The streamlines  are then given
by 
\be
\frac{\dif{\mu} }{ v_\mu} = \frac{\dif{\nu} }{ v_\nu}
\ee
which can be integrated to give
\be
\nu(\mu) = \nu_\tsh  \sqrt{  \frac{1- \mu_\tsh^2 / \mu_\cd^2 }{ 1- \mu^2 / \mu_\cd^2 }}
\label{nu}
\ee
where $\nu_\tsh$ is a value of the variable $\nu$ on the \TS 
($\nu_\tsh$ parameterizes different streamlines), see Fig.~\ref{head}. 

Next we add kinematically  a toroidal \Bf. We assume that  at the \TS
the  \Bf is weak and its influence on dynamics can be neglected. 
From the conservations of magnetic flux
$
\curl ( \v \ts \B)= 0
$ 
we find
\be
B_\phi = \Phi(P) \mu \nu 
\ee
where 
$ \Phi(P)$ is a magnetic flux function that parameterizes spacial 
dependence of the \Bf. 
Using the expression for streamlines 
we can then find
how  \Bf evolves along any given streamline:
\be
B_\phi = \Phi(P) \mu  \nu_\tsh  \sqrt{\frac{  1- \mu_\tsh^2 / \mu_\cd^2 }{ 1- \mu^2 / \mu_\cd^2 }}
\ee
Which immediately shows that  \Bf diverges  close to the \CD, 
$\mu \rightarrow \mu_\cd$.
As a function of spatial coordinates
 \Bf diverges at the turnaround point given by
 \be
 \begin{split}
\nu &= \sqrt{\nu_s} \left( \mu_\cd^2 - \mu_\tsh^2 \right)^{1/4},\\
%\hskip .3 truein
r &=\sqrt{x^2+y^2} = \mu_\cd^2 / 2 = {z_\cd },
\end{split}
\ee
see Fig.~\ref{head}.
 At this point the dynamic effects of the magnetic field on the flow
evolution cannot be neglected. 
In particular our assumption of incompressible, irrotational flow
will be broken.

Thus, we demonstrated  that however small magnetic field is in the pulsar 
wind, it will become dynamically important approximately half way
through the head part of \Rs. 
Close to the \CD magnetic field will dominated over the plasma pressure. 
On the \ISM side of the \CD similar effect will happen: magnetic field
will be compressed, so that {\it  on the \CD the pressure is communicated by the 
magnetic stresses on both sides}. 

This is an example of the so called  Cranfill effect \citep{cran71}; its  relativistic generalization
has served as a basis of the  
\cite{kc84} model of static \R; see also \cite{2002PhFl...14..963L,Lyutikovdraping}.
Qualitatively, magnetized wind creates \Ef $E_\theta \propto v_{w,r} B_{w,\phi}$ ($v_{w,r}$ and $ B_{w,\phi}$ are the corresponding components of the velocity and \Bf). In the steady case the curl of \Ef vanishes; hence $ v_{w,r} B_{w,\phi}  r=$constant. Since on the \CD the normal component of the velocity goes to zero, the \Bf is amplified. 
Thus, the pulsar wind becomes strongly magnetized inside the sheath  even if it was only
weakly magnetized at the \TS. This explains the formation of magnetized layer seen in low-sigma simulations, \citep{2005A&A...434..189B}.

\end{document}